\documentclass[a4paper,11pt]{article}
\usepackage{graphicx,epsfig,pstricks,fancybox,fancyhdr,epic,rotating,color}
\usepackage{amsfonts}

\usepackage{graphicx,epsfig,pstricks,fancybox,fancyhdr,epic,rotating,color}
\usepackage{amsfonts}

\begin{document}


\begin{center}

{\color{blue} \bf 
Canonical (anti-)commutation rules in QCD and unbroken gauge invariance

{\bf QCD - the two central local anomalies and canonical structure}
}
\vspace*{0.3cm}

\color{blue}
{\bf Peter Minkowski}
\\
{\bf Albert Einstein Center for Fundamental Physics - 
ITP, University of Bern}
\\
\vspace{0.5cm}

{\color{magenta} \bf
Abstract} 
\vspace*{0.1cm} \\

{\color{blue}
\begin{tabular}{l}
The regularities at large distances of complete gauge invariance 
\vspace*{-0.0cm} \\
in QCD
are shown to bear nontrivial consequences for the selection 
\vspace*{-0.0cm} \\
among inequivalent
representations of canonical commutation
\vspace*{-0.0cm} \\
(anticommutation) rules for gauge boson (quark) fields. 
\vspace*{-0.0cm} \\
{\color{cyan} The trace anomaly forces a modification of the
gauge boson}
\vspace*{-0.0cm} \\
{\color{cyan} Lagrangean
and by this of the entire associated canonical stucture .}
\end{tabular}
}
\vspace*{0.2cm}

{\color{green} 
Lectures prepared for the '1. IAS-CERN School on Particle Physics and
Cosmology and Implications for Technology, 9. to 31. January 2012,
Nanyang Executive Centre/NTU, Singapore .

}

\end{center}




{\color{blue} 

\vspace*{-0.3cm}
\begin{center}
{\color{red} List of contents}
\end{center}
\vspace*{0.0cm}

{\color{red}
\begin{tabular}{@{\hspace*{-0.8cm}}l@{\hspace*{-0.1cm}}l@{\hspace*{-0.5cm}}r}
1 & \begin{tabular}[t]{l}
Introduction
\end{tabular}
& ~\pageref{'1'}
\vspace*{-0.1cm} \\
2 & \begin{tabular}[t]{l}
How unique is a temporal axial gauge ?
\end{tabular}
& ~\pageref{'2'}
\vspace*{-0.1cm} \\
{\color{cyan} 2a} & \begin{tabular}[t]{l}
{\color{cyan} Integration of eq. \ref{eq:2-5} , interval by
interval}
\end{tabular}
& {\color{cyan} ~\pageref{'2a'}}
\vspace*{-0.1cm} \\
{\color{cyan} 2b} & \begin{tabular}[t]{l}
{\color{cyan}
Reparametrization of straight line by 
a tangent vector $\ ( \ \mbox{\underline{v}}^{\ \mu} \ ) \ $}
\end{tabular}
& {\color{cyan} ~\pageref{'2b'}}
\vspace*{-0.1cm} \\
{\color{cyan} 2c} & \begin{tabular}[t]{l} 
{\color{cyan}
QCD $\ {\cal{D}} \ ( {\cal{G}} \ ) \ - \ $
strings : 
building blocks of bi-local gauge}
\vspace*{-0.1cm} \\
{\color{cyan}
covariant parallel transports
$\ x \ \stackrel{C}{\leftarrow} \ y \ $,
relative to the irreducible
}
\vspace*{-0.1cm} \\
{\color{cyan}
representations
$\ {\cal{D}} \ \ \leftrightarrow \ x \ , \ \overline{{\cal{D}}}
\ \leftrightarrow \ y \ $ of the gauge group $\ {\cal{G}} \ $}
\end{tabular}
& {\color{cyan} ~\pageref{'2c'}}
\vspace*{-0.1cm} \\
3 & \begin{tabular}[t]{l} 
Canonical quantization in an axial gauge
\end{tabular}
& ~\pageref{'3'}
\vspace*{-0.1cm} \\
{\color{cyan} 3a} & \begin{tabular}[t]{l} 
{\color{cyan} Bare Lagrangean density
and equations of motion in}
\vspace*{-0.1cm} \\
{\color{cyan} unconstrained gauges}
\end{tabular}
& {\color{cyan} ~\pageref{'3a'}}
\vspace*{-0.1cm} \\
{\color{cyan} 3${\cal{A}}$} & \begin{tabular}[t]{l}
{\color{cyan} Canonical commutation rules for
gauge fields in axial gauges from}
\vspace*{-0.1cm} \\
{\color{cyan} bare Lagrangean density and residual fixed time gauge invariance}
\end{tabular}
& {\color{cyan} ~\pageref{'3A'}}
\vspace*{-0.1cm} \\
{\color{cyan} 3I} & \begin{tabular}[t]{l} 
{\color{cyan} Insertion : Energy momentum tensor density
as a conserved} 
\vspace*{-0.1cm} \\
{\color{cyan} generlized N\"{o}ther current , restricted 
to $\ \overline{{\cal{L}}} \ $ in the absence} 
\vspace*{-0.1cm} \\
{\color{cyan} of matter fields, i.e. neglecting 
$\ {\cal{L}}_{\ \left \lbrace q \right \rbrace} \ $} 
\end{tabular}
& {\color{cyan} ~\pageref{'3A-I'}}
\vspace*{-0.1cm} \\
3C & \begin{tabular}[t]{l} 
Remarks and consequences arising from
derivations in the
\vspace*{-0.1cm} \\
last subsection : {\color{cyan} 3I}
\end{tabular}
& ~\pageref{'3C-I'}
\vspace*{-0.1cm} \\
{\color{cyan} 3C1} & \begin{tabular}[t]{l} 
{\color{cyan}
Equations of motion and canonically conjugate variables}
\vspace*{-0.1cm} \\
{\color{cyan} pertaining to
$\ \overline{{\cal{L}}} \ +
\ {\cal{L}}_{\ \left \lbrace q \right \rbrace} \ $ 
}
\end{tabular}
& {\color{cyan} ~\pageref{'3C-1'}}
\vspace*{-0.1cm} \\
{\color{cyan} 3CD} & \begin{tabular}[t]{l} 
{\color{cyan} 
Equations of motion modify the canonically conjugate variables}
\vspace*{-0.1cm} \\
{\color{cyan} pertaining to
$\ \overline{{\cal{L}}} \ $,}
\vspace*{-0.1cm} \\
{\color{cyan}
beyond the reduction to consider 
exclusively the composite}
\vspace*{-0.1cm} \\
{\color{cyan} local field $\ {\cal{X}} \ ( \ x \ ) \ $
}
\end{tabular}
& {\color{cyan} ~\pageref{'3C-D'}}
\vspace*{-0.1cm} \\
3D & \begin{tabular}[t]{l}   
Remarks and consequences arising from
derivations 
\vspace*{-0.1cm} \\
in the last subsections : {\color{cyan} 3I} , 3C , 
{\color{cyan} 3C1 , 3CD}
\end{tabular}
&  ~\pageref{'3D'}
\vspace*{-0.1cm} \\
{\color{cyan} 3b} & \begin{tabular}[t]{l} 
{\color{cyan} Dimension 
$\ \left \lbrack \  M^{\ 4} \ \right \rbrack \ $ equations for field
strengths}
\vspace*{-0.1cm} \\
{\color{cyan} in unconstrained gauges}
\end{tabular}
& {\color{cyan} ~\pageref{'3b'}}
\vspace*{-0.1cm} \\
3c & \begin{tabular}[t]{l}  
QED : bare Lagrangean density
and equations of motion in 
\vspace*{-0.1cm} \\
unconstrained -- abelian -- gauges
\end{tabular}
& ~\pageref{'3c'}
\end{tabular}

}

\vspace*{0.15cm}

}



\newpage

{\color{blue} 

\begin{center}
\vspace*{-0.3cm}
{\color{red} 1 - Introduction
 }
\label{'1'}
\end{center}
\vspace*{0.0cm}

Let me limit this introductory section to characterizations of the topics 
presented .

These topics form three parts :

\begin{description} \item First part 

\begin{tabular}{l}
section {\color{red} 2 - How unique is an axial temporal gauge ?} 
\vspace*{-0.0cm} \\
subsections {\color{cyan} 2a , 2b , 2c}
\vspace*{-0.0cm} \\
first part of section {\color{red} 3 Canonical quantization in an axial gauge}
\vspace*{-0.0cm} \\
subsections {\color{cyan} 3a , 3 {\cal{A}}}
\end{tabular}

\item Second and main part, devoted to the essential modificatins
of the trace anomaly restricted to QCD , beyond the leading operator(s)
in the perturbative , i.e. asymptotically free environment 

\begin{tabular}{l}
central part of section {\color{red} 3} 
\vspace*{-0.0cm} \\
subsections {\color{cyan} 3I} , {\color{red} 3C} , {\color{cyan} 3C1 , 3CD}
, {\color{red} 3D}
\end{tabular}

\item Third part , remaining subsections of section {\color{red} 3}

\begin{tabular}{l}
subsections {\color{cyan} 3b} , {\color{red} 3c}
\end{tabular}

\end{description}

}



{\color{blue} 

\begin{center}
\vspace*{-0.5cm}
{\bf \color{red} 2 - How unique is a temporal axial gauge ?
 }
\label{'2'}
\end{center}
\vspace*{0.0cm}

\noindent
We begin studying the problem of parallel transport and
the equations determining a gauge in which the time component of
the connection vanishes, for complete connections with respect to a simple
local compact structure group $\ {\cal{G}} \ $

\vspace*{-0.3cm}
\begin{equation}
\label{eq:2-1}
\begin{array}{l}
\left ( \ {\cal{W}}_{\ \mu} \ ( \ {\cal{D}} \ ) \ \right )_{\ \alpha \beta}
\ ( \ x \ )
\ = \ {\cal{W}}^{\ r}_{\ \mu} \ ( \ x \ ) \ \left ( \ d_{\ r} 
\ \right )_{\ \alpha \beta}
\vspace*{0.1cm} \\
d_{\ r} \ = \ - \ d_{\ r}^{\ \dagger} \ = \ \frac{1}{i} \ J_{\ r}
\ \in \ Lie \ ( \ {\cal{D}} \ )
\hspace*{0.2cm} ; \hspace*{0.2cm}
\left \lbrack \ d_{\ p} \ , \ d_{\ q} \ \right \rbrack \ = \ f_{\ p q r}
\ d_{\ r}
\vspace*{0.1cm} \\
r \ , \ p \ , \ q \ = \ 1, \ \cdots , \ dim \ {\cal{G}} 
\hspace*{0.2cm} ; \hspace*{0.2cm}
\alpha \ , \ \beta \ = \ 1, \ \cdots , \ dim \ {\cal{D}}
\end{array}
\end{equation}

\noindent
The quantities defined in eq. \ref{eq:2-1} follow the notation in the
notefile \cite{phaseQCD2011}, recapitulated below

\vspace*{-0.5cm}
\begin{equation}
\label{eq:2-2}
\begin{array}{cl@{\hspace*{-0.1cm}}l}
\left ( \ {\cal{W}}_{\ \mu} \ ( \ {\cal{D}} \ ) \ \right )_{\ \alpha \beta}
\ ( \ x \ )
& : &
\begin{array}[t]{l}
\mbox{local} \ \mbox{operator} \ \times \ {\cal{D}}- 
\mbox{representation valued}
\vspace*{-0.1cm} \\
\mbox{connection over flat space time} \ x
\end{array}
\vspace*{0.1cm} \\
d_{\ r} \ \in \ Lie \ ( \ {\cal{D}} \ )
& : &
\begin{array}[t]{l}
\mbox{basis of antihermitian matrices forming}
\vspace*{-0.1cm} \\
\mbox{an irreducible representation of the}
\vspace*{-0.1cm} \\
\ \mbox{Lie algebra of} \ {\cal{G}}
\end{array}
\vspace*{0.1cm} \\
{\cal{W}}^{\ r}_{\ \mu} \ ( \ x \ )
& : &
\begin{array}[t]{l}
4 \ \times \ dim \ {\cal{G}} \ \mbox{components of hermitian local}
\vspace*{-0.1cm} \\
\mbox{connection fields}
\end{array}
\end{array}
\end{equation}

\noindent
A connection is called complete, if all regularity- , differentiability- and
integrability conditions extend to the complete ring of representations
$ \left . {\cal{R}} \ \right |_{\ {\cal{G}}} \ $, pertaining to 
$\ {\cal{G}} \ $ \cite{Segal,phaseQCD2011} .

}



{\color{blue} 

\noindent
The notion of complete connection is straightforwardly defined for classical
connection-field configurations, whence entering a path integral description,
but much less so directly for 
$4 \ \times \ dim \ {\cal{G}} \ $ components
of hermitian local connection fields, as defined in eq. \ref{eq:2-2} .
At this stage we do not attempt to substitute a precise definition.
\vspace*{0.1cm}

\noindent
Instead we look for an operator valued assembly of local gauge
transformations denoted $\ {\cal{A}} \ \leftarrow \ \Omega_{\ general} \ $
which achieve an axial gauge condition

\vspace*{-0.3cm}
\begin{equation}
\label{eq:2-3}
\begin{array}{l}
\left \lbrace \ \left ( \ {\cal{A}} \ ( \ {\cal{D}} \ ) 
\ \right )_{\ \alpha 
\ \beta} \ ( \ x \ ) \ \right \rbrace 
\hspace*{0.2cm} \mbox{with} \hspace*{0.2cm}
n^{\ \mu} \ {\cal{W}}_{\ \mu} \ ( \ {\cal{D}} \ )^{\ {\cal{A}}} \ = \ 0
\vspace*{0.1cm} \\
\left \lbrace \ {\cal{W}}_{\ \mu} \ ( \ {\cal{D}} \ )^{\ {\cal{A}}} \ = 
\ {\cal{A}} \ {\cal{W}}_{\ \mu} \ ( \ {\cal{D}} \ ) \ {\cal{A}}^{\ -1}
\ + \ {\cal{A}} \ \partial_{\ \mu} \ {\cal{A}}^{\ -1}
\ \right \rbrace \ ( \ x \ )
\vspace*{0.2cm} \\
\mbox{e.g. :} \ n^{\ \mu} \ = \ ( \ 1 \ , \ \vec{0} \ )
\ ; \ \mbox{independent of } \ x
\end{array}
\end{equation}

\noindent
Eq. \ref{eq:2-3} takes the form of a first order differential equation
for the quantity $\ {\cal{A}}^{\ '} \ = \ {\cal{A}}^{\ -1} \ ( \ x ) \ $ 

\vspace*{-0.3cm}
\begin{equation}
\label{eq:2-4}
\begin{array}{l}
\left . - \ n^{\ \mu} \ \partial_{\ \mu} \ {\cal{A}}^{\ '} \ ( \ x \ )
\ = \ n^{\ \mu} \ {\cal{W}}_{\ \mu} \ ( \ x \ ) \ {\cal{A}}^{\ '} \ ( \ x \ )
\ \right |_{\ {\cal{D}}}
\hspace*{0.2cm} ; \hspace*{0.2cm}
{\cal{A}}^{\ '} \ = \ {\cal{A}}^{\ -1} \ ( \ x )
\vspace*{0.1cm} \\
\mbox{with the unitarity condition }
\ ( \ {\cal{A}}^{\ '} \ )^{\ \dagger} \ {\cal{A}}^{\ '} \ = 
\ \P_{\ dim \ {\cal{D}} \ \times \ dim \ {\cal{D}}}
\end{array}
\end{equation}

\noindent
assuming the connection component 
$\ n^{\ \mu} \ {\cal{W}}_{\ \mu} \ ( \ x \ ) \ = \ w \ ( \ x \ )$
as given $\ \forall \ x \ $.

}



{\color{blue} 

\noindent
Adapting the four vector $\ n^{\ \mu} \ $ to the special form in eq.
\ref{eq:2-3} , treating the spacelike part of 
$\ x \ = \ ( \ t \ , \ \vec{x} \ ) \ $ as a parameter and substituting
$\ U \ \equiv \ {\cal{A}}^{\ '} \ = \ {\cal{A}}^{\ -1} \ $, eq. \ref{eq:2-4} 
takes the form

\vspace*{-0.3cm}
\begin{equation}
\label{eq:2-5}
\begin{array}{l}
\begin{array}{c}
d
\vspace*{0.2cm} \\
\hline  \vspace*{-0.3cm} \\
d \ t
\end{array} \hspace*{0.1cm}
\ \left . U \ ( \ t \ ) \ \right |_{\ \vec{x}}
\ = \ - \ \left . \ w \ ( \ t \ ) \ \right |_{\ \vec{x}} 
\ \left . U \ \right |_{\ \vec{x}}
\hspace*{0.2cm} \rightarrow \hspace*{0.2cm}
\dot{\mbox{}} \ = 
\ \begin{array}{c}
d
\vspace*{0.2cm} \\
\hline  \vspace*{-0.3cm} \\
d \ t
\end{array} \hspace*{0.1cm}
\vspace*{0.1cm} \\
\dot{U} \ ( \ t \ ) \ = \ - \ w \ ( \ t \ ) \ U \ ( \ t \ )
\hspace*{0.2cm} ; \hspace*{0.2cm}
U^{\ \dagger} \ U \ =
\ \P_{\ dim \ {\cal{D}} \ \times \ dim \ {\cal{D}}}
\vspace*{0.1cm} \\
U \ = \ \left ( \ \Omega^{\ '} \ \right )^{\ -1} \ ( \ x \ )
\end{array}
\end{equation}

\begin{center}
\vspace*{-0.0cm}
{\bf \color{cyan} 2 a - Integration of eq. \ref{eq:2-5} , interval by
interval
 }
 \label{'2a'}
 \end{center}

\noindent
The matrix valued operator extension implicit in eq. \ref{eq:2-5}
poses no problem to its integration , on any forward and/or backward 
time interval 
$\ I \ ( \ \tau_{\ 1} \ , \ \tau_{\ 0} \ ) \ :
\hspace*{0.2cm} \left \lbrace \ t \ | 
\ \tau_{\ 1} \ \geq \ \tau \ \geq \ \tau_{\ 0} \ \right \rbrace \ $, 
the backward
time intervals resulting from a time reversal operation. 
To see this we rewrite eq. \ref{eq:2-5} 

\vspace*{-0.3cm}
\begin{equation}
\label{eq:2-6}
\begin{array}{l}
\dot{U} \ ( \ t \ ) \ = \ - \ w \ ( \ t \ ) \ U \ ( \ t \ )
\hspace*{0.2cm} ; \hspace*{0.2cm}
\begin{array}[t]{l}
w^{\ \dagger}_{\ \alpha \ \beta} \ = \ W^{\ r}_{\ 0} 
\ \left ( \ \overline{d}_{\ r} \ \right )_{\ \beta \ \alpha} \ = 
\ - \ w_{\ \alpha \ \beta} 
\vspace*{0.1cm} \\
\left ( \ W_{\ \mu}^{\ r} \ \right )^{\ op. \ \dagger} \ = 
\ W_{\ \mu}^{\ r}
\end{array}
\end{array}
\end{equation}

\noindent
The suffix $^{\ op. \ \dagger}$ in eq. \ref{eq:2-6} characterizes the
local connection fields $\ W_{\ \mu}^{\ r} \ $ as beeing 
real, i.e. hermitian or selfadjoint local fields , here with incomplete 
association with actual selfadjoint operators in a 
-- connection extended -- Hilbert space.

}



{\color{blue} 

\noindent
We endow the differential equation in eq. \ref{eq:2-6} with the interval
associated initial condition

\vspace*{-0.3cm}
\begin{equation}
\label{eq:2-7}
\begin{array}{l}
U \ ( \ t \ ) \ \rightarrow \ U \ ( \ \tau_{\ 1} \ , \ \tau_{\ 0} \ ; \ t \ )
\vspace*{0.1cm} \\
\dot{U} \ ( \ t \ ) \ = \ - \ w \ ( \ t \ ) \ U \ ( \ t \ )
\hspace*{0.2cm} ; \hspace*{0.2cm}
U \ ( \ t \ = \ \tau_{\ 0} \ ) \ = \ \P
\hspace*{0.1cm} , \hspace*{0.1cm} \ t \ \in 
\ I \ ( \ \tau_{\ 1} \ , \ \tau_{\ 0} \ )
\end{array}
\end{equation}

\noindent
In eq. \ref{eq:2-7} -- $\ \P \ $ -- stands for the unit operator in the direct
product $\ Lie \ {\cal{D}} \ \times \ {\cal{H}} \ ( \ x \ ) \ $, where
$\ {\cal{H}} \ ( \ x \ ) \ $ shall represent the Hilbert space dissociated from
color, spanned by all gauge invariant (local) fields generated from the
(complete) connections and color carrying matter fields 
(quark and antiquark flavors) .

\noindent
The structure of eq. \ref{eq:2-7} but stripped of the Lie algebra structure
is well known from the differential equation for the time evolution operator

\vspace*{-0.5cm}
\begin{equation}
\label{eq:2-8}
\begin{array}{l}
U \ ( \ t \ ) \ \rightarrow \ E \ ( \ t \ ) \ = \ \exp \ ( \ i \ H_{\ 0} \ t \ )
\exp \ ( \ - \ i \ H \ t \ )
\vspace*{0.1cm} \\
H \ = \ H_{\ 0} \ + \ H_{\ I} \ : \ \mbox{\begin{tabular}[t]{l}
conserved
\vspace*{-0.1cm} \\
Hamilton operator
\end{tabular}
}
\hspace*{0.1cm} ; \hspace*{0.1cm}
\begin{array}{c}
H \ , \ H_{\ 0} \ , \ H_{\ I}
\vspace*{-0.1cm} \\
\mbox{independent of time}
\end{array}
\vspace*{0.1cm} \\
\dot{E} \ ( \ t \ ) \ = - \ \left \lbrace \ i \ {\cal{H}}_{\ I} \ ( \ t \ )
\ \right \rbrace \ E \ ( \ t \ )
\vspace*{0.1cm} \\
{\cal{H}}_{\ I} \ ( \ t \ ) \ = 
\ \exp \ ( \ i H_{\ 0} \ t \ ) \ {\cal{H}}_{\ I}
\ \exp \ ( \ - \ i \ H_{\ 0} \ t \ )
\end{array}
\end{equation}

\noindent
Despite this similarity, the charge-like gauge structure inherent to eq.
\ref{eq:2-7} develops its own subtleties. The solution is obtained by iteration
and is represented by the infinite sum

\vspace*{-0.5cm}
\begin{equation}
\label{eq:2-9}
\begin{array}{l}
U \ ( \ t \ - \ \tau_{\ 0} \ ) \ = \ \P \ + 
\vspace*{0.1cm} \\
+ \ \sum_{\ n=1}^{\ \infty} \ (-1)^{\ n}
\ {\displaystyle{\int}}_{\ 0}^{\ \Delta}  d \ t_{\ n}
\ {\displaystyle{\int}}_{\ 0}^{\ t_{\ n}}  d \ t_{\ n-1} \ \cdots
\ {\displaystyle{\int}}_{\ 0}^{\ t_{\ 2}} d \ t_{\ 1} \ \times
\vspace*{0.1cm} \\
\times \ \ w \ ( \ t_{ n} \ ) \ w \ ( \ t_{ n-1} \ ) \ \cdots
\ w \ ( \ t_{ 1} \ )
\vspace*{0.1cm} \\
\Delta \ = \ \tau_{\ 1} \ - \ \tau_{\ 0} 
\hspace*{0.2cm} \rightarrow \hspace*{0.2cm}
\mbox{for} \hspace*{0.2cm} \Delta \ > \ 0 \ :
\ \tau_{ n} \ \geq \ \tau_{ n-1} \ \geq \ \cdots \ \geq \ t_{\ 1}
\end{array}
\end{equation}

\noindent
As distinguished before, in the restricted case, where 
$\ w \ ( \ t \ , \ \vec{x} \ ) \ $ are given classical field configurations
the convergence and range of the matrix valued functions
$\ U \ ( \ \tau_{\ 1} \ , \ \tau_{\ 0} \ ; \ t \ ) \ $ resulting from the
form of the solution to eq. \ref{eq:2-7} presents conditions on admitted
given classical field configurations within complete connections 
$\ w \ ( \ t \ , \ \vec{x} \ ) \ $ .

\noindent
The interval associated solutions to eq. \ref{eq:2-7} ,
$\ U \ ( \ \tau_{\ 1} \ , \ \tau_{\ 0} \ ; \ t \ ) \ $, as defined in 
eq. \ref{eq:2-9} are best subsummed by path ordered specified integrals of the
connection 1-forms -- {\it not} satisfying any gauge fixing conditions --

\vspace*{-0.4cm} 
\begin{equation}
\label{eq:2-10}
\begin{array}{l}
\left ( \ {\cal{W}}^{\ (1)} \ ( \ {\cal{D}} \ \right )_{\ \alpha \beta}
\ = \ \left ( \ {\cal{W}}_{\ \mu} \ ( \ {\cal{D}} \ ) \ \right
)_{\ \alpha \beta}
\ ( \ x \ ) \ d \ x^{\ \mu}
\hspace*{0.2cm} \longrightarrow \hspace*{0.2cm}
{\cal{W}}^{\ (1)} \ ( \ {\cal{D}} \ )
\end{array}
\end{equation}

\noindent
over oriented (straight) lines associated to these intervals 

\vspace*{-0.3cm}
\begin{equation}
\label{eq:2-11}
\begin{array}{l}
C \ = \ C \ \left \lbrace \ \overline{x} \ \right \rbrace \ =
\ \left \lbrace \ \overline{x} \ | \ \overline{x}^{\ \mu} \ ( \ \tau \ ) \ = 
\ x_{\ 0}^{\ \mu} \ + \ \left ( \ x_{\ 1} \ - \ x_{\ 0} \ \right )^{\ \mu}
\ \tau \ \right \rbrace
\vspace*{0.1cm} \\
0 \ \leq \ \tau \ \leq \ 1
\end{array}
\end{equation}
\vspace*{-0.3cm}

}

{\color{blue}

\vspace*{-0.3cm}
\begin{equation}
\label{eq:2-12}
\begin{array}{l}
U \ \left ( \ x \ \stackrel{C}{\leftarrow} \ y \ \right )
\ = \ P \ \exp \ \left ( \ - \ {\displaystyle{\int}}_{ C} \ W^{\ (1)}
\ ( \ \overline{x} \ )
\ \right )
\vspace*{0.1cm} \\
\ \rightarrow \ \left ( \ U \ ( \ x \ , C \ , \ y \ ) 
\ \right )_{\ \alpha \beta}
\ \in \ {\cal{D}} \ ( \ {\cal{G}} \ )
\vspace*{0.2cm} \\ \hline \vspace*{-0.2cm} \\
C \ = \ C \ \left \lbrace \ \overline{x} \ \right \rbrace
\hspace*{0.2cm} :
\vspace*{0.1cm} \\
\begin{array}[t]{lll}
\overline{x} \ ( \ \tau \ ) 
& ;  & 
1 \ \geq \ \tau \ \geq \ 0 
\hspace*{0.2cm} ; \hspace*{0.2cm} 
\tau \ : \ \mbox{path parameter}
\vspace*{0.1cm} \\
& & \overline{x} \ ( \ \tau \ = \ 1 \ ) \ = \ x \ \leftarrow \ x_{\ 1}
\hspace*{0.2cm} ; \hspace*{0.2cm} 
\overline{x} \ ( \ \tau \ = \ 0 \ ) \ = \ y \ \leftarrow \ x_{\ 0}
\end{array}
\end{array}
\end{equation}

\noindent
In eq. \ref{eq:2-12}  P denotes matrix as well as operator
ordering along the path C since $\ W^{\ (1)} \ $ is matrix and operator valued.

\noindent
Since we want to keep differentiability criteria, we abstain for the moment
from considering
polygon like convolutions , of -- a finite number -- of segmentwise 
straight lines, but allow only
continuation of segments along one straight line to arbitrary but finite
extension. This can be done maintaining the parameter segment
$ 0 \ \leq \ \tau \ \leq \ 1 $ for individual segments, assuming that 
the series expansion of the quantities  
$\ U \ \left ( \ x_{\ r} \ \stackrel{C_{\ r}}{\leftarrow} \ y_{\ r} \ \right )
\ ; \ r \ = \ 1,2,\cdots,M \ < \infty \ $ of the form given in eq. \ref{eq:2-12}
are well defined for individual parallel segments along one straight line
-- in configuration space.

\noindent
The convolution of 2 so restricted segments continuing each other

}



{\color{blue} 

\noindent
then takes the form

\vspace*{-0.3cm}
\begin{equation}
\label{eq:2-13}
\begin{array}{l}
U \ \left ( \ x_{\ 2} \ \stackrel{C_{\ 2}}{\leftarrow} \ y_{\ 2} \ = \ x_{\ 1}
\ \right )
\ U \ \left ( \ x_{\ 1} \ \stackrel{C_{\ 1}}{\leftarrow} \ y_{\ 1} \ \right )
\ =
U \ \left ( \ x_{\ 2} \ \stackrel{C_{\ 2\&1}}{\leftarrow} \ y_{\ 1} \ \right )
\vspace*{0.1cm} \\
\begin{array}{cll}
C_{\ 1} & = & \left \lbrace \ \overline{x} \ ( \ \tau \ )
\ = \ y_{\ 1} \ + \
\ + \ \left ( \ x_{\ 1} \ - \ y_{\ 1} \ \right ) \ \tau \ \right \rbrace
\vspace*{0.1cm} \\
C_{\ 2} & = & \left \lbrace \ \overline{x} \ ( \ \tau \ )
\ = \ x_{\ 1} \ + \
\ + \ \left ( \ x_{\ 2} \ - \ x_{\ 1} \ \right ) \ \tau \ \right \rbrace
\vspace*{0.1cm} \\
C_{\ 2\&1} & = & \left \lbrace \ \overline{x} \ ( \ \tau \ )
\ = \ y_{\ 1} \ + \
\ + \ \left ( \ x_{\ 2} \ - \ y_{\ 1} \ \right ) \ \tau \ \right \rbrace
\end{array}
\hspace*{0.2cm} ; \hspace*{0.2cm}
0 \ \leq \ \tau \ \leq \ 1
\vspace*{0.2cm} \\
\mbox{with} 
\hspace*{0.2cm} : \hspace*{0.2cm}
x_{\ 1} \ - \ y_{\ 1} \ // \ x_{\ 2} \ - \ x_{\ 1} 
\ // \ x_{\ 2} \ - \ y_{\ 1}
\end{array}
\end{equation}

\noindent
The family of unitary transformations, solutions to eq. \ref{eq:2-7},
corresponds to the choice of segments along the straigt line with fixed
$\ \vec{x} \ $ coordinates

\vspace*{-0.1cm}
\begin{equation}
\label{eq:2-14}
\begin{array}{l}
U \ ( \ \tau_{\ 1} \ , \ \tau_{\ 0} \ ; \ \Delta \ t \ )
\ = 
\ U \ \left ( \ x \ \stackrel{C}{\leftarrow} \ y \ \right )
\hspace*{0.2cm} ; \hspace*{0.0cm}
\vspace*{0.1cm} \\
\begin{array}[t]{l@{\hspace*{0.1cm}}c@{\hspace*{0.1cm}}l@{\hspace*{0.1cm}}l}
x \ = \ ( & \tau_{\ 0} \ + \ \Delta \ t & , & \vec{x} \ )
\vspace*{0.1cm} \\
y \ = \ ( & \tau_{\ 0} & , & \vec{x} \ )
\end{array} 
\hspace*{0.1cm} \begin{array}[t]{l}
\begin{array}[t]{c}
\ ; \ \Delta \ t \ = \ t \ - \ \tau_{\ 0}
\vspace*{0.1cm} \\
\ ; \ \tau_{\ 1} \ \geq \ t \ \geq \ \tau_{\ 0}
\end{array}
\end{array}
\end{array}
\end{equation}

}



\newpage

{\color{blue} 

\begin{center}
\vspace*{-0.0cm}
{\bf \color{cyan} 2 b - Reparametrization of straight line using
a tangent vector $\ \mbox{\underline{v}} \ = 
\ ( \ \mbox{\underline{v}}^{\ \mu} \ ) \ $
}
 \label{'2b'}
 \end{center}

\noindent
We want to use a variant of the parametrizations of the parallel 
transport operators 
$\ U \ \left ( \ x \ \stackrel{C}{\leftarrow} \ y \ \right )
\ = \ P \ \exp \ \left ( \ - \ {\displaystyle{\int}}_{ C} \ W^{\ (1)}
\ ( \ \overline{x} \ )
\ \right ) \ $ introduced in eq. \ref{eq:2-12} , following
ref. \cite{phaseQCD2011} ( eqs. 77-79, op.cit. )

\vspace*{-0.3cm}
\begin{equation}
\label{eq:2-15}
\begin{array}{l}
U \ \left ( \ x \ \stackrel{C}{\leftarrow} \ y \ \right )
\ = \ P \ \exp \ \left ( \ - \ {\displaystyle{\int}}_{ C} \ W^{\ (1)}
\ ( \ \overline{x} \ )
\ \right )
\ \rightarrow 
\vspace*{0.1cm} \\
\hspace*{1.0cm} \rightarrow 
\ \left ( \ U \ ( \ x \ , C \ , y \ ) \ \right )_{\ \alpha \beta}
\ \in \ {\cal{D}} \ ( \ {\cal{G}} \ )
\vspace*{0.1cm} \\
C \ = \ C \ \left \lbrace \ \overline{x} \ \right \rbrace
\hspace*{0.2cm} : \hspace*{0.2cm} 
\begin{array}[t]{lll}
\overline{x} \ = \ \overline{x} \ ( \ s \ ) 
& ;  & 
\tau \ \geq \ s \ \geq \ 0 
\hspace*{0.1cm} ; \hspace*{0.1cm} 
s \ : \ \mbox{path parameter}
\vspace*{0.1cm} \\
& & \overline{x} \ ( \ s \ = \ \tau \ ) \ = \ x
\end{array}
\vspace*{0.1cm} \\
\overline{x} \ ( \ s \ = \ 0 \ ) \ = \ y
\end{array}
\end{equation}

\noindent
Next we use the explicit parametric dependence $\ x \ = \ x \ ( \ \tau \ ) \ $
and assume that there exists a continous extension of the functional dependence
as defined in eq. \ref{eq:2-15} into a small neighbourhood in all of space-time
around the point $\ x \ ( \ \tau \ ) \ $ defined e.g. with
a Lorentz noninvariant Euclidean norm 
$\ || \ \xi \ ||^{\ 2} \ = \ ( \ \xi^{\ 0} \ )^{\ 2} \ + 
\ ( \ \vec{\xi} \ )^{\ 2} \ $

\vspace*{-0.1cm}
\begin{equation}
\label{eq:2-16}
\begin{array}{l}
U \ \left ( \ x \ ( \ \tau \ ) \ \stackrel{C}{\leftarrow} \ y \ \right )
\hspace*{0.2cm} \rightarrow \hspace*{0.2cm}
\widetilde{U} \ \left ( \ x \ ( \ \tau \ ) \ + \ \xi \ \right )
\hspace*{0.2cm} ; \hspace*{0.2cm}
|| \ \xi \ || \ < \ \delta 
\end{array}
\end{equation}

}



{\color{blue} 

\noindent
The differential equation ( eqs. \ref{eq:2-5} , \ref{eq:2-6} )
takes the form

\vspace*{-0.3cm}
\begin{equation}
\label{eq:2-17}
\begin{array}{@{\hspace*{-0.2cm}}l}
\begin{array}{c}
d
\vspace*{0.2cm} \\
\hline  \vspace*{-0.3cm} \\
d \ \tau
\end{array} \hspace*{0.2cm}
U \ \left ( \ x \ ( \ \tau \ ) \ \stackrel{C}{\leftarrow} \ y \ \right )
\ = \ - \ \mbox{\underline{v}}^{\ \mu} \ {\cal{W}}_{\ \mu} 
\ \left ( \ x \ ( \ \tau \ ) \ \right ) \ U \ \left ( \ x \ ( \ \tau \ ) \
\stackrel{C}{\leftarrow} \ y \ \right )
\vspace*{0.1cm} \\
\begin{array}{c}
d
\vspace*{0.2cm} \\
\hline  \vspace*{-0.3cm} \\
d \ \tau
\end{array} \hspace*{0.2cm}
\widetilde{U} \ \left ( \ x \ ( \ \tau \ ) \ \stackrel{C}{\leftarrow} \ y 
\ \right )
\ = \ \mbox{\underline{v}}^{\ \mu} \ \partial_{\ \xi \ \mu} 
\ \left . \widetilde{U} \ \left ( \ x \ ( \ \tau \ ) \ + \ \xi \ \right )
\ \right |_{\ \xi=0}
\hspace*{0.5cm} \longrightarrow
\vspace*{0.2cm} \\ \hline \vspace*{-0.2cm} \\
\longrightarrow \hspace*{0.2cm}
\left . \mbox{\underline{v}}^{\ \mu} 
\ \left ( \ \partial_{\ x \ \mu} \ + \ {\cal{W}}_{\ \mu} \ ( \ x \ ) 
\ \right )
\ \widetilde{U} \ \left ( \ x \ \ \right ) \ \right |_{\ x = x(\tau)} 
\ = \ 0
\vspace*{0.2cm} \\
\widetilde{U} \ = \widetilde{U} \ \left ( \ x \ ; \ \left \lbrace \
{\cal{W}}^{\ (1)} \ \left ( \ {\cal{D}} \ \right ) \ \right \rbrace 
\ , \ y \ \right )
\hspace*{0.2cm} ; \hspace*{0.2cm}
\vspace*{0.1cm} \\
\widetilde{U} \ \left ( \ y \ ; \ \left \lbrace \ {\cal{W}}^{\ (1)} 
\ \left ( \ {\cal{D}} \ \right ) 
\ \right \rbrace \ , \ y \ \right ) \ = \ \P
\end{array}
\end{equation}

\noindent
The notation of the arguments 
$\ x \ ; \ \left \lbrace \ {\cal{W}}^{\ (1)} \ \left ( \ {\cal{D}} \ \right ) 
\ \right \rbrace , \ y \ $ of 
the parallel transport operators $\ \widetilde{U} \ $ in the last line of eq.
\ref{eq:2-17} shall indicate that x is understood as space time variable,
whereas the entire connection $\ \left \lbrace \ {\cal{W}}^{\ (1)} 
\ \left ( \ {\cal{D}} \ \right ) \ \right \rbrace \ $
as well as the base point y shall be understood in a parametric sense.

\noindent
The relations in the last two lines of eq. \ref{eq:2-17} show the intrinsic 
connection 

}



{\color{blue} 

\noindent
between boundary - and integrability conditions pertaining to
complete connections and the associated determination of an axial gauge.

\noindent
The 'variable-' point x ( 'Aufpunkt' in german terminology )
and base point y represented by the endpoints along the straight line of
parallel transport through the quantity

\vspace*{-0.3cm}
\begin{equation}
\label{eq:2-18}
\begin{array}{l}
\widetilde{U} \ \left ( \ x \ ; \ \left \lbrace \
{\cal{W}}^{\ (1)} \ \left ( \ {\cal{D}} \ \right ) \ \right \rbrace 
\ , \ y \ \right )
\end{array}
\end{equation}

\noindent
separate the adjoint representation of the local limit into the product
of two distinct local representations of the gauge group , 
one each at x and y , transforming under relatively complex conjugate
but singly arbitrary irreducible rerpresentations $\ {\cal{D}} 
\ \leftrightarrow \ x \ , 
\ \overline{\cal{D}} \ \leftrightarrow \ y \ $ of
the gauge group $\ {\cal{G}} \ $ -- subject to restrictions from continuity of
local gauge group representations -- for complete connections.

\noindent
This gives rise to the $\ \left ( \ x \ , \ y \ \right ) \ $ associated
bi-local transformation properties under an -- operator valued -- local family
of gauge transformations discssed in the next subsection below.

\begin{center}
\vspace*{-0.0cm}
{\bf \color{cyan} 2 c - QCD $\ {\cal{D}} \ ( {\cal{G}} \ ) \ - \ $
strings \vspace*{-0.0cm} \protect \\ 
building blocks of bi-local gauge covariant parallel transports
$\ x \ \stackrel{C}{\leftarrow} \ y \ $,
relative to the associated irreducible representations 
$\ {\cal{D}} \ \ \leftrightarrow \ x \ , \ \overline{{\cal{D}}}
\ \leftrightarrow \ y \ $ of the gauge group $\ {\cal{G}} \ $
}
 \label{'2c'}
 \end{center}

\noindent
In the following we drop the $\ \widetilde{U} \ \longrightarrow \ U \ $
symbol explained in eqs. \ref{eq:2-17} - \ref{eq:2-18} for simplicity and
repeat the differential 'parallel-transport'- equation ( eq. \ref{eq:2-17} )

}



{\color{blue} 

\vspace*{-0.3cm}
\begin{equation}
\label{eq:2-19}
\begin{array}{l}
\left . \mbox{\underline{v}}^{\ \mu} 
\ \left ( \ \partial_{\ x \ \mu} \ + \ {\cal{W}}_{\ \mu} \ ( \ x \ ) 
\ \right )
\ U \ \left ( \ x \ \ \right ) \ \right |_{\ x = x(\tau)} 
\ = \ 0
\vspace*{0.2cm} \\
U \ ( \ x \ ) \ \longleftarrow
\ U \ \left ( \ x \ ; \ \left \lbrace \
{\cal{W}}^{\ (1)} \ \left ( \ {\cal{D}} \ \right ) \ \right \rbrace
\ , \ y \ \right )
\vspace*{0.1cm} \\
U \ \left ( \ y \ ; \ \left \lbrace \ {\cal{W}}^{\ (1)}
\ \left ( \ {\cal{D}} \ \right )
\ \right \rbrace \ , \ y \ \right ) \ = \ \P
\end{array}
\end{equation}

\noindent
Next we consider local gauge transformations $\ \left \lbrace \ \Omega \ \right
\rbrace \ $, substitute the identity 
$\ \Omega^{\ -1} \ ( \ x \ ) \ \Omega \ ( \ x \ ) \ = \ \P \ $ and use
the chain rule for partial derivatives

\vspace*{-0.4cm}
\begin{equation}
\label{eq:2-20}
\begin{array}{l}
\left . \mbox{\underline{v}}^{\ \mu}
\ \left ( \ \partial_{\ x \ \mu} \ + \ {\cal{W}}_{\ \mu} \ ( \ x \ )
\ \right )
\ \left ( \ \Omega^{\ -1} \ ( \ x \ ) \ \Omega \ ( \ x \ ) \ \right )
\ U \ \left ( \ x \ \ \right ) \ \right |_{\ x = x(\tau)}
\ = \ 0
\vspace*{0.2cm} \\
\left .
\mbox{\underline{v}}^{ \mu} \ \Omega^{ -1} \ (  x  )
\ \left ( \begin{array}{c} \Omega \ (  x ) 
\ \left \lbrack \hspace*{0.05cm} 
\partial_{ x \ \mu} \ + \ {\cal{W}}_{ \mu} \ ( x )
\hspace*{0.05cm} \right \rbrack  \ \Omega^{ -1} \ ( x )
\vspace*{0.1cm} \\
\ + \ \partial_{\ x \ \mu}
\end{array} \right )
\ \Omega \ ( x ) \ U \ ( x )  
 \right |_{ x = x(\tau)} 
\vspace*{0.2cm} \\
\hspace*{2.3cm} = \ 0
\end{array}
\end{equation}

\noindent
By eq. \ref{eq:2-3} the quantity in $\ \left \lbrack \ \right \rbrack \ $
brackets in the second relation of eq. \ref{eq:2-20} represents the gauge
transformed connection 

\vspace*{-0.3cm}
\begin{equation}
\label{eq:2-21}
\begin{array}{l}
\Omega \ ( \ x \ )
\ \left \lbrack \ \partial_{\ x \ \mu} \ + \ {\cal{W}}_{\ \mu} \ ( \ x \ )
\ \right \rbrack  \ \Omega^{\ -1} \ ( \ x \ )
\ = \ {\cal{W}}_{\ \mu}^{\ \Omega} \ ( \ x \ )
\end{array}
\end{equation}

\noindent
And so we derive the bi-locally transformed 

}



{\color{blue} 

\noindent
parallel transport operators, from the associated differential equation
and initial conditions ( eqs. \ref{eq:2-19} , \ref{eq:2-20} )

\vspace*{-0.3cm}
\begin{equation}
\label{eq:2-22}
\begin{array}{l}
\left . \begin{array}{@{\hspace*{-0.3cm}}l}
\left . \mbox{\underline{v}}^{\ \mu}
\ \left ( \ \partial_{\ x \ \mu} \ + \ {\cal{W}}_{\ \mu}^{\ \Omega} \ (  x  )
\ \right )
\ U \ \left (  x  ;  \left \lbrace
 \left \lbrace \ {\cal{W}}^{\ (1)} \ \left ( \ {\cal{D}} \ \right )
\ \right \rbrace^{\ \Omega}
 \right \rbrace \ , \ y \ \right )
 \right |_{\ x = x(\tau)}
\vspace*{0.1cm} \\
\ = \ 0
\vspace*{0.1cm} \\
U \ \left ( \ y \ ; \ \left \lbrace
\ \left \lbrace \ {\cal{W}}^{\ (1)} \ \left ( \ {\cal{D}} \ \right )
\ \right \rbrace^{\ \Omega}
\ \right \rbrace \ , \ y \ \right ) \ = \ \P
\end{array} \right \rbrace  \rightarrow
\vspace*{0.2cm} \\ \hline \vspace*{-0.2cm} \\
\ U \ \left ( \ x \ ; \ \left \lbrace 
\ \left \lbrace \ {\cal{W}}^{\ (1)} \ \left ( \ {\cal{D}} \ \right ) 
\ \right \rbrace^{\ \Omega}
\ \right \rbrace \ , \ y \ \right )
\ = 
\vspace*{0.1cm} \\
\ = \ \Omega \ ( \ x \ ) 
\ U \ \left ( \ x \ ; \ 
\ \left \lbrace \ {\cal{W}}^{\ (1)} \ \left ( \ {\cal{D}} \ \right )
\ \right \rbrace \ , \ y \ \right )
\ \Omega^{\ -1} \ ( \ y \ )
\vspace*{0.1cm} \\
\hspace*{1.0cm} \forall \hspace*{0.2cm} \mbox{irreducible representations}
\hspace*{0.2cm} {\cal{D}} \ ( \ {\cal{G}} \ ) \ \mbox{and} \ W^{\ (1)} \ ,
\ \left \lbrace \ \Omega \ \right \rbrace
\vspace*{0.1cm} \\
U \ \left ( \ x \ ; \ \left \lbrace
\ \left \lbrace \ {\cal{W}}^{\ (1)} \ \left ( \ {\cal{D}} \ \right )
\ \right \rbrace
\ \right \rbrace \ , \ y \ \right )
\ = \ P \ \exp \ \left ( \ - \ {\displaystyle{\int}}_{ C} \ W^{\ (1)}
\ ( \ x \ )
\ \right )
\end{array}
\end{equation}

\noindent
We emphasize that existence and uniqueness of the operator valued families
$\ U \ ( \ . \ ) \ $ , necessary for the derivation 
of the transformation
properties , displayed in eqs. \ref{eq:2-15} and \ref{eq:2-22} , 
constitute regularity conditions imposed on complete connections
\cite{phaseQCD2011} . Connections 
$\ {\cal{W}}^{\ (1)} \ $ as well as local gauge transformations $\ \left
\lbrace \ \Omega \ \right \rbrace \ $ shall denote general quantities, not
related to gauge fixing.

}



{\color{blue} 

\begin{center}
\vspace*{-0.3cm}
{\bf \color{red} 3 - Canonical quantization in an axial gauge
 }
\label{'3'}
\end{center}
\vspace*{0.0cm}

\noindent
We postpone all discussion of multiplicative renormalization factors
and identify field strengths , imposing the axial gauge condition
( eqs. \ref{eq:2-3} - \ref{eq:2-5} ) . Gauge fixed quantities are always 
singled out by a superfix $\ ^{{\cal{A}}} \ $ contrasting with 
connections and field strengths in a general gauge

\vspace*{-0.3cm}
\begin{equation}
\label{eq:3-1}
\begin{array}{l}
n^{\ \mu} \ \left ( \ W_{\ \mu} \ \right )^{\ {\cal{A}}} \ ( \ x \ ) \ = \ 0
\hspace*{0.2cm} ; \hspace*{0.2cm}
n^{\ \mu} \ = \ \left ( \ 1 \ , \ \vec{0} \ \right )
\end{array}
\end{equation}

\noindent
We follow the notation used in ref. \cite{phaseQCD2011} 
( eqs. (65) - (72) , op.cit. ) , with the exception to
use $\ {\cal{W}}^{\ (n)} \ ( \ {\cal{D}} \ ) \ $ 
for the Lie algebra matrix valued  n-forms 
$\ \left ( \ n \ = \ 1,2 \ \right ) \ $ .

\vspace*{-0.3cm}
\begin{equation}
\label{eq:3-2}
\begin{array}{l}
\left ( \ {\cal{W}}^{\ (1)} \ ( \ {\cal{D}} \ ) \ \right )_{\ \alpha \beta} \ = 
\ W^{\ r}_{\ \mu} \ ( \ x \ ) \ \left ( \ d_{\ r} \ ( \ {\cal{D}} \ ) 
\ \right )_{\ \alpha \beta} \ d \ x^{\ \mu}  
\vspace*{0.1cm} \\
W^{\ r}_{\ \mu} \ ( \ x \ ) \ : \ \mbox{real}
\hspace*{0.2cm} ; \hspace*{0.2cm} 
r \ = \ 1, \ \cdots, \ G \ = \ dim \ ( \ {\cal{G}} \ )
\end{array}
\end{equation}

\vspace*{-0.3cm}
\begin{equation}
\label{eq:3-3}
\begin{array}{l}
{\cal{W}}^{\ (2)} \ ( \ {\cal{D}} \ ) \ \rightarrow \ {\cal{W}}^{\ (2)}
\ = \ \partial \ {\cal{W}}^{\ (1)} \ + \ \left ( \ {\cal{W}}^{\ (1)} 
\ \right )^{\ 2}
\hspace*{0.2cm} ; \hspace*{0.2cm}
\partial \ \equiv \ d \ x^{\ \mu} \ \partial_{\ x \ \mu}
\vspace*{0.1cm} \\
\left ( \ {\cal{W}}^{\ (2)} \ \right )_{\ \alpha \beta}
\ = \ \frac{1}{2} \ W^{\ r}_{\ \mu \nu} \ \left ( \ d_{\ r} 
\ \right )_{\ \alpha \beta} \ d \ x^{\ \mu} \ \wedge \ d \ x^{\ \nu} 
\hspace*{0.2cm} ; \hspace*{0.2cm} d_{\ r} \ \in \ Lie \ ( \ {\cal{D}} \ )
\vspace*{0.1cm} \\
\rightarrow \ 
{\cal{W}}^{\ (2)}_{\ \mu \nu} \ = \ \partial_{\ \mu} 
\ {\cal{W}}^{\ (1)}_{\ \nu}
\ - \ \partial_{\ \nu} \ {\cal{W}}^{\ (1)}_{\ \mu}
\ + \ \left \lbrack \ {\cal{W}}^{\ (1)}_{\ \mu} \ , \ {\cal{W}}^{\ (1)}_{\ \nu} 
\ \right \rbrack
\vspace*{0.1cm} \\
W^{\ r}_{\ \mu \nu} \ = \ - \ W^{\ r}_{\ \nu \mu} \ =
\ \partial_{\ \mu} \ W^{\ r}_{\ \nu} \ - \ \partial_{\ \nu} \ W^{\ r}_{\ \mu}
\ + \ f_{r p q} \ W^{\ p}_{\ \mu} \ W^{\ q}_{\ \nu}
\vspace*{0.2cm} \\ \hline \vspace*{-0.1cm}
{\cal{W}}^{ (2)} \ ( {\cal{D}} ) \ \equiv \ {\cal{B}}^{\ (2)} 
\ ( \ {\cal{D}} \ )
\hspace*{0.1cm} ; \hspace*{0.1cm}
W^{\ r}_{\ \mu \nu} \ \equiv \ B^{\ r}_{\ \mu \nu}
\hspace*{0.1cm} \mbox{\begin{tabular}{c} 
field strength components 
\vspace*{-0.15cm} \\
independent of $\ {\cal{D}} \ $
\end{tabular}
} 
\end{array}
\vspace*{-0.4cm}
\end{equation}

}



{\color{blue} 

\vspace*{-0.3cm}
\noindent
We decompose the field strength vectors and tensors, in axial gauge,
with the conventions

\vspace*{-0.5cm}
\begin{equation}
\label{eq:3-4}
\begin{array}{l}
W^{\ r \ {\cal{A}}}_{\ \mu} \ = \ \eta_{\ \mu \nu} \ W^{\ r \ \nu \ {\cal{A}}}
\hspace*{0.2cm} ; \hspace*{0.2cm} 
W^{\ r \ 0 \ {\cal{A}}} \ = \ 0 \ , \ W^{\ r \ k \ {\cal{A}} } \ = 
\ \left ( \ \begin{array}[b]{l}
\vspace*{-0.5cm} \\
\rightarrow
\vspace*{-0.2cm} \\
W^{\ r} 
\end{array}
\ \right )^{\ k \ {\cal{A}}}
\vspace*{0.1cm} \\
\eta_{\ \mu \nu} \ = \ diag \ ( \ 1 \ , \ -1 \ , \ -1 \ , \ -1 \ )
\hspace*{0.1cm} : \hspace*{0.1cm} 
\mbox{flat space metric}
\hspace*{0.1cm} ; \hspace*{0.1cm}
k=1,2,3
\vspace*{0.1cm} \\
B^{\ r \ {\cal{A}}}_{\ 0 k} \ = \ - \ \left ( \ \begin{array}[b]{l}
\vspace*{-0.5cm} \\
\rightarrow
\vspace*{-0.2cm} \\
E^{\ r}
\end{array}
\ \right )^{\ k \ {\cal{A}}}
\hspace*{0.05cm} ; \hspace*{0.05cm}
B^{\ r \ {\cal{A}}}_{\ jl} \ = \ \varepsilon_{\ kjl} 
\ \left ( \ \begin{array}[b]{l}
\vspace*{-0.5cm} \\
\rightarrow
\vspace*{-0.2cm} \\
B^{\ r}
\end{array}
\ \right )^{\ k \ {\cal{A}}}
\hspace*{0.0cm} ; \hspace*{0.0cm} 
\vspace*{0.1cm} \\
k,j,l \ = \ 1,2,3
\vspace*{0.2cm} \\
\begin{array}{l}
\vspace*{-0.5cm} \\
\rightarrow
\vspace*{-0.2cm} \\
w^{\ r \ {\cal{A}}}
\vspace*{0.3cm}
\end{array} \ \equiv
\ -
\ \begin{array}[b]{l}
\vspace*{-0.5cm} \\
\rightarrow
\vspace*{-0.2cm} \\
W^{\ r \ {\cal{A}}}
\end{array}
\hspace*{0.1cm} ; \hspace*{0.1cm} \mbox{covariantly  :} \hspace*{0.2cm}
w_{\ \mu}^{\ r \ {\cal{A}}} \ = \ - \ W_{\ \mu}^{\ r \ {\cal{A}}}  
\hspace*{0.1cm} ; \hspace*{0.1cm}
^{\bullet} \ = \ \partial_{\ t}
\end{array}
\vspace*{-0.0cm}
\end{equation}

\noindent
Then the field strengths are expressed through the potentials

\vspace*{-0.4cm}
\begin{equation}
\label{eq:3-5}
\begin{array}{l}
\begin{array}[b]{l}
\vspace*{-0.5cm} \\
\rightarrow
\vspace*{-0.2cm} \\
E^{\ r \ {\cal{A}}}
\end{array}
\ =  
\ - \begin{array}[b]{l}
\vspace*{-0.5cm} \\
\stackrel{\bullet}
{\rightarrow}
\vspace*{-0.2cm} \\
w^{\ r \ {\cal{A}}}
\end{array}
\hspace*{0.0cm} ; \hspace*{0.0cm}
\begin{array}[b]{l}
\vspace*{-0.5cm} \\
\rightarrow
\vspace*{-0.2cm} \\
B^{\ r \ {\cal{A}}}
\end{array}
\ = 
\ \nabla \ \wedge 
\hspace*{-0.1cm} \begin{array}[b]{l}
\vspace*{-0.5cm} \\
\rightarrow
\vspace*{-0.2cm} \\
w^{\ r \ {\cal{A}}}
\end{array}
\ + \ f_{\ rpq}
\ \begin{array}[b]{l}
\vspace*{-0.5cm} \\
\rightarrow
\vspace*{-0.2cm} \\
w^{\ p  {\cal{A}}}
\end{array}
 \wedge 
\hspace*{-0.1cm} \begin{array}[b]{l}
\vspace*{-0.5cm} \\
\rightarrow
\vspace*{-0.2cm} \\
w^{\ q  {\cal{A}}}
\end{array}
\vspace*{0.1cm} \\
\left ( \begin{array}{c}
\vspace*{-0.5cm} \\
\rightarrow
\vspace*{-0.3cm} \\
\mbox{w}
\vspace*{0.2cm}
\end{array} \right )^{\ {\cal{A}}}
 = \hspace*{-0.1cm} \left ( \begin{array}{l}
\vspace*{-0.5cm} \\
\rightarrow
\vspace*{-0.2cm} \\
w^{\ r}
\vspace*{0.1cm}
\end{array} \right )^{ {\cal{A}}} \ d_{\ r} 
\hspace*{0.1cm} ; \hspace*{0.1cm}
d_{\ r} \ \rightarrow \ \left ( \ d_{\ r} \ \right )_{\ \alpha \beta}
\end{array}
\end{equation}

\noindent
It is advantageous to use systematically in parallel the Lie algebra 
$\ {\cal{D}} \ $ matrix valued quantities completing the one given on the
last line of eq. \ref{eq:3-5}

}



{\color{blue} 

\vspace*{-0.3cm}
\begin{equation}
\label{eq:3-6}
\begin{array}{l}
{\cal{A}} \ :
\begin{array}{lll}
{\cal{B}}^{\ (2)}_{\ 0 k} \ = 
\hspace*{0.2cm} - \ \left ( 
\begin{array}[b]{c}
\vspace*{-0.5cm} \\
\rightarrow
\vspace*{-0.25cm} \\
\mbox{E}
\end{array}
 \right )^{\ k}
& ; &
\begin{array}[b]{c}
\vspace*{-0.5cm} \\
\rightarrow
\vspace*{-0.25cm} \\
\mbox{E}
\end{array} \ =
\ \begin{array}[b]{l}
\vspace*{-0.5cm} \\
\rightarrow
\vspace*{-0.2cm} \\
E^{\ r}
\end{array} \ d_{\ r}
\vspace*{0.2cm} \\
{\cal{B}}^{\ (2)}_{\ jl} \ = \ \varepsilon_{\ kjl}
\left ( \begin{array}[b]{c}
\vspace*{-0.5cm} \\
\rightarrow
\vspace*{-0.2cm} \\
\mbox{B}
\end{array}
\right )^{\ k}
& ; &
\begin{array}[b]{c}
\vspace*{-0.5cm} \\
\rightarrow
\vspace*{-0.25cm} \\
\mbox{B}
\end{array}
\ = \ \begin{array}[b]{l}
\vspace*{-0.5cm} \\
\rightarrow
\vspace*{-0.2cm} \\
B^{\ r}
\end{array}
\ d_{\ r}
\end{array}
\end{array}
\end{equation}

\noindent
The axial gauge fields are indicated by $\ {\cal{A}} \ : \ $ in front of
their appearance.
Then eq. \ref{eq:3-5} becomes

\vspace*{-0.3cm}
\begin{equation}
\label{eq:3-7}
\begin{array}{l}
{\cal{A}} \ :
\ \left .
\begin{array}[b]{c}
\vspace*{-0.5cm} \\
\rightarrow
\vspace*{-0.25cm} \\
\mbox{E}
\end{array} \ =
\ - \ \begin{array}[b]{c}
\vspace*{-0.5cm} \\
\stackrel{\bullet}
{\rightarrow}
\vspace*{-0.25cm} \\
\mbox{w}
\end{array}
\hspace*{0.2cm} ; \hspace*{0.2cm}
\begin{array}[b]{c}
\vspace*{-0.5cm} \\
\rightarrow
\vspace*{-0.25cm} \\
\mbox{B}
\end{array}
 =
 \nabla \ \wedge 
\hspace*{-0.1cm} \begin{array}[b]{c}
\vspace*{-0.5cm} \\
\rightarrow
\vspace*{-0.25cm} \\
\mbox{w}
\end{array}
\ +
\ \begin{array}[b]{l}
\vspace*{-0.5cm} \\
\rightarrow
\vspace*{-0.25cm} \\
\mbox{w}
\end{array}
\wedge 
\hspace*{-0.0cm} \begin{array}[b]{l}
\vspace*{-0.5cm} \\
\rightarrow
\vspace*{-0.25cm} \\
\mbox{w} 
\end{array}
\hspace*{0.5cm} \right | \ \left ( \ {\cal{D}} \ \right )
\end{array}
\end{equation}

\noindent
In line with the $\ {\cal {D}} \ $ projection we specify the
$\ {\cal{D}} \ - \ $ dependent trace normalization 

\vspace*{-0.4cm}
\begin{equation}
\label{eq:3-8}
\begin{array}{l}
- \ tr_{\ {\cal{D}} \ } \ d^{\ p} \ d^{\ q} \ = 
\ \delta^{\ p q} \ T_{\ 2} \ ( \ {\cal{D}} \ )
\vspace*{0.1cm} \\
- \ \sum_{\ r} \ \left ( \ d^{\ r} \ \right )^{\ 2} \ = 
\ C_{\ 2} \ ( {\cal{D}} \ ) \ \P_{\ dim \ {\cal{D}} \ \times 
\ dim \ {\cal{D}}}
\hspace*{0.2cm} \rightarrow
\vspace*{-0.1cm} \\
\hspace*{0.0cm}
T_{\ 2} \ ( \ {\cal{D}} \ ) \ = 
\ \begin{array}{c}
dim \ ( \ {\cal{D}} \ )
\vspace*{0.2cm} \\
\hline  \vspace*{-0.3cm} \\
dim \ ( \ {\cal{G}} \ )
\end{array} \hspace*{0.1cm}
\ C_{\ 2} \ ( \ {\cal{D}} \ )
\hspace*{0.2cm} ; \hspace*{0.1cm}
\mbox{for} \left . \begin{array}{l}
{\cal{G}} \ = \ SU3 
\vspace*{0.1cm} \\
{\cal{D}} \ = \ 3 \ \mbox{or} \ \overline{3}
\end{array} \right \rbrace
\vspace*{0.0cm}  : \vspace*{0.0cm} \\
 \vspace*{-0.1cm} \\
dim \ {\cal{D}} \ = 3 \ , \ dim \ {\cal{G}} \ = \ 8
\ , \ T_{\ 2} \ ( \ {\cal{D}} \ ) \ = \ \frac{1}{2} 
\ \ C_{\ 2} \ ( \ {\cal{D}} \ ) \ = \ \frac{4}{3}
\end{array}
\end{equation}

}



{\color{blue} 

\vspace*{-0.1cm}
\noindent
In eq. \ref{eq:3-8} $\ C_{\ 2} \ ( \ {\cal{D}} \ ) \ $ denotes
the eigenvalue of the second order Casimir operator projected on 
the irreducible representation $\ {\cal{D}} \ $

\vspace*{-0.3cm}
\begin{equation}
\label{eq:3-9}
\begin{array}{l}
- \ \sum_{\ r} \ \left ( \ d^{\ r} \ \right )^{\ 2} \ = 
\ C_{\ 2} \ ( \ {\cal{D}} \ ) \ \P_{\ dim \ {\cal{D}} \ \times 
\ dim \ {\cal{D}}}
\end{array}
\end{equation}

\begin{center}
\vspace*{-0.0cm}
{\bf \color{cyan} 3 a - Bare Lagrangean density
and equations of motion in unconstrained gauges
}
 \label{'3a'}
 \end{center}

\noindent
The bare local Lagrangean density -- postponing the discssion of quark and 
antiquark
(matter-) fields -- is formed by the bare field strengths
using the relations in eq. \ref{eq:3-8}

\vspace*{-0.5cm}
\begin{equation}
\label{eq:3-10}
\begin{array}{l}
{\cal{L}} \ ( x ) \ = 
\ \begin{array}{c}
1
\vspace*{0.2cm} \\
\hline  \vspace*{-0.3cm} \\
4 \ T_{\ 2} \ ( \ {\cal{D}} \ ) \ g^{\ 2}
\end{array} \hspace*{0.1cm}
\ tr_{\ {\cal{D}}} \ \left \lbrack
 {\cal{B}}_{\ \mu \ \nu} \ {\cal{B}}^{\ \mu \nu} \right \rbrack
\ ( \ {\cal{D}} \ ) \ ( x ) \ +
\ {\cal{L}}_{\ \left \lbrace q \right \rbrace} \ ( x )
\vspace*{0.1cm} \\
{\cal{L}}_{\ \left \lbrace q \right \rbrace} \ ( \ x \ )
\ = 
\vspace*{0.1cm} \\
 \sum_{\ q-fl} \ \overline{q}^{\ \dot{c}'}
 \left \lbrace \frac{i}{2}
\ \gamma^{\ \mu} \ \left \lbrack
\begin{array}{c}
 \left ( \ 
\begin{array}{c}
\vspace*{-0.5cm} \\
\rightarrow
\vspace*{-0.25cm} \\
D
\end{array}_{\ \mu} \ \left ( \ 3 \ \right ) \ \right )_{\ c' \dot{c}}
\ - 
\vspace*{0.1cm} \\
- \ \left ( \
\begin{array}{c}
\vspace*{-0.5cm} \\
\leftarrow
\vspace*{-0.25cm} \\
D
\end{array}_{\ \mu} \ \left ( \ \overline{3} \ \right ) 
\ \right )_{\ \dot{c} c'}
\end{array} \right \rbrack
\ - \ m_{\ q} \ \delta_{\ c' \dot{c}} \right \rbrace
 q^{\ c} \ (  x  )
\end{array}
\end{equation}

}



{\color{blue} 

\vspace*{-0.1cm}
\noindent
The Euler equations
generating an extremum of the bare action take the form

\vspace*{-0.3cm}
\begin{equation}
\label{eq:3-11}
\begin{array}{l}
\partial_{\ \varrho}
\ \left \lbrace 
\begin{array}{c}
\delta \ {\cal{L}}
\vspace*{0.2cm} \\
\hline  \vspace*{-0.3cm} \\
\delta \ B_{\ \mu \ \nu}^{\ r} 
\end{array} 
\hspace*{0.1cm} \begin{array}{c}
\delta \ B_{\ \mu \nu}^{\ r}
\vspace*{0.2cm} \\
\hline  \vspace*{-0.3cm} \\
\delta \ \left ( \ \partial_{\ \varrho} \ W_{\ \sigma}^{\ s} \ \right )
\end{array} \right \rbrace
\ - \ \begin{array}{c}
\delta \ {\cal{L}}
\vspace*{0.2cm} \\
\hline  \vspace*{-0.3cm} \\
\delta \ B_{\ \mu \ \nu}^{\ r}
\end{array}
\hspace*{0.1cm} \begin{array}{c}
\delta \ B_{\ \mu \nu}^{\ r}
\vspace*{0.2cm} \\
\hline  \vspace*{-0.3cm} \\
\delta \ W_{\ \sigma}^{\ s}
\end{array}
\ = \ 0
\vspace*{0.1cm} \\
\begin{array}{c}
\delta \ {\cal{L}}
\vspace*{0.2cm} \\
\hline  \vspace*{-0.3cm} \\
\delta \ B_{\ \mu \ \nu}^{\ r}
\end{array}
\ = \ \begin{array}{c}
1
\vspace*{0.2cm} \\
\hline  \vspace*{-0.3cm} \\
2 \ g^{\ 2}
\end{array} \hspace*{0.1cm} B^{\ \nu \mu \ r}
\end{array}
\end{equation}

\noindent
The extremum is to be taken over variations of the connection components
$\ \delta \ W_{\ \sigma}^{\ s} \ ( \ x \ ) \ $ restricted to vanish on a 
limiting finite or infinite space time domain .

\noindent
Substitting the variations of the field strengths 

\vspace*{-0.3cm}
\begin{equation}
\label{eq:3-12}
\begin{array}{l}
\begin{array}{c}
\delta \ B_{\ \mu \nu}^{\ r}
\vspace*{0.2cm} \\
\hline  \vspace*{-0.3cm} \\
\delta \ \left ( \ \partial_{\ \varrho} \ W_{\ \sigma}^{\ s} \ \right )
\end{array} \ =
\ \delta^{\ rs} \ \left ( \ \delta_{\ \mu}^{\ \varrho} 
\ \delta_{\ \nu}^{\ \sigma} \ - \ \delta_{\ \nu}^{\ \varrho}
\ \delta_{\ \mu}^{\ \sigma} \ \right )
\vspace*{0.1cm} \\
\begin{array}{c}
\delta \ B_{\ \mu \nu}^{\ r}
\vspace*{0.2cm} \\
\hline  \vspace*{-0.3cm} \\
\delta \ W_{\ \sigma}^{\ s}
\end{array} \ = \ f_{\ s r t} \ \left ( \ \delta_{\ \nu}^{\ \sigma} 
\ W_{\ \mu}^{\ t} \ - \ \delta_{\ \mu}^{\ \sigma} \ W_{\ \nu}^{\ t}
\ \right )
\end{array}
\vspace*{-0.3cm}
\end{equation}

}



{\color{blue} 

\noindent
the equations of motion become

\vspace*{-0.5cm}
\begin{equation}
\label{eq:3-13}
\hspace*{0.3cm} \begin{array}{l}
\partial_{\ \varrho} 
\ \left \lbrace
\ \begin{array}{c}
1
\vspace*{0.2cm} \\
\hline  \vspace*{-0.3cm} \\
g^{\ 2}
\end{array} \hspace*{0.1cm} B^{\ \sigma \varrho \ s} \ \right \rbrace
\ +
\ \begin{array}{c}
1
\vspace*{0.2cm} \\
\hline  \vspace*{-0.3cm} \\
g^{\ 2}
\end{array} \hspace*{0.1cm}
f_{\ s t r } \ \ W_{\ \varrho}^{\ t}
\ \left \lbrace
\ B^{\ \sigma \varrho \ r} \ \right \rbrace
\vspace*{0.1cm} \\
\left ( \ d_{\ t} \ ( \ {\cal{D}} \ = \ \mbox{adjoint representation} \ )
\ \right )_{\ s r} \ = \ \left ( \ ad_{\ t} \ \right )_{\ s r}
\ = \ f_{\ s t r}
\hspace*{0.2cm} \longrightarrow
\vspace*{0.1cm} \\
\left ( \ D_{\ \varrho} \ ( \ ad \ ) \ \right )_{\ s r} 
\ \left \lbrace \begin{array}{c}
1
\vspace*{0.2cm} \\
\hline  \vspace*{-0.3cm} \\
g^{\ 2}
\end{array} \hspace*{0.1cm}
B^{\ \sigma \varrho \ r} \right \rbrace \ = \ 0
\vspace*{0.1cm} \\
\left ( \ D_{\ \varrho} \ ( \ ad \ ) \ \right )_{\ s r} \ =
\ \partial_{\ \varrho} \ \delta_{\ s r} \ +
\ W_{\ \varrho}^{\ t} \ \left ( \ ad_{\ t} \ \right )_{\ s r}
\end{array}
\end{equation}

\noindent
Here we repeat the Lagrangean density pertaining to quark-antiquark flavors
as defined in eq. \ref{eq:3-10}

}



{\color{blue} 

\vspace*{-0.3cm}
\begin{equation}
\label{eq:3-14}
\begin{array}{l}
{\cal{L}}_{\ \left \lbrace q \right \rbrace}
\ = 
\vspace*{0.1cm} \\
= \ \sum_{\ q-fl} \ \overline{q}^{\ \dot{c}'}
\ \left \lbrace \frac{i}{2}
\ \gamma^{\ \mu} \ \left \lbrack
\begin{array}{c}
\ \left ( \ 
\begin{array}{c}
\vspace*{-0.5cm} \\
\rightarrow
\vspace*{-0.25cm} \\
D
\end{array}_{\ \mu} \ \left ( \ 3 \ \right ) \ \right )_{\ c' \dot{c}}
\ - 
\vspace*{0.1cm} \\
- \ \left ( \
\begin{array}{c}
\vspace*{-0.5cm} \\
\leftarrow
\vspace*{-0.25cm} \\
D
\end{array}_{\ \mu} \ \left ( \ \overline{3} \ \right ) 
\ \right )_{\ \dot{c} c'}
\end{array} \right \rbrack
\ - \ m_{\ q} \ \delta_{\ c' \dot{c}} \right \rbrace
\ q^{\ c}
\vspace*{0.1cm} \\
\begin{array}{lll}
\left ( \
\begin{array}{c}
\vspace*{-0.5cm} \\
\rightarrow
\vspace*{-0.25cm} \\
D
\end{array}_{\ \mu} \ \left ( \ 3 \ \right ) \ \right )_{\ c' \dot{c}}
& = & \begin{array}{c}
\vspace*{-0.5cm} \\
\rightharpoonup
\vspace*{-0.25cm} \\
\partial_{\ \mu}
\end{array} \ \delta_{\ c' \dot{c}} \ + \ W_{\ \mu}^{\ r} \ \frac{1}{i} 
\ \left ( \ \frac{1}{2} \ \lambda^{\ r} \ \right )_{\ c' \dot{c}}
\vspace*{0.1cm} \\
\left ( \
\begin{array}{c}
\vspace*{-0.5cm} \\
\leftarrow
\vspace*{-0.25cm} \\
D
\end{array}_{\ \mu} \ \left ( \ \overline{3} \ \right )
\ \right )_{\ \dot{c} c'} & =
& \begin{array}{c}
\vspace*{-0.5cm} \\
\leftharpoonup
\vspace*{-0.25cm} \\
\partial_{\ \mu}
\end{array} \ \delta_{\ \dot{c} c'} \ - \ W_{\ \mu}^{\ r} \ \frac{1}{i}
\ \left ( \ \frac{1}{2} \ \overline{\lambda}^{\ r} \ \right )_{\ \dot{c} c'}
\vspace*{0.1cm} \\
& = &
\begin{array}{c}
\vspace*{-0.5cm} \\
\leftharpoonup
\vspace*{-0.25cm} \\
\partial_{\ \mu}
\end{array} \ \delta_{\ c' \dot{c}} \ - \ W_{\ \mu}^{\ r} \ \frac{1}{i}
\ \left ( \ \frac{1}{2} \ \lambda^{\ r} \ \right )_{\ c' \dot{c}}
\end{array}
\end{array}
\end{equation}

\noindent
Substituting the relative to each other complex conjugate representation
matrices of $\ Lie \ ( \ SU3_{\ c} \ ) \ $ in eq. \ref{eq:3-14} it follows

\vspace*{-0.3cm}
\begin{equation}
\label{eq:3-15}
\begin{array}{l}
{\cal{L}}_{\ \left \lbrace q \right \rbrace}
\ = 
\vspace*{0.1cm} \\
= \ \sum_{\ q-fl} \ \overline{q}^{\ \dot{c}'}
\ \ \left \lbrace
\ \gamma^{\ \mu} 
\ \left \lbrack 
\begin{array}{l} \frac{i}{2} \ \begin{array}{c}
\vspace*{-0.5cm} \\
\rightleftharpoons
\vspace*{-0.20cm} \\
\partial \vspace*{-0.08cm}
\end{array}_{\mu} \hspace*{0.1cm}  \delta_{\ c' \dot{c}} \ +
\vspace*{0.1cm} \\
+ \hspace*{0.1cm} W_{\ \mu}^{\ r}
\ \left ( \ \frac{1}{2} \ \lambda^{\ r} \ \right )_{\ c' \dot{c}}
\end{array} \right \rbrack
\ - \ m_{\ q} \ \ \delta_{\ c' \dot{c}}
\ \right \rbrace \ q^{\ c}
\end{array}
\end{equation}

}



{\color{blue} 

\noindent
In eqs. \ref{eq:3-14} and \ref{eq:3-15} 
$\ \lambda^{\ r} \ ; \ r \ = \ 1,\cdot,8 \ $ denote the eight Gell-Mann
matrices defining the Lie algebra representation ( 3 ) of $\ SU3 \ $ 
and associated structure constants, with the identifications

\vspace*{-0.3cm}
\begin{equation}
\label{eq:3-16}
\begin{array}{l}
{\cal{G}} \ = \ SU3_{\ c} 
\hspace*{0.2cm} ; \hspace*{0.2cm}
{\cal{D}} \ = 3 \ :
\vspace*{0.1cm} \\
\left ( \ d_{\ r} \ \right )_{\ \alpha \beta}
\ = \ \frac{1}{i} 
\ \left ( \ \frac{1}{2} \ \lambda^{\ r} \ \right )_{\ \alpha \beta}
\hspace*{0.2cm} ; \hspace*{0.2cm}
\left \lbrace \ \alpha \ = \ c' \ , \ \beta \ = \ \dot{c}
\ \right \rbrace \ = \ 1,2,3
\vspace*{0.1cm} \\
tr_{\ ({\cal{D}} \ = \ 3)} \ \left ( \ \frac{1}{2} \ \lambda^{\ r} \ \right ) 
\ \left ( \ \frac{1}{2} \ \lambda^{\ s} \ \right ) \ = \ T_{\ 2} \ ( \ 3 \ ) 
\ \delta_{\ r s} 
\hspace*{0.2cm} ; \hspace*{0.2cm} 
T_{\ 2} \ ( \ 3 \ ) \ = \ \frac{1}{2}
\vspace*{0.2cm} \\
\left \lbrack \ \frac{1}{2} \ \lambda^{r} \ , \ \frac{1}{2} \ \lambda^{s} 
\ \right \rbrack \ = \ i \ f_{\ r st} \ \frac{1}{2} \ \lambda^{\ t}
\vspace*{0.1cm} \\
f_{\ r u v} \ f_{\ s u v} \ = \ C_{\ 2} \ ( \ \mbox{ad} \ ) 
\ \delta_{\ r s} 
\hspace*{0.2cm} ; \hspace*{0.2cm}
C_{\ 2} \ ( \ \mbox{ad} \ ) \ = \ 3
\end{array}
\end{equation}

\noindent
The equations of motion ( eqs. \ref{eq:3-11} - \ref{eq:3-13} ) are modified
by the quark-antiquark current

\vspace*{-0.3cm}
\begin{equation}
\label{eq:3-17}
\begin{array}{l}
\partial_{\ \varrho}
\ \left \lbrace 
\begin{array}{c}
\delta \ {\cal{L}}
\vspace*{0.2cm} \\
\hline  \vspace*{-0.3cm} \\
\delta \ B_{\ \mu \ \nu}^{\ r} 
\end{array} 
\hspace*{0.1cm} \begin{array}{c}
\delta \ B_{\ \mu \nu}^{\ r}
\vspace*{0.2cm} \\
\hline  \vspace*{-0.3cm} \\
\delta \ \left ( \ \partial_{\ \varrho} \ W_{\ \sigma}^{\ s} \ \right )
\end{array} \right \rbrace
\ - \ \begin{array}{c}
\delta \ {\cal{L}}
\vspace*{0.2cm} \\
\hline  \vspace*{-0.3cm} \\
\delta \ B_{\ \mu \ \nu}^{\ r}
\end{array}
\hspace*{0.1cm} \begin{array}{c}
\delta \ B_{\ \mu \nu}^{\ r}
\vspace*{0.2cm} \\
\hline  \vspace*{-0.3cm} \\
\delta \ W_{\ \sigma}^{\ s}
\end{array}
\ = \ \begin{array}{c}
\delta \ {\cal{L}}_{\ \left \lbrace q \right \rbrace}
\vspace*{0.2cm} \\
\hline  \vspace*{-0.3cm} \\
\delta \ W_{\ \sigma}^{\ s}
\end{array}
\vspace*{0.1cm} \\
\begin{array}{c}
\delta \ {\cal{L}}_{\ \left \lbrace q \right \rbrace}
\vspace*{0.2cm} \\
\hline  \vspace*{-0.3cm} \\
\delta \ W_{\ \sigma}^{\ s}
\end{array} \hspace*{0.1cm} = 
\ \sum_{\ q-fl} \ \overline{q}^{\ \dot{c}'}
\ \left \lbrace \ \gamma^{\ \sigma}
\ \left ( \ \frac{1}{2} \ \lambda^{\ s} \ \right )_{\ c' \dot{c}}
\ \right \rbrace \ q^{\ c}
\ = \ \left ( \hspace*{0.1cm} j^{\ \sigma \ s} 
\ \right )_{\ \left \lbrace q \right \rbrace}
\end{array}
\end{equation}

}



{\color{blue} 

\vspace*{-0.3cm}
\noindent
Eq. \ref{eq:3-13} becomes

\vspace*{-0.3cm}
\begin{equation}
\label{eq:3-18}
\begin{array}{l}
\left ( \ D_{\ \varrho} \ ( \ ad \ ) \ \right )_{\ s r} 
\ \left \lbrace \begin{array}{c}
1
\vspace*{0.2cm} \\
\hline  \vspace*{-0.3cm} \\
g^{\ 2}
\end{array} \hspace*{0.1cm}
B^{\ \sigma \varrho \ r} \right \rbrace \ =
\ \left ( \hspace*{0.1cm} j^{\ \sigma \ s}
\ \right )_{\ \left \lbrace q \right \rbrace}
\vspace*{0.2cm} \\
\left ( \ D_{\ \varrho} \ ( \ ad \ ) \ \right )_{\ s r} \ =
\ \partial_{\ \varrho} \ \delta_{\ s r} \ +
\ W_{\ \varrho}^{\ t} \ \left ( \ ad_{\ t} \ \right )_{\ s r}
\hspace*{0.2cm} ; \hspace*{0.2cm}
\left ( \ ad_{\ t} \ \right )_{\ s r}
\ = \ f_{\ s t r}
\vspace*{0.2cm} \\
\left ( \hspace*{0.1cm} j^{\ \sigma \ s} 
\ \right )_{\ \left \lbrace q \right \rbrace} \ =
\ \sum_{\ q-fl} \ \overline{q}^{\ \dot{c}'}
\ \left \lbrace \ \gamma^{\ \sigma}
\ \left ( \ \frac{1}{2} \ \lambda^{\ s} \ \right )_{\ c' \dot{c}}
\ \right \rbrace \ q^{\ c}
\end{array}
\end{equation}

\begin{center}
\vspace*{-0.0cm}
{\bf \color{cyan} 3 ${\cal{A}}$ - Canonical commutation rules for
gauge fields in axial gauges from bare Lagrangean density \\
and residual fixed time gauge invariance
}
 \label{'3A'}
 \end{center}

\noindent
Because we use the metric 
$\ \eta_{\ \mu \nu} \ = \ diag \ ( \ 1 \ , \ -1 \ , \ -1 \ , \ -1 \ ) \ $
as given in eq. \ref{eq:3-4}, choosing as field coordinates the
covariant space components $\ W_{\ \mu}^{\ t \ {\cal{A}}} \ \rightarrow
\ W_{\ m}^{\ t \ {\cal{A}}} \ ; \ m \ = \ 1,2,3 \ $, some care must be taken
with respect to the difference of sign between contra- and covariant space 
vectors. The above choice corresponds to coordinates 
$\ \widehat{q}_{\ m} \ = \ - \ \widehat{q}^{\ m} \ , 
\ \widehat{p}^{\ n} \ = \ L_{\ \partial_{\ t} 
\ \widehat{q}_{\ n}} \ $ for quantum mechanical conjugate variables in uncurved
space-time, with the commutation rules

\vspace*{-0.3cm}
\begin{equation}
\label{eq:3A-1}
\begin{array}{l}
\left \lbrack \ \widehat{p}^{\ n} \ , \ \widehat{q}_{\ m} \ \right \rbrack
\ = \ - 
\left \lbrack \ \widehat{p}^{\ n} \ , \ \widehat{q}^{\ m} \ \right \rbrack
\ = \ \frac{1}{i} \ \delta^{\ n}_{\ m} \ \P
\end{array}
\end{equation}

}



{\color{blue} 

\noindent
In order to show things step by step, I repeat two relations from eqs.
\ref{eq:3-11} and \ref{eq:3-12}

\vspace*{-0.3cm}
\begin{equation}
\label{eq:3A-2}
\begin{array}{l}
\begin{array}{c}
\delta \ {\cal{L}}
\vspace*{0.2cm} \\
\hline  \vspace*{-0.3cm} \\
\delta \ \left ( \ \partial_{\ \varrho} \ W_{\ \sigma}^{\ s} \ \right )
\end{array}
\ =
\ \begin{array}{c}
\delta \ {\cal{L}}
\vspace*{0.2cm} \\
\hline  \vspace*{-0.3cm} \\
\delta \ B_{\ \mu \ \nu}^{\ r}
\end{array}
\ \begin{array}{c}
\delta \ B_{\ \mu \nu}^{\ r}
\vspace*{0.2cm} \\
\hline  \vspace*{-0.3cm} \\
\delta \ \left ( \ \partial_{\ \varrho} \ W_{\ \sigma}^{\ s} \ \right )
\end{array}
\ {\color{magenta}
\ = \ \begin{array}{c}
1
\vspace*{0.2cm} \\
\hline  \vspace*{-0.3cm} \\
g^{\ 2}
\end{array} \hspace*{0.1cm} B^{\ \sigma \varrho \ s}
}
\vspace*{0.1cm} \\
\begin{array}{c}
\delta \ {\cal{L}}
\vspace*{0.2cm} \\
\hline  \vspace*{-0.3cm} \\
\delta \ B_{\ \mu \ \nu}^{\ r}
\end{array}
\ = \ \begin{array}{c}
1
\vspace*{0.2cm} \\
\hline  \vspace*{-0.3cm} \\
2 \ g^{\ 2}
\end{array} \hspace*{0.1cm} B^{\ \nu \mu \ r}
\vspace*{0.1cm} \\
\begin{array}{c}
\delta \ B_{\ \mu \nu}^{\ r}
\vspace*{0.2cm} \\
\hline  \vspace*{-0.3cm} \\
\delta \ \left ( \ \partial_{\ \varrho} \ W_{\ \sigma}^{\ s} \ \right )
\end{array} \ =
\ \delta^{\ rs} \ \left ( \ \delta_{\ \mu}^{\ \varrho} 
\ \delta_{\ \nu}^{\ \sigma} \ - \ \delta_{\ \nu}^{\ \varrho}
\ \delta_{\ \mu}^{\ \sigma} \ \right )
\end{array}
\end{equation}

\noindent
In the temporal axial gauges we are using it follow from eq. \ref{eq:3A-2}

\vspace*{-0.5cm}
\begin{equation}
\label{eq:3A-3}
\begin{array}{l}
{\cal{A}} \ :
\ \begin{array}[t]{l}
\Pi^{\ s \ m} \ =
\ \begin{array}[t]{lll}
\ \begin{array}{c}
\delta \ {\cal{L}}
\vspace*{0.2cm} \\
\hline  \vspace*{-0.3cm} \\
\delta \ \left ( \ \partial_{\ \varrho \ \rightarrow \ 0} 
\ W_{\ \sigma \ \rightarrow \ m}^{\ s} \ \right )
\end{array}
& = & \begin{array}{c}
1
\vspace*{0.2cm} \\
\hline  \vspace*{-0.3cm} \\
g^{\ 2}
\end{array} \hspace*{0.1cm} B^{\ m 0 \ s}
\vspace*{0.1cm} \\
& = & - 
\ \begin{array}{c}
1
\vspace*{0.2cm} \\
\hline  \vspace*{-0.3cm} \\
g^{\ 2}
\end{array}
\hspace*{0.15cm} E^{\ s \ m} 
\end{array}
\vspace*{0.2cm} \\
\hline  \vspace*{-0.3cm} \\
\left .
\Pi^{\ s \ m} \ ( t , \vec{x} ) 
\ \sim \ \widehat{p}^{\ s \ m} \ ( \ \vec{x} \ )
\hspace*{0.1cm} ; \hspace*{0.1cm}
\begin{array}{l}
W^{\ r}_{\ n} \ ( \ t , \vec{x} \ ) 
\ \sim \ \widehat{q}^{\ r}_{\ n} \ ( \ \vec{x} \ ) 
\vspace*{0.1cm} \\
\widehat{q}^{\ r}_{\ n} \ ( \ \vec{x} \ )
\ \equiv
\ - \ \widehat{q}^{\ r \ n} \ ( \ \vec{x} \ )
\vspace*{-0.0cm}
\end{array}
\ \right |_{ \begin{array}{l} 
{\scriptscriptstyle gauge}
\vspace*{-0.1cm} \\
{\scriptscriptstyle fields}
\vspace*{0.5cm}
\end{array}
}
\vspace*{-0.1cm} \\
- \ E^{\ s \ m} \ = \ \left ( \ W^{\ r}_{\ m} \ \right )^{\ \bullet} 
\end{array}
\end{array}
\vspace*{0.2cm}
\end{equation}

}



{\color{blue} 

\vspace*{-0.1cm}
\noindent
We cast eq. \ref{eq:3A-3} into a 'canonical' form, adapted to a plane with
fixed time , whereas $\ \vec{x} \ , \ \vec{y} \ , \ \cdots \ $,
varying on the chosen plane, become
continuous labels of canonical variables

\vspace*{-0.2cm}
\begin{equation}
\label{eq:3A-4}
\begin{array}{c}
\begin{array}{|l|l|l|}
\hline  \vspace*{-0.48cm} \\
 & & \vspace*{-0.3cm} \\
{\cal{A}} \ , \ t & \Pi^{\ s \ m} \ ( \ \vec{x} \ ) \ = 
\  {\cal{L}}_{\ , \ \left . {\cal{Q}}^{\ \bullet} 
\right .^{ s}_{ m} \ ( \ \vec{x} \ )} 
& {\cal{Q}}^{\ r}_{\ n} \ ( \ \vec{y} \ ) \ = 
\ W^{\ r}_{\ n} \ ( \ t \ , \ \vec{y} \ )
\vspace*{-0.3cm} \\ & &  \\ \hline \vspace*{-0.5cm} \\
& \multicolumn{2}{c@{\hspace*{0.2cm}}|}
{\Pi^{\ s \ m} \ ( \ \vec{x} \ ) \ = 
\ \ {\cal{L}}_{\ , \ \left . {\cal{Q}}^{\ \bullet}
\right .^{ s}_{ m} \ ( \ \vec{x} \ )}
\ = \ - \ \begin{array}{r}
1
\vspace*{0.2cm} \\
\hline  \vspace*{-0.3cm} \\
g^{\ 2}
\end{array}
\ E^{\ s \ m} \ ( \ \vec{x} \ )
\hspace*{0.15cm} {\color{red} *}
\vspace*{0.4cm} \hspace*{0.2cm} } \hspace*{-0.065cm} 
\vspace*{-0.4cm} \\ \hline \vspace*{-0.5cm} \\
& \multicolumn{2}{c@{\hspace*{0.2cm}}|}
{- \ E^{\ s \ m} \ ( \ \vec{x} \ ) \ = 
\ \left . {\cal{Q}}^{\ \bullet}
\right .^{ s}_{ m} \ ( \ \vec{x} \ ) \ =
\ \left . W^{\ \bullet} \right .^{\ s}_{\ m} \ ( \ t \ , \ \vec{x} \ )
\vspace*{0.3cm} \hspace*{0.2cm} } \hspace*{-0.065cm} 
\vspace*{-0.3cm} \\ \hline 
\end{array}
\vspace*{0.2cm} \\
{\color{red} * : \mbox{\begin{tabular}[t]{l}
as stated previously , we do not allow any dependence of the
\vspace*{-0.15cm} \\
bare coupling constant g on a scale, which enters upon
\vspace*{-0.15cm} \\
renormalization giving rise to the
trace anomaly .
\end{tabular} 
}
}
\end{array}
\end{equation}

\noindent
With canonical variables for gauge fields defined in eq. \ref{eq:3A-4}
the associated equal time commutators take the form, for one selected
common time $\ t \ $

\vspace*{-0.3cm}
\begin{equation}
\label{eq:3A-5}
\begin{array}{ll}
{\cal{A}} \ : &
\left \lbrack 
\ W^{\ r}_{\ m} \ ( \ t \ , \ \vec{x} \ ) \ ,
\ \Pi^{\ s \ n} \ ( \ t \ , \ \vec{y} \ ) \ \right \rbrack
\ = \ i \ \delta^{\ r \ s} \ \delta_{\ m}^{\ n} 
\ \delta^{\ (3)} \ ( \ \vec{x} \ - \ \vec{y} \ ) \ \P  
\vspace*{0.1cm} \\
& \left \lbrack
\ W^{\ r}_{\ m} \ ( \ t \ , \ \vec{x} \ ) \ ,
\ W^{\ s}_{\ n} \ ( \ t \ , \ \vec{y} \ ) \ \right \rbrack \ = \ 0
\vspace*{0.1cm} \\
& \left \lbrack
\ \Pi^{\ r \ m} \ ( \ t \ , \ \vec{x} \ ) \ ,
\ \Pi^{\ s \ n} \ ( \ t \ , \ \vec{y} \ ) \ \right \rbrack \ = \ 0
\end{array}
\end{equation}

\noindent
In eq. \ref{eq:3A-5} $\ \P \ $ denotes the unit operator in a suitable
extension of the Hilbert space of fully gauge invariant states, allowing
for the gauge variant canonical variables to be fully represented .

}



{\color{blue} 

\noindent
In the follwing we concentrate on the nonvanishing commutation rules of the 
canonical conjugate variables as displayed in the first relation of 
eq. \ref{eq:3A-5}, which transforms into

\vspace*{-0.3cm}
\begin{equation}
\label{eq:3A-7}
\begin{array}{ll}
{\cal{A}} \ : &
\left \lbrack 
\  W^{\ r}_{\ m} \ ( \ t \ , \ \vec{x} \ ) \ ,
\ E^{\ s \ n} \ ( \ t , \vec{y} \ ) \ \right \rbrack
\ = \ \frac{1}{i} \ g^{\ 2} \ \delta^{\ r \ s} \ \delta_{\ m}^{\ n} 
\ \delta^{\ (3)} \ ( \vec{x} \ - \ \vec{y} ) \ \P  
\vspace*{0.1cm} \\
& E^{\ s \ n} \ ( \ t \ , \ \vec{y} \ ) \ = 
 -  \left . W^{\ \bullet} \right .^{\ s}_{\ m} \ ( \ t  , \vec{y} \ )
\end{array}
\end{equation}

\begin{center}
\vspace*{-0.0cm}
{\bf \color{cyan} 3 ${\cal{A}}$ R - Residual fixed time gauge invariance
}
 \label{'3A-R'}
 \end{center}

\noindent
The conjugate variables in eq. \ref{eq:3A-5} allow further gauge
transformations, i.e. are not fixed by the choice 

\vspace*{-0.3cm}
\begin{equation}
\label{eq:3A-8}
\begin{array}{ll}
{\cal{A}} \ : & W^{\ r}_{\ 0} \ ( \ t' \ , \ \vec{x} \ ) \ = \ 0
\hspace*{0.2cm} ; \hspace*{0.2cm} 
\forall \ t'
\end{array}
\end{equation}

}



{\color{blue} 

\vspace*{-0.0cm}
\noindent
We repeat the covariant derivative for the adjoint representation eq. 
\ref{eq:3-13} for general gauges

\vspace*{-0.3cm}
\begin{equation}
\label{eq:3A-9}
\begin{array}{l}
\left ( \ d_{\ t} \ ( \ {\cal{D}} \ = \ \mbox{adjoint representation} \ )
\ \right )_{\ s r} \ = \ \left ( \ ad_{\ t} \ \right )_{\ s r}
\ = \ f_{\ s t r}
\vspace*{0.1cm} \\
\left \lbrack \  ad_{\ p} \ , \  ad_{\ q} \ \right \rbrack \ = \ f_{\ p q r}
\  ad_{\ r}
\vspace*{0.2cm} \\
e.g. \hspace*{0.1cm} : \begin{array}[t]{l}
\ \left ( \ D_{\ \varrho} \ ( \ ad \ ) \ \right )_{\ s r} 
\ \left \lbrace 
\ B_{\ \sigma \tau}^{\ r} \right \rbrace \ =
\vspace*{0.2cm} \\
\hspace*{1.0cm} = 
\ \partial_{\ \varrho} 
\ \left \lbrace
\hspace*{0.1cm} B_{\ \sigma \tau}^{\ s} \ \right \rbrace
\ +
\ f_{\ s t r } \ \ W_{\ \varrho}^{\ t}
\ \left \lbrace
\ B_{\ \sigma \tau}^{\ r} \ \right \rbrace 
\end{array}
\end{array}
\end{equation}

\noindent
We also retrace gauge transformations of the connection fields 
$\ {\cal{W}}_{\ \mu} \ ( \ {\cal{D}} \ ) \ $ relative to the irreducible
representation $\ {\cal{D}} \ $ and its Lie algebra $\ Lie \ {\cal{D}} \ $
as defined in eqs. \ref{eq:2-1} and \ref{eq:2-2}

\vspace*{-0.5cm}
\begin{equation}
\label{eq:3A-10}
\begin{array}{l}
\left ( \ {\cal{W}}_{\ \mu} \ ( \ {\cal{D}} \ ) \ \right )_{\ \alpha \beta}
\ ( \ x \ )
\ = \ {\cal{W}}^{\ r}_{\ \mu} \ ( \ x \ ) \ \left ( \ d_{\ r} 
\ \right )_{\ \alpha \beta}
\vspace*{0.1cm} \\
d_{\ r} \ = \ - \ d_{\ r}^{\ \dagger} \ = \ \frac{1}{i} \ J_{\ r}
\ \in \ Lie \ ( \ {\cal{D}} \ )
\hspace*{0.2cm} ; \hspace*{0.2cm}
\left \lbrack \ d_{\ p} \ , \ d_{\ q} \ \right \rbrack \ = \ f_{\ p q r}
\ d_{\ r}
\vspace*{0.1cm} \\
r \ , \ p \ , \ q \ = \ 1, \ \cdots , \ dim \ {\cal{G}} 
\hspace*{0.2cm} ; \hspace*{0.2cm}
\alpha \ , \ \beta \ = \ 1, \ \cdots , \ dim \ {\cal{D}}
\vspace*{0.1cm} \\
\begin{array}{cll}
\left ( \ {\cal{W}}_{\ \mu} \ ( \ {\cal{D}} \ ) \ \right )_{\ \alpha \beta}
\ ( \ x \ )
& : &
\begin{array}[t]{l}
\mbox{local operator} \ \times
\vspace*{-0.1cm} \\
\ {\cal{D}}- 
\mbox{representation valued}
\vspace*{-0.1cm} \\
\mbox{connection over}
\vspace*{-0.1cm} \\
\mbox{flat space time} \ x
\end{array}
\vspace*{-0.1cm} \\
d_{\ r} \ \in \ Lie \ ( \ {\cal{D}} \ )
& : &
\begin{array}[t]{l}
\mbox{basis of antihermitian matrices forming}
\vspace*{-0.1cm} \\
\mbox{an irreducible representation of the} 
\vspace*{-0.1cm} \\
\mbox{Lie algebra of} \ {\cal{G}}
\end{array}
\vspace*{-0.1cm} \\
{\cal{W}}^{\ r}_{\ \mu} \ ( \ x \ )
& : &
\begin{array}[t]{l}
4 \ \times \ dim \ {\cal{G}} \ \mbox{components of hermitian} 
\vspace*{-0.1cm} \\
\mbox{local connection fields}
\end{array}
\end{array} 
\vspace*{0.1cm} \\
\hspace*{-0.3cm}
{\color{cyan}
\begin{array}{@{\hspace*{12.0cm}}c} 
\hline \\
\end{array}
}
\hspace*{0.3cm}
\vspace*{-1.5cm} 
\end{array}
\vspace*{1.2cm}
\end{equation}

}



{\color{blue} 

\vspace*{-0.3cm}
\noindent
{\color{red} * :} Replacing the gauge field Lagrangean density according to
the trace anomaly \cite{PMtran} 

\vspace*{-0.3cm}
\begin{equation}
\label{eq:3A-6}
\begin{array}{l}
{\cal{L}}_{\ gauge} \ = \ - 
\ \begin{array}{c}
1
\vspace*{0.2cm} \\
\hline  \vspace*{-0.3cm} \\
g^{\ 2}
\end{array} \hspace*{0.1cm} {\cal{X}}
\hspace*{0.1cm} \rightarrow \hspace*{0.1cm}
\overline{{\cal{L}}} \ = \ - 
\ \left \lbrack \ \begin{array}{c}
1
\vspace*{0.2cm} \\
\hline  \vspace*{-0.3cm} \\
\overline{g}^{\ 2} \ ( \ {\cal{X}} \ )
\end{array} \hspace*{0.1cm} - \ J 
\ \right \rbrack \ {\cal{X}}
\vspace*{0.1cm} \\
{\cal{X}} \ = \ \frac{1}{4} 
\ B_{\ \mu \nu}^{\ r} \ B^{\ \mu \nu \ r}
\end{array}
\end{equation}

\noindent
In eq. \ref{eq:3A-6} $\ J \ $ stands for a suitable constant,
while $\ \overline{g} \ ( \ \overline{l} \ ) \ $
denotes the scale dependent coupling constant, as discussed also in ref. 
\cite{PagTomboulis} , 
established initially in the perturbative domain of QCD  

\vspace*{-0.5cm}
\begin{equation}
\label{eq:3A-6a}
\begin{array}{l}
\overline{l} \ = \ \log \ ( \ \overline{\mu} \ / \ \mu \ )
\ = 
\ \frac{1}{4} \ \log \ \left ( \ \overline{\mu}^{\ 4} 
\ / \ \mu^{\ 4} \ \right )
\hspace*{0.1cm} \rightarrow \hspace*{0.1cm}
\frac{1}{4} \ \log \ \left ( \ {\cal{X}} \ / \ \mu^{\ 4} \ \right )
\vspace*{0.1cm} \\
\hspace*{1.1cm} \rightarrow \hspace*{0.1cm}
\frac{1}{8} \ \log \ \left ( \ \left ( {\cal{X}} \ / \ \mu^{\ 4} 
\right )^{\ 2} \ \right )
\vspace*{0.1cm} \\
\overline{g} \ ( \ {\cal{X}} \ ) \ = \ \overline{g} 
\ \left ( \ \overline{l} \ =
\ \frac{1}{8} \ \log \ \left ( \ \left ( \ {\cal{X}} \ / \ \mu^{\ 4} 
\ \right )^{\ 2}  
\ \right ) \ \right )
\vspace*{0.1cm} \\
\begin{array}{c}
d
\vspace*{+0.1cm} \\
\hline  \vspace*{-0.3cm} \\
d \ \overline{l}
\end{array} \hspace*{0.1cm} 
\overline{g} \ = \ \beta \ ( \ \overline{g} \ ) \ =
\ \overline{g} \ \overline{\kappa} 
\left \lbrack \ - \ \left ( \ b_{\ 0} \ + \ b_{\ 1} \
\overline{\kappa}^{\ 1} \ + \ \cdots \ \right ) \ \right \rbrack
\vspace*{0.1cm} \\
\overline{\kappa} \ = 
\ \begin{array}{c}
\overline{g}^{\ 2}
\vspace*{0.2cm} \\
\hline  \vspace*{-0.3cm} \\
16 \ \pi^{\ 2}
\end{array}
\hspace*{0.1cm} ; \hspace*{0.1cm}
\overline{\alpha}_{\ s} \ = \ 4 \ \pi \ \overline{\kappa}
\vspace*{-0.1cm} \\
\begin{array}{c}
d
\vspace*{+0.1cm} \\
\hline  \vspace*{-0.3cm} \\
d \ \overline{l}
\end{array} \hspace*{0.1cm} 
\overline{g}^{\ -2} \ =
\ \begin{array}{c}
1
\vspace*{0.1cm} \\
\hline  \vspace*{-0.3cm} \\
8 \ \pi^{\ 2}
\end{array} \hspace*{0.1cm} \ b \ ( \ \overline{\kappa} \ )
\vspace*{0.1cm} \\
b \ (  \overline{\kappa}  ) \ = 
\ - \ \left ( \ \overline{g} \ \overline{\kappa} \ \right )^{\ -1}
\ \beta \ ( \ \overline{g} \ ) \ =
\ \left (  b_{\ 0} \ + \ b_{\ 1} \
\overline{\kappa}^{\ 1} \ + \ \cdots \right )
\end{array}
\end{equation}

}



{\color{blue} 

\vspace*{-0.3cm}
\noindent
At this point we recall eq. \ref{eq:3-12}

\vspace*{-0.3cm}
\begin{equation}
\label{eq:3A-6b}
\begin{array}{l}
\begin{array}{c}
\delta \ B_{\ \mu \nu}^{\ r}
\vspace*{0.2cm} \\
\hline  \vspace*{-0.3cm} \\
\delta \ \left ( \ \partial_{\ \varrho} \ W_{\ \sigma}^{\ s} \ \right )
\end{array} \ =
\ \delta^{\ rs} \ \left ( \ \delta_{\ \mu}^{\ \varrho} 
\ \delta_{\ \nu}^{\ \sigma} \ - \ \delta_{\ \nu}^{\ \varrho}
\ \delta_{\ \mu}^{\ \sigma} \ \right )
\vspace*{0.1cm} \\
\begin{array}{c}
\delta \ B_{\ \mu \nu}^{\ r}
\vspace*{0.2cm} \\
\hline  \vspace*{-0.3cm} \\
\delta \ W_{\ \sigma}^{\ s}
\end{array} \ = \ f_{\ s r t} \ \left ( \ \delta_{\ \nu}^{\ \sigma} 
\ W_{\ \mu}^{\ t} \ - \ \delta_{\ \mu}^{\ \sigma} \ W_{\ \nu}^{\ t}
\ \right )
\end{array}
\end{equation}

\noindent
and adapt eq. \ref{eq:3A-2} obtained from eq. \ref{eq:3-11} to the quantity
$\ {\cal{X}} \ $ defined in eq. \ref{eq:3A-6}

\vspace*{-0.3cm}
\begin{equation}
\label{eq:3A-6c}
\begin{array}{l}
\begin{array}{c}
\delta \ {\cal{X}}
\vspace*{0.2cm} \\
\hline  \vspace*{-0.3cm} \\
\delta \ \left ( \ \partial_{\ \varrho} \ W_{\ \sigma}^{\ s} \ \right )
\end{array}
\ =
\ \begin{array}{c}
\delta \ {\cal{X}}
\vspace*{0.2cm} \\
\hline  \vspace*{-0.3cm} \\
\delta \ B_{\ \mu \ \nu}^{\ r}
\end{array}
\ \begin{array}{c}
\delta \ B_{\ \mu \nu}^{\ r}
\vspace*{0.2cm} \\
\hline  \vspace*{-0.3cm} \\
\delta \ \left ( \ \partial_{\ \varrho} \ W_{\ \sigma}^{\ s} \ \right )
\end{array}
\ = 
\ B^{\ \varrho \ \sigma \ s}
\vspace*{0.1cm} \\
\begin{array}{c}
\delta \ {\cal{X}}
\vspace*{0.2cm} \\
\hline  \vspace*{-0.3cm} \\
\delta \ B_{\ \mu \ \nu}^{\ r}
\end{array}
\ = \ \frac{1}{2} \ B^{\ \mu \nu \ r}
\hspace*{0.1cm} ; \hspace*{0.1cm}
\ \begin{array}{c}
\delta \ {\cal{X}} 
\vspace*{0.2cm} \\
\hline  \vspace*{-0.3cm} \\
\delta \ \left ( \ W_{\ \sigma}^{\ s} \ \right )
\end{array} \ =
\ \begin{array}{c}
\delta \ {\cal{X}}
\vspace*{0.2cm} \\
\hline  \vspace*{-0.3cm} \\
\delta \ B_{\ \mu \ \nu}^{\ r}
\end{array}
\ \begin{array}{c}
\delta \ B_{\ \mu \nu}^{\ r}
\vspace*{0.2cm} \\
\hline  \vspace*{-0.3cm} \\
\delta \ \left ( \ W_{\ \sigma}^{\ s} \ \right )
\end{array}
\vspace*{0.1cm} \\
{\cal{X}} \ = \ \frac{1}{4} 
\ B_{\ \mu \nu}^{\ r} \ B^{\ \mu \nu \ r} \ = 
\ - \ g^{\ 2} \ {\cal{L}}_{\ gauge}
\end{array}
\end{equation}

}



{\color{blue} 

\vspace*{-0.3cm}
\noindent
The variation of the trace anomaly induced Lagrangean density
$\ \overline{{\cal{L}}} \ $, defined in eq. \ref{eq:3A-6}, is obtained
by repeated use of the chain rule

\vspace*{-0.2cm}
\begin{equation}
\label{eq:3A-6d}
\begin{array}{|l|}
\hline \vspace*{-0.2cm} \\
\begin{array}{lll}
\begin{array}{c}
\delta \ \overline{{\cal{L}}}
\vspace*{0.2cm} \\
\hline  \vspace*{-0.3cm} \\
\delta \ \left ( \ \partial_{\ \varrho} \ W_{\ \sigma}^{\ s} \ \right )
\end{array}
& = &
\left \lbrack \begin{array}{c}
\delta \ \overline{{\cal{L}}}
\vspace*{0.2cm} \\
\hline  \vspace*{-0.3cm} \\
\delta \ {\cal{X}} 
\end{array}
\hspace*{0.1cm} \begin{array}{c}
\delta \ {\cal{X}}
\vspace*{0.2cm} \\
\hline  \vspace*{-0.3cm} \\
\delta \ B_{\ \mu \nu}^{\ r}
\end{array} \right \rbrack
\hspace*{0.1cm} \begin{array}{c}
\delta \ B_{\ \mu \nu}^{\ r}
\vspace*{0.2cm} \\
\hline  \vspace*{-0.3cm} \\
\delta \ \left ( \ \partial_{\ \varrho} \ W_{\ \sigma}^{\ s} \ \right )
\end{array}
\vspace*{0.2cm} \\
\begin{array}{c}
\delta \ \overline{{\cal{L}}} 
\vspace*{0.2cm} \\
\hline  \vspace*{-0.3cm} \\
\delta \ \left ( \ W_{\ \sigma}^{\ s} \ \right )
\end{array} 
& = &
\left \lbrack \begin{array}{c}
\delta \ \overline{{\cal{L}}}
\vspace*{0.2cm} \\
\hline  \vspace*{-0.3cm} \\
\delta \ {\cal{X}}
\end{array}
\ \begin{array}{c}
\delta \ {\cal{X}} 
\vspace*{0.2cm} \\
\hline  \vspace*{-0.3cm} \\
\delta \ B_{\ \mu \ \nu}^{\ r} 
\end{array} \right \rbrack
\hspace*{0.1cm} \begin{array}{c}
\delta \ B_{\ \mu \nu}^{\ r}
\vspace*{0.2cm} \\
\hline  \vspace*{-0.3cm} \\
\delta \ \left ( \ W_{\ \sigma}^{\ s} \ \right )
\end{array}
\vspace*{-0.3cm} \\
& &
\end{array}
\vspace*{-0.0cm} \\
\hline  \vspace*{-0.3cm} \\
\begin{array}{c}
\delta \ B_{\ \mu \nu}^{\ r}
\vspace*{0.2cm} \\
\hline  \vspace*{-0.3cm} \\
\delta \ \left ( \ \partial_{\ \varrho} \ W_{\ \sigma}^{\ s} \ \right )
\end{array} \ =
\ \delta^{\ rs} \ \left ( \ \delta_{\ \mu}^{\ \varrho} 
\ \delta_{\ \nu}^{\ \sigma} \ - \ \delta_{\ \nu}^{\ \varrho}
\ \delta_{\ \mu}^{\ \sigma} \ \right )
\vspace*{-0.2cm} \\ \\
\begin{array}{c}
\delta \ B_{\ \mu \nu}^{\ r}
\vspace*{0.2cm} \\
\hline  \vspace*{-0.3cm} \\
\delta \ W_{\ \sigma}^{\ s}
\end{array} \ = \ f_{\ s r t} \ \left ( \ \delta_{\ \nu}^{\ \sigma} 
\ W_{\ \mu}^{\ t} \ - \ \delta_{\ \mu}^{\ \sigma} \ W_{\ \nu}^{\ t}
\ \right )
\vspace*{-0.3cm} \\ \\
\hline
\end{array}
\end{equation}

}



{\color{blue} 

\noindent
We proceed to evaluate the derivatives in eq. \ref{eq:3A-6d} using
eq, \ref{eq:3A-6a} .

\noindent
The quantities in $\ \left \lbrack \ . \ \right \rbrack \ $ brackets in the
first two relations of eq. \ref{eq:3A-6d} contain expressions wherein both
canonically conjugate variables appear. Hence an ordering is necessary. This
is straightforward in all such situations and amounts to neglecting all local
terms arising from the noncommutativity of the latter.
 
\vspace*{-0.4cm}
\begin{equation}
\label{eq:3A-6e}
\begin{array}{l}
\begin{array}{c}
\delta \ \overline{{\cal{L}}}
\vspace*{0.2cm} \\
\hline  \vspace*{-0.3cm} \\
\delta \ {\cal{X}} 
\end{array}
\ = \ \overline{{\cal{L}}} \ / \ {\cal{X}} 
\ - 
\ \left ( \ \begin{array}{c}
d
\vspace*{+0.2cm} \\
\hline  \vspace*{-0.3cm} \\
d \ \overline{l}
\end{array} \hspace*{0.2cm} 
\overline{g}^{\ -2} \ \right )
\hspace*{0.1cm} 
\begin{array}{c}
\delta \ \overline{l}
\vspace*{+0.2cm} \\
\hline  \vspace*{-0.3cm} \\
\delta \ {\cal{X}}
\end{array} \hspace*{0.1cm} \ {\cal{X}}
\vspace*{+0.1cm} \\
\overline{l} \ =
\ \frac{1}{8} \ \log \ \left ( \ \left ( \ {\cal{X}} \ / \ \mu^{\ 4} 
\ \right )^{\ 2} \ \right )
\vspace*{0.2cm} \\ \hline \vspace*{-0.3cm} \\
\begin{array}{c}
d
\vspace*{+0.2cm} \\
\hline  \vspace*{-0.3cm} \\
d \ \overline{l}
\end{array} \hspace*{0.1cm} 
\overline{g}^{\ -2} \ =
\ \begin{array}{c}
1
\vspace*{0.1cm} \\
\hline  \vspace*{-0.3cm} \\
8 \ \pi^{\ 2}
\end{array} \hspace*{0.1cm} \ b \ ( \ \overline{\kappa} \ )
\vspace*{0.1cm} \\
b \ (  \overline{\kappa}  ) \ = 
\ - \ \left ( \ \overline{g} \ \overline{\kappa} \ \right )^{\ -1}
\ \beta \ ( \ \overline{g} \ ) \ =
\ - \ \left (  b_{\ 0} \ + \ b_{\ 1} \
\overline{\kappa}^{\ 1} \ + \ \cdots \right )
\end{array}
\end{equation}

\noindent
We substitute the derivatives from eq. \ref{eq:3A-6c}

\vspace*{-0.3cm}
\begin{equation}
\label{eq:3A-6f}
\begin{array}{l}
\begin{array}{c}
\delta \ {\cal{X}}
\vspace*{0.2cm} \\
\hline  \vspace*{-0.3cm} \\
\delta \ B_{\ \mu \ \nu}^{\ r}
\end{array}
\ = \ \frac{1}{2} \ B^{\ \mu \nu \ r}
\end{array}
\end{equation}

}



{\color{blue} 

\vspace*{-0.3cm}
\noindent
which gives the derivatives of $\ {\cal{X}} \ $, using eq. \ref{eq:3A-6d}

\vspace*{-0.3cm}
\begin{equation}
\label{eq:3A-6g}
\begin{array}{l}
\begin{array}{c}
\delta \ {\cal{X}}
\vspace*{0.2cm} \\
\hline  \vspace*{-0.3cm} \\
\delta \ \left ( \ \partial_{\ \varrho} \ W_{\ \sigma}^{\ s} \ \right )
\end{array}
\ = \ B^{\ \varrho \ \sigma \ s}
\hspace*{0.2cm} ; \hspace*{0.2cm}
\begin{array}{c}
\delta \ {\cal{X}}
\vspace*{0.2cm} \\
\hline  \vspace*{-0.3cm} \\
\delta \ W_{\ \sigma}^{\ s}
\end{array}
\ = \ \frac{1}{2} \ B^{\ \mu \nu \ r} 
\hspace*{0.15cm} \begin{array}{c}
\delta \ B_{\ \mu \nu}^{\ r}
\vspace*{0.2cm} \\
\hline  \vspace*{-0.3cm} \\
\delta \ W_{\ \sigma}^{\ s}
\end{array}
\vspace*{0.2cm} \\
\begin{array}{lll}
\begin{array}{c}
\delta \ {\cal{X}}
\vspace*{0.2cm} \\
\hline  \vspace*{-0.3cm} \\
\delta \ W_{\ \sigma}^{\ s}
\end{array}
& = &
- \ f_{\ s t r} \ W_{\ \varrho}^{\ t} \ B^{\ \varrho \ \sigma \ r}
\vspace*{-0.1cm} \\
& = &
\frac{1}{2} \ B^{\ \mu \nu \ r} 
\ \ f_{\ s r t} \ \left ( \ \delta_{\ \nu}^{\ \sigma}
\ W_{\ \mu}^{\ t} \ - \ \delta_{\ \mu}^{\ \sigma} \ W_{\ \nu}^{\ t}
\ \right )
\hspace*{0.3cm} {\color{red} (\uparrow)}
\vspace*{0.0cm} \\
& = & 
\frac{1}{2} \ f_{\ s r t} \ 
\ \left ( \ W_{\ \mu}^{\ t} \ B^{\ \mu \ \sigma \ r} 
\ - \ W_{\ \nu}^{\ t} \ B^{\ \sigma \ \nu \ r} \ \right )
\hspace*{0.2cm} {\color{red} (\uparrow)}
\end{array}
\vspace*{0.2cm} \\ \hline \vspace*{-0.3cm} \\
\begin{array}{c}
\delta \ B_{\ \mu \nu}^{\ r}
\vspace*{0.2cm} \\
\hline  \vspace*{-0.3cm} \\
\delta \ \left ( \ \partial_{\ \varrho} \ W_{\ \sigma}^{\ s} \ \right )
\end{array} \ =
\ \delta^{\ rs} \ \left ( \ \delta_{\ \mu}^{\ \varrho} 
\ \delta_{\ \nu}^{\ \sigma} \ - \ \delta_{\ \nu}^{\ \varrho}
\ \delta_{\ \mu}^{\ \sigma} \ \right )
\vspace*{0.2cm} \\
\begin{array}{c}
\delta \ B_{\ \mu \nu}^{\ r}
\vspace*{0.2cm} \\
\hline  \vspace*{-0.3cm} \\
\delta \ W_{\ \sigma}^{\ s}
\end{array} \ = \ f_{\ s r t} \ \left ( \ \delta_{\ \nu}^{\ \sigma} 
\ W_{\ \mu}^{\ t} \ - \ \delta_{\ \mu}^{\ \sigma} \ W_{\ \nu}^{\ t}
\ \right )
\end{array}
\end{equation}

}



{\color{blue} 

\begin{center}
\vspace*{-0.0cm}
{\bf \color{cyan} 3 I - Insertion : Energy momentum tensor density
as a conserved generalized N\"{o}ther current , restricted 
to $\ \overline{{\cal{L}}} \ $ in the absence of matter fields,
i.e. neglecting 
$\ {\cal{L}}_{\ \left \lbrace q \right \rbrace} \ $ 
}
 \label{'3A-I'}
 \end{center}

\noindent
The Lagrangean density pertaining to quark-antiquark flavors
$\ {\cal{L}}_{\ \left \lbrace q \right \rbrace} \ $ 
is defined in eqs. \ref{eq:3-10} , \ref{eq:3-14} ,
the one for pure gauge fields 
$\ \overline{{\cal{L}}} \ $, induced by the trace anomaly ,
in eq. \ref{eq:3A-6} .

\noindent
We digress here, to derive the energy momentum density tensor
$\ \vartheta_{\ \nu}^{\hspace*{0.2cm} \mu} \ ( \ x \ ) \ $ as a local
functional of the Lagrangean density 
$\ \overline{{\cal{L}}} \ $ and the gauge field variables
$\ \partial_{\ \varrho} \ W_{\ \sigma}^{\ s} \ \longleftrightarrow 
\ B_{\ \varrho \sigma}^{\ s} \ , \ W_{\ \sigma}^{\ s} \ $,
as they appear e.g. in eq. \ref{eq:3A-6g} in variational techniques going back
to Emmy N\"{o}ther \cite{EmmaN} .

\noindent
First we consider general variations of the base variables, understood 
to depend continuously on a family parameter, denoted $\ f \ $. This relates to
the equations of motion

\vspace*{-0.3cm}
\begin{equation}
\label{eq:3I-1}
\begin{array}{l}
W_{\ \sigma}^{\ s} \ = \ W_{\ \sigma}^{\ s} \ ( \ f \ ; \ x \ )
\hspace*{0.2cm} ; \hspace*{0.2cm}
\delta \ W_{\ \sigma}^{\ s} \ ( \ x \ ) \ = 
\ \partial_{\ f} \ \left . W_{\ \sigma}^{\ s} \ ( \ f \ ; \ x \ ) 
\ \right |_{\ f=0}
\hspace*{0.2cm} \rightarrow
\vspace*{0.1cm} \\
\delta \ \partial_{\ \varrho} \ W_{\ \sigma}^{\ s} \ ( \ x \ ) \ = 
\ \partial_{\ \varrho} \ \delta \ W_{\ \sigma}^{\ s} \ ( \ x \ )
\end{array}
\end{equation}

\noindent
and determine the conditions for the f-dependent action integral
over a general 4-dimensional volume V

\vspace*{-0.3cm}
\begin{equation}
\label{eq:3I-2}
\begin{array}{l}
S \ ( \ V \ , \ f \ )
\ = \ {\displaystyle{\int}}_{\ V} \ d^{\ 4} \ x
\ \overline{{\cal{L}}} \ \left ( \ W_{\ \sigma}^{\ s} \ ( \ f \ , \ x \ )
\ ; \ \partial_{\ \varrho} \ W_{\ \sigma}^{\ s} \ ( \ f \ ; \ x \ )
\ \right )
\end{array}
\end{equation}

\noindent
to acquire an extremal value for the particular member of the family of
base fields corresponding to $\ f \ = \ 0 \ $. 

}



{\color{blue} 

\vspace*{0.3cm}
\noindent
This gives rise to the condition

\vspace*{-0.3cm}
\begin{equation}
\label{eq:3I-3}
\begin{array}{l}
\delta \ S \ = \ \partial_{\ f} \ \left . S \ ( \ V \ , \ f \ )
\ \right |_{\ f=0} \ = \ 0
\hspace*{0.2cm} \rightarrow
\vspace*{0.1cm} \\
\delta \ S \ =
\vspace*{0.1cm} \\
\ = \ {\displaystyle{\int}}_{\ V} \ d^{\ 4} \ x
\ \left ( 
\ \delta \ W_{\ \sigma}^{\ s} 
\hspace*{0.15cm} \left . \overline{{\cal{L}}}_{\ , \ W_{\ \sigma}^{\ s}} 
\ \right |_{\ f=0}
\ +
\ \delta \ \partial_{\ \varrho} \ W_{\ \sigma}^{\ s}
\hspace*{0.15cm} \left .  
\overline{{\cal{L}}}_{\ , \ \partial_{\ \varrho} \ W_{\ \sigma}^{\ s}}
\ \right |_{\ f=0}
\ \right ) 
\vspace*{0.1cm} \\
\ = \ 0
\vspace*{0.2cm} \\ \hline \vspace*{-0.3cm} \\
\left . \overline{{\cal{L}}}_{\ , \ W_{\ \sigma}^{\ s}}
\ \right |_{\ f=0} \ \rightarrow
\ \overline{{\cal{L}}}_{\ , \ W_{\ \sigma}^{\ s}}
\ =
\ \begin{array}{c}
\partial \ \overline{{\cal{L}}}
\vspace*{0.2cm} \\
\hline  \vspace*{-0.3cm} \\
\partial \ W_{\ \sigma}^{\ s} \ ( \ f \ = \ 0 \ ; \ x \ )
\end{array}
\vspace*{0.1cm} \\
\left . \overline{{\cal{L}}}_{\ , \ \partial_{\ \varrho} \ W_{\ \sigma}^{\ s}}
\ \right |_{\ f=0} \ \rightarrow
\ \overline{{\cal{L}}}_{\ , \ \partial_{\ \varrho} \ W_{\ \sigma}^{\ s}}
\ =
\ \begin{array}{c}
\partial \ \overline{{\cal{L}}}
\vspace*{0.2cm} \\
\hline  \vspace*{-0.3cm} \\
\partial \ ( \ \partial_{\ \varrho} \ W_{\ \sigma}^{\ s} \ ) 
\ ( \ f \ = \ 0 \ ; \ x \ )
\end{array}
\end{array}
\end{equation}

\noindent
For simplicity of notation we neglect the symbol $\ |_{\ f=0} \ $
in the quantities $\ \overline{{\cal{L}}}_{\ , \ W_{\ \sigma}^{\ s}} \ , 
\ \overline{{\cal{L}}}_{\ , \ \partial_{\ \varrho} \ W_{\
\sigma}^{\ s}} \ $ in the following as indicated in eq. \ref{eq:3I-3}

\noindent
$\ \delta \ S \ $ ( eq. \ref{eq:3I-3} ) becomes using eq. \ref{eq:3I-1}

\vspace*{-0.3cm}
\begin{equation}
\label{eq:3I-4}
\begin{array}{l}
\delta \ S \ =
\ {\displaystyle{\int}}_{\ V} \ d^{\ 4} \ x
\ \left (
\ \delta \ W_{\ \sigma}^{\ s}
\hspace*{0.15cm} \overline{{\cal{L}}}_{\ , \ W_{\ \sigma}^{\ s}}
\ +
\ ( \ \partial_{\ \varrho} 
\ \delta \ W_{\ \sigma}^{\ s} \ )
\hspace*{0.15cm}
\overline{{\cal{L}}}_{\ , \ \partial_{\ \varrho} \ W_{\ \sigma}^{\ s}}
\ \right ) 
\hspace*{0.2cm} \rightarrow
\vspace*{0.1cm} \\
( \ \partial_{\ \varrho}
\ \delta \ W_{\ \sigma}^{\ s} \ )
\hspace*{0.15cm}
\overline{{\cal{L}}}_{\ , \ \partial_{\ \varrho} \ W_{\ \sigma}^{\ s}}
\ = 
\vspace*{0.1cm} \\
\ = \ \partial_{\ \varrho} 
\ \left (
\ \delta \ W_{\ \sigma}^{\ s} 
\hspace*{0.15cm}
\overline{{\cal{L}}}_{\ , \ \partial_{\ \varrho} \ W_{\ \sigma}^{\ s}}
\ \right )
\ -
\ \delta \ W_{\ \sigma}^{\ s} \hspace*{0.15cm} 
\ \partial_{\ \varrho} \hspace*{0.15cm} 
\overline{{\cal{L}}}_{\ , \ \partial_{\ \varrho} \ W_{\ \sigma}^{\ s}}
\end{array}
\end{equation}

}



{\color{blue} 

\noindent
Substituting the second relation in eq. \ref{eq:3I-4} we obtain

\vspace*{-0.5cm}
\begin{equation}
\label{eq:3I-5}
\begin{array}{l}
\delta \ S \ =
\ {\displaystyle{\int}}_{\ V} \ d^{\ 4} \ x
\ \left \lbrack
\begin{array}{l}
\ \delta \ W_{\ \sigma}^{\ s}
\hspace*{0.15cm} \left ( \ \overline{{\cal{L}}}_{\ , \ W_{\ \sigma}^{\ s}}
\ - 
\ \partial_{\ \varrho} \ \overline{{\cal{L}}}_{\ , \ \partial_{\ \varrho}
\ W_{\ \sigma}^{\ s}}
\ \right ) 
\ +
\vspace*{0.1cm} \\
\ + \ \partial_{\ \varrho}
\ \left (
\ \delta \ W_{\ \sigma}^{\ s}
\hspace*{0.15cm}  \overline{{\cal{L}}}_{\ , \ \partial_{\ \varrho}
\ W_{\ \sigma}^{\ s}}
\ \right )
\end{array}
\ \right \rbrack
\end{array}
\end{equation}

\noindent
The variational setting is incomplete without conditions on the boundary
of the volme $\ V \ $ -- boundary conditions --
which arise from the integration of the
divergence in the last term in eq. \ref{eq:3I-5} 

\vspace*{-0.5cm}
\begin{equation}
\label{eq:3I-6}
\begin{array}{@{\hspace*{0.3cm}}l}
{\displaystyle{\int}}_{ V} \ d^{\ 4} \ x
\hspace*{0.15cm} \partial_{\ \varrho}
\ \left (
\ \delta \ W_{\ \sigma}^{\ s}
\hspace*{0.15cm}  \overline{{\cal{L}}}_{\ , \ \partial_{\ \varrho}
\ W_{\ \sigma}^{\ s}}
\ \right ) \ =
\vspace*{0.1cm} \\
\ = \ {\displaystyle{\int}}_{ \partial \ V} 
\hspace*{0.15cm} d^{\ 3} \sigma_{\ \varrho \ , \ \partial \ V}
\ \ \left (
\ \delta \ W_{\ \sigma}^{\ s}
\hspace*{0.15cm}  \overline{{\cal{L}}}_{\ , \ \partial_{\ \varrho}
\ W_{\ \sigma}^{\ s}}
\ \right )
\hspace*{0.1cm} \rightarrow
\vspace*{0.1cm} \\
\longrightarrow \hspace*{0.15cm} \mbox{boundary conditions :}
\hspace*{0.2cm}
\delta \ W_{\ \sigma}^{\ s} \ ( \ x \ ) \ = \ 0 
\hspace*{0.2cm} \mbox{for} \hspace*{0.2cm}
x \ \in \ \partial \ V
\end{array}
\end{equation}

\noindent
With the boundary conditions as defined in eq. \ref{eq:3I-6}
satisfied, the extremum condition for the action integral 
( for all volumes $\ V \ $ ) and {\it otherwise} arbitrary variations
$ \ \delta \ W_{\ \sigma}^{\ s} \ ( \ x \ ) \ $ are equivalent to the
local Euler-Lagrange equations of motion ( eq. \ref{eq:3I-5} )

\vspace*{-0.3cm}
\begin{equation}
\label{eq:3I-7}
\begin{array}{l}
\left (
\ \partial_{\ \varrho} \ \overline{{\cal{L}}}_{\ , \ \partial_{\ \varrho}
\ W_{\ \sigma}^{\ s}}
\ - 
\hspace*{0.15cm} \overline{{\cal{L}}}_{\ , \ W_{\ \sigma}^{\ s}}
\ \right ) \ ( \ x \ )
\ = \ 0
\end{array}
\end{equation}

\noindent
Eqs. \ref{eq:3I-1} - \ref{eq:3I-5} together with the boundary conditions in 
eq. \ref{eq:3I-6} generate the Euler-Lagrange equations 
( eq. \ref{eq:3I-7} )  .

\noindent
Together this sets the stage for deriving associating with every additional
symmetry of the Lagrangean density a conserved N\"{o}ther 'current' , but with
general spin, depending on the symmetry involved.

}



{\color{blue} 

\noindent
We apply the variations as generally given in eq. \ref{eq:3I-1} to the
special case relating rigid space-time translations to the canonical energy
momentum tensor

\vspace*{-0.3cm}
\begin{equation}
\label{eq:3I-8}
\begin{array}{l}
x^{\ \nu} \ \rightarrow \ f \ a^{\ \nu} 
\hspace*{0.2cm} \longleftrightarrow \hspace*{0.2cm}
\begin{array}[t]{lll}
\delta_{\ a} \ W_{\ \sigma}^{\ s} 
& = &
a^{\ \nu} \ \partial_{\ \nu} \ W_{\ \sigma}^{\ s}
\vspace*{0.1cm} \\
\delta_{\ a} \ \partial_{\ \varrho} \ W_{\ \sigma}^{\ s}
& = &
a^{\ \nu} \ \partial_{\ \nu} \ \partial_{\ \varrho} \ W_{\ \sigma}^{\ s}
\end{array}
\vspace*{0.2cm} \\ \hline \vspace*{-0.3cm} \\
\delta_{\ a} \ \overline{{\cal{L}}} \ =
\ a^{\ \nu} 
\ \left ( \begin{array}{l}
\left ( \ \partial_{\ \nu} \ W_{\ \sigma}^{\ s} \ \right )
\ \overline{{\cal{L}}}_{\ , \ W_{\ \sigma}^{\ s}}
\ +
\vspace*{0.1cm} \\
\ + \ \left ( \ \partial_{\ \varrho} 
\ \partial_{\ \nu} \ W_{\ \sigma}^{\ s} \ \right )
\ \overline{{\cal{L}}}_{\ , \ \partial_{\ \varrho} \ W_{\ \sigma}^{\ s}}
\end{array} \right ) \ =
\vspace*{0.1cm} \\
\ = \ a^{\ \nu} \ \partial_{\ \nu} \ \overline{{\cal{L}}}
\end{array}
\end{equation}

\noindent
The underlying translation symmetry reveals itself subtracting the last term in
the expression for $\ \delta_{\ a} \ \overline{{\cal{L}}} \ $ 
in eq. \ref{eq:3I-8} from the second to yield

\vspace*{-0.3cm}
\begin{equation}
\label{eq:3I-9}
\begin{array}{l}
a^{\ \nu} \ \left ( \ \begin{array}{l}
\partial_{\ \varrho} 
\ \left \lbrack
\ \left ( \ \partial_{\ \nu} \ \ W_{\ \sigma}^{\ s} \ \right )
\ \overline{{\cal{L}}}_{\ , \ \partial_{\ \varrho} \ W_{\ \sigma}^{\ s}}
\ - \ \delta_{\ \nu}^{\ \varrho} \ \overline{{\cal{L}}}
\ \right \rbrack \ -
\vspace*{0.1cm} \\
\ - \ \left ( \ \partial_{\ \nu} \ W_{\ \sigma}^{\ s} \ \right )
\ {\cal{E}}^{\ \sigma \ s}
\end{array} 
\ \right ) \ = \ 0
\vspace*{0.1cm} \\
{\cal{E}}^{\ \sigma \ s} \ ( \ x \ ) \ =
\ \left (
\ \partial_{\ \varrho} \ \overline{{\cal{L}}}_{\ , \ \partial_{\ \varrho}
\ W_{\ \sigma}^{\ s}}
\ - 
\hspace*{0.15cm} \overline{{\cal{L}}}_{\ , \ W_{\ \sigma}^{\ s}}
\ \right ) \ ( \ x \ ) \ \rightarrow \ 0
\vspace*{0.1cm} \\
\hspace*{0.3cm}
\mbox{\begin{tabular}[t]{l} 
Euler-Lagrange
\vspace*{-0.1cm} \\
equations
\end{tabular}}
\end{array}
\end{equation}

\noindent
From eq. \ref{eq:3I-9} we read off the canonical energy momentum density tensor
in mixed components

\vspace*{-0.3cm}
\begin{equation}
\label{eq:3I-10}
\begin{array}{l}
T_{\ \nu}^{\hspace*{0.2cm} \varrho}
\ = \ \left ( \ \partial_{\ \nu} \ \ W_{\ \sigma}^{\ s} \ \right )
\ \overline{{\cal{L}}}_{\ , \ \partial_{\ \varrho} \ W_{\ \sigma}^{\ s}}
\ - \ \delta_{\ \nu}^{\ \varrho} \ \overline{{\cal{L}}}
\hspace*{0.2cm} ; \hspace*{0.2cm} 
\partial_{\ \varrho} \ T_{\ \nu}^{\hspace*{0.2cm} \varrho} \ = 0
\vspace*{0.1cm} \\
T_{\ \nu \ \varrho} \ = \ \eta_{\ \rho \ \tau} 
\ T_{\ \nu}^{\hspace*{0.2cm} \tau}
\hspace*{0.2cm} ; \hspace*{0.2cm} 
T_{\ \nu \ \varrho} \ - \ T_{\ \varrho \ \nu} \ \neq \ 0
\end{array}
\end{equation}

\noindent
$\ T_{\ \nu \ \varrho} \ $ is neither symmetric nor gauge invariant.

}



{\color{blue} 

\noindent
Nevertheless a symmetric and gauge invariant energy momentum density tensor can
always be achieved, by also considering variations of 
$\ \overline{{\cal{L}}} \ $ under Lorentz transformations.

\noindent
The consistensy of the gravitational coupling of 
$\ \sqrt{\ g \ } \hspace*{0.15cm} \overline{{\cal{L}}} 
\ ( \ g_{\ \mu \nu} \ ; \ B_{\ \sigma \ \tau}^{\ s} \ ) \ $,
with $\ g \ = \ - \ Det \ g_{\ \mu \nu} \ $,
allows to simplify the explicit symmetrization procedure , due to
Belinfante \cite{Belinf} , in the limit $\ g_{\ \mu \ \nu} \ \rightarrow \
\eta_{\ \mu \nu} $ , i.e. in uncurved space time , after considering
a variation of the metric to first order .

\noindent
Here I follow my derivation in ref. \cite{PMtran} , denoting the symmetric
energy momentum tensor $\ \vartheta_{\ \nu \ \varrho} \ $.

\vspace*{-0.1cm}
\begin{equation}
\label{eq:3I-11}
\begin{array}{l}
\left . 
2 \ \delta \ \left ( \ \sqrt{\ g \ } \hspace*{0.15cm} \overline{{\cal{L}}}
\ \right ) \ = \ \sqrt{\ g \ } \hspace*{0.15cm} \vartheta_{\ \mu \ \nu}
\ \left ( \ \delta \ g^{\ \mu \ \nu} \ \right ) 
\ \right |_{\ g_{\ \alpha \ \beta} \ \rightarrow \ \eta_{\ \alpha \ \beta}}
\vspace*{0.2cm} \\ \hline \vspace*{-0.3cm} \\
\delta \ \left ( \ \sqrt{\ g \ } \hspace*{0.15cm} \overline{{\cal{L}}} 
\ \right ) \ = ( \ \delta \ \sqrt{\ g \ } \ ) \ \hspace*{0.15cm}
\overline{{\cal{L}}} \ + \ \sqrt{\ g \ } \hspace*{0.15cm} 
\delta \ \overline{{\cal{L}}}
\vspace*{0.1cm} \\
\delta \ : \ g^{\ \mu \ \nu} \ \rightarrow \ g^{\ \mu \ \nu} \ +
\ \delta \ g^{\ \mu \ \nu}
\hspace*{0.2cm} ; \hspace*{0.2cm}
\left \lbrack \ g_{\ \mu \nu} \ : \ \mbox{base metric} \ \right \rbrack
\vspace*{0.1cm} \\
g^{\ \mu \ \nu} \ = \ \left ( \ g^{\ -1} \ \right )_{\ \mu \ \nu} 
\vspace*{0.1cm} \\
\delta \ \sqrt{\ g \ } \ = \ - \ \frac{1}{2} \ \sqrt{\ g \ } 
\ g_{\ \mu \ \nu} \ \left ( \ \delta \ g^{\ \mu \ \nu} \ \right )
\end{array}
\end{equation}

\noindent
In eq. \ref{eq:3I-11} $\ \overline{{\cal{L}}} \ $ is, for general metric,
a scalar quantity and $\ \vartheta_{\ \mu \ \nu} \ $ a symmetric tensor .

\noindent
The variation of $\ \overline{{\cal{L}}} \ $ becomes
using eq. \ref{eq:3A-6e}

}



{\color{blue} 

\vspace*{-0.5cm}
\begin{equation}
\label{eq:3I-12}
\begin{array}{l}
\delta \ \overline{{\cal{L}}} \ = \ \overline{{\cal{L}}}_{\ , \ X}
\ \delta \ X
\hspace*{0.2cm} ; \hspace*{0.2cm}
X \ = \ \frac{1}{4} \ B_{\ \mu \sigma}^{\ s} \ B_{\ \nu \tau} \ g^{\ \mu \ \nu}
\ g^{\ \sigma \ \tau}
\vspace*{0.25cm} \\
\overline{{\cal{L}}}_{\ , \ X} \ =
\ \begin{array}{c}
\delta \ \overline{{\cal{L}}}
\vspace*{0.2cm} \\
\hline  \vspace*{-0.3cm} \\
\delta \ {\cal{X}} 
\end{array}
\ = \ \overline{{\cal{L}}} \ / \ {\cal{X}} 
\ - 
\ \left ( \ \begin{array}{c}
d
\vspace*{+0.2cm} \\
\hline  \vspace*{-0.3cm} \\
d \ \overline{l}
\end{array} \hspace*{0.2cm} 
\overline{g}^{\ -2} \ \right )
\hspace*{0.1cm} 
\begin{array}{c}
\delta \ \overline{l}
\vspace*{+0.2cm} \\
\hline  \vspace*{-0.3cm} \\
\delta \ {\cal{X}}
\end{array} \hspace*{0.1cm} \ {\cal{X}}
\vspace*{0.1cm} \\
\overline{l} \ =
\ \frac{1}{8} \ \log \ \left (  \left ( \ {\cal{X}} \ / \ \mu^{\ 4} 
\ \right )^{\ 2} 
 \right )
\end{array}
\end{equation}

\noindent
Proceeding step by step we first substitute 

\vspace*{-0.3cm}
\begin{equation}
\label{eq:3I-13}
\begin{array}{l}
\begin{array}{c}
\delta \ \overline{l}
\vspace*{+0.2cm} \\
\hline  \vspace*{-0.3cm} \\
\delta \ {\cal{X}}
\end{array} \hspace*{0.1cm} \ {\cal{X}} \ = \ \frac{1}{4}
\hspace*{0.2cm} ; \hspace*{0.2cm} 
\overline{{\cal{L}}} \ / \ {\cal{X}} 
\ = \ - \ \left ( \ \overline{g}^{\ -2} \ - J \ \right )
\end{array}
\end{equation}

\noindent
in the second relation in eq. \ref{eq:3I-12}

\vspace*{-0.3cm}
\begin{equation}
\label{eq:3I-14}
\begin{array}{l}
\delta \ \overline{{\cal{L}}} \ = 
\ \left \lbrack
\ - \ \left ( \ \overline{g}^{\ -2} \ - J \ \right )
\ - \ \frac{1}{4}
\ \left ( \ \begin{array}{c}
d
\vspace*{+0.2cm} \\
\hline  \vspace*{-0.3cm} \\
d \ \overline{l}
\end{array} \hspace*{0.2cm} 
\overline{g}^{\ -2} \ \right )
\ \right \rbrack
\ \delta \ {\cal{X}}
\end{array}
\end{equation}

}



{\color{blue} 

\noindent
Next we recall eqs. \ref{eq:3A-6} - \ref{eq:3A-6a} and \ref{eq:3A-6e}

\vspace*{-0.7cm}
\begin{equation}
\label{eq:3I-15}
\begin{array}{l}
\begin{array}{l}
\begin{array}{c}
d
\vspace*{+0.2cm} \\
\hline  \vspace*{-0.3cm} \\
d \ \overline{l}
\end{array} \hspace*{0.2cm} 
\overline{g}^{\ -2}
\ = \ 2 \ \left ( \ - \ \beta \ ( \ \overline{g} \ )
\ / \ \overline{g}^{\ 3} \ \right )
\ = \ \frac{1}{8 \ \pi^{\ 2}} \ b_{\ 0} \ B \ ( \ \overline{\kappa} \ )
\vspace*{0.1cm} \\ 
\overline{\kappa} \ = 
\ \begin{array}{c}
\overline{g}^{\ 2}
\vspace*{0.2cm} \\
\hline  \vspace*{-0.3cm} \\
16 \ \pi^{\ 2}
\end{array}
\hspace*{0.1cm} ; \hspace*{0.1cm}
\overline{\alpha}_{\ s} \ = \ 4 \pi \ \overline{\kappa}
\end{array}
\vspace*{0.2cm} \\ \hline \vspace*{-0.3cm} \\
B \ ( \ \overline{\kappa} \ ) \ = \ b \ ( \ \overline{\kappa} \ ) \ / \ b_{\ 0}
\ = \ \left \lbrack 
\ - \ \beta \ ( \ \overline{g} \ ) \ / \ \overline{g} \ \right \rbrack
\ ( \ b_{\ 0} \ \overline{\kappa} \ )^{\ -1}
\vspace*{0.1cm} \\ 
b_{\ 0} \ = \ 11 \ - \ \frac{2}{3} \ N_{\ fl} 
\vspace*{0.1cm} \\
B \ ( \ \overline{\kappa} \ ) \ = \ B_{\ 0} \ + \ B_{\ 1} \ \overline{\kappa}
\ + \ \cdots \ )
\vspace*{0.1cm} \\
B_{\ 0} \ = \ 1 
\hspace*{0.2cm} , \hspace*{0.2cm}
B_{\ n} \ = \ b_{\ n} \ / \ b_{\ 0} 
\hspace*{0.2cm} ; \hspace*{0.2cm} 
n \ = \ 0.1.2 \ \cdots
\end{array}
\end{equation}

\noindent
The differential equation in eq. \ref{eq:3I-15} is to be solved for
suitable initial conditions as if $\ l \ , \ \overline{\kappa} \ $ were
c-numbers and the functional dependence 
as indicated in eq. \ref{eq:3I-16} below evaluated with the 
operator valued (local field valued) substitution, defined
in eq. \ref{eq:3A-6e}

\vspace*{-0.3cm}
\begin{equation}
\label{eq:3I-16}
\begin{array}{l}
\overline{\kappa} \ = \ \overline{\kappa} \ ( \ l \ ) \ = \ F \ ( \ l \ )
\hspace*{0.2cm} ; \hspace*{0.2cm}
l \ \longrightarrow 
\ \frac{1}{8} \ \log \ \left ( \ \left ( \ {\cal{X}} \ / \ \mu^{\ 4} 
\ \right )^{\ 2} \ \right )
\end{array}
\end{equation}

\noindent
In the substition, defined in eqs. \ref{eq:3A-6e} and \ref{eq:3I-16}, 
local products of local operators 
are involved, which again requires a regularisation in the sense of normal
ordering of canonical variables .

}



{\color{blue} 

\noindent
We insert the relation in eq. \ref{eq:3I-15} in eq. \ref{eq:3I-14} and obtain

\vspace*{-0.3cm}
\begin{equation}
\label{eq:3I-17}
\begin{array}{l}
\delta \ \overline{{\cal{L}}} \ = 
\ \left \lbrack
\ - \ \left ( \ \overline{g}^{\ -2} \ - J \ \right )
\ - \ \frac{1}{32 \ \pi^{\ 2}} \ b_{\ 0} \ B \ ( \ \overline{\kappa} \ )
\ \right \rbrack
\ \delta \ {\cal{X}}
\end{array}
\end{equation}

\noindent
Next we evaluate $\ \delta \ {\cal{X}} \ $ using the first relation in 
eq. \ref{eq:3A-6e} 

\vspace*{-0.3cm}
\begin{equation}
\label{eq:3I-18}
\begin{array}{l}
\delta \ \overline{{\cal{L}}} \ = \ \overline{{\cal{L}}}_{\ , \ X}
\ \delta \ X
\hspace*{0.2cm} ; \hspace*{0.2cm}
X \ = \ \frac{1}{4} \ B_{\ \mu \sigma}^{\ s} \ B_{\ \nu \tau} \ g^{\ \mu \ \nu}
\ g^{\ \sigma \ \tau}
\hspace*{0.2cm} \rightarrow
\vspace*{0.1cm} \\
\begin{array}{lll}
\delta \ X & = & \frac{1}{2} 
\left . \ B_{\ \mu \sigma}^{\ s} \ g^{\ \sigma \ \tau} \ B_{\ \nu \tau}^{\ s}
\ \right |_{\ g^{\ \alpha \ \beta} \ \rightarrow \ \eta^{\ \alpha \ \beta}}
\ \delta \ g^{\ \mu \ \nu}
\vspace*{0.1cm} \\
& = & - \ \frac{1}{2} 
\ \left ( \ B_{\ \mu \sigma}^{\ s} \ B_{\hspace*{0.2cm} \nu}^{\ \sigma \ s}
\ \right ) \ \delta \ g^{\ \mu \ \nu}
\end{array}
\end{array}
\end{equation}

\noindent
The - sign in the last expression in eq. \ref{eq:3I-18} together with a
transposition of the tensor indices in 
$\ B_{\hspace*{0.2cm} \nu}^{\ \sigma \ s} \ $ is chosen to offset the two -
signs in the expression inside [.] brackets in eq. \ref{eq:3I-17} as well as to
comply with the analogous expressions in QED \cite{PMtracean} .

\noindent
Inserting the expressions in eq. \ref{eq:3I-18} in eq. \ref{eq:3I-17}
the final form of $\ \delta \ \overline{{\cal{L}}} \ $ becomes

\vspace*{-0.3cm}
\begin{equation}
\label{eq:3I-19}
\begin{array}{l}
\delta \ \overline{{\cal{L}}} \ = 
\ \frac{1}{2} 
\ \left \lbrack
\ \left ( \ \overline{g}^{\ -2} \ - J \ \right )
\ + \ \frac{1}{32 \ \pi^{\ 2}} \ b_{\ 0} \ B \ ( \ \overline{\kappa} \ )
\ \right \rbrack
\ \left ( \ B_{\ \mu \sigma}^{\ s} \ B_{\hspace*{0.2cm} \nu}^{\ \sigma \ s}
\ \right ) \ \delta \ g^{\ \mu \ \nu}
\end{array}
\end{equation}

\noindent
Finally we complete the expression for the symmetric
energy momentum tensor 
$\ \vartheta_{\ \mu \ \nu} \ $
in eq. \ref{eq:3I-11} ,
which is conserved in the limit of uncurved space time

}



{\color{blue} 

\vspace*{-0.3cm}
\begin{equation}
\label{eq:3I-20}
\begin{array}{l}
2 \ \delta \ \left ( \ \sqrt{\ g \ } \hspace*{0.15cm} \overline{{\cal{L}}}
\ \right ) \ = 
\vspace*{0.1cm} \\
\ = \ \left . \sqrt{\ g \ } \hspace*{0.15cm} \vartheta_{\ \mu \ \nu}
\ \left ( \ \delta \ g^{\ \mu \ \nu} \ \right ) 
\ \right |_{\ g_{\ \alpha \ \beta} \ \rightarrow \ \eta_{\ \alpha \ \beta}}
\hspace*{0.2cm} \rightarrow
\vspace*{0.2cm} \\ \hline \vspace*{-0.3cm} \\
2 \ \delta \ \left ( \ \sqrt{\ g \ } \hspace*{0.15cm} \overline{{\cal{L}}}
\ \right ) \ = 
\vspace*{0.1cm} \\
= \ \sqrt{\ g \ } \hspace*{0.15cm} 
\left \lbrack \begin{array}{l}
\ \left \lbrack
\ \left ( \ \overline{g}^{\ -2} \ - J \ \right )
\ + \ \frac{1}{32 \ \pi^{\ 2}} \ b_{\ 0} \ B \ ( \ \overline{\kappa} \ )
\ \right \rbrack
\ B_{\ \mu \sigma}^{\ s} \ B_{\hspace*{0.2cm} \nu}^{\ \sigma \ s}
\vspace*{0.1cm} \\
- \ \frac{1}{4} \ g_{\ \mu \nu} 
\ \left ( \ \overline{g}^{\ -2} \ - J \ \right )
\ B_{\ \varrho \sigma}^{\ s} \ B^{\ \sigma \ \varrho \ s}
\end{array} \right \rbrack
\ \delta \ g^{\ \mu \ \nu}
\end{array}
\end{equation}

\noindent
Combining the two terms proportional to $\ \overline{g}^{\ -2} \ - J \ $ in
eq. \ref{eq:3I-20} we obtain two characteristic contributions
to $\ \sqrt{\ g \ } \hspace*{0.15cm} \vartheta_{\ \mu \ \nu}
\ \left ( \ \delta \ g^{\ \mu \ \nu} \ \right ) \ $,
even before the limit $\ g_{\ \mu \ \nu} \ \rightarrow \ \eta_{\ \mu \ \nu} \ $
is taken .

\noindent
It follows

\vspace*{-0.3cm}
\begin{equation}
\label{eq:3I-21}
\begin{array}{l}
\vartheta_{\ \mu \ \nu} \ =
\ \left \lbrack \begin{array}{c}
\ \left ( \ \overline{g}^{\ -2} \ - J \ \right )
\ \left \lbrack 
\ B_{\ \mu \sigma}^{\ s} \ B_{\hspace*{0.2cm} \nu}^{\ \sigma \ s}
- \ \frac{1}{4} \ g_{\ \mu \nu} 
\ B_{\ \varrho \sigma}^{\ s} \ B^{\ \sigma \ \varrho \ s}
\ \right \rbrack 
\vspace*{0.1cm} \\
+
\hspace*{0.15cm} 
\frac{1}{8 \ \pi^{\ 2}} \ b_{\ 0} \ B \ ( \ \overline{\kappa} \ )
\ \left \lbrack \ \frac{1}{4}
\ B_{\ \mu \sigma}^{\ s} \ B_{\hspace*{0.2cm} \nu}^{\ \sigma \ s}
\ \right \rbrack
\end{array} \ \right \rbrack
\end{array}
\end{equation}

\noindent
The upper term in the outer [.] brackets in eq. \ref{eq:3I-21} , 
proportional to $\ \overline{g}^{\ -2} \ - J \ $,  is
traceless , whereas the anomalous trace is contained in the lower term ,
proportional to 
$\ \frac{1}{8 \ \pi^{\ 2}} \ b_{\ 0} \ B \ ( \ \overline{\kappa} \ ) \ $.

}



{\color{blue} 

\noindent
In the limit of vanishing gravitational interactions , 
i.e. $\ g_{\ \alpha \ \beta} \ \rightarrow \ \eta_{\ \alpha \ \beta} \ $,
the energy momentum tensor $\ \vartheta_{\ \mu \ \nu} \ $, defined in eq.
\ref{eq:3I-21} , is not only symmetric but conserved , provided 
the {\it modified} equations of motion can be enforced, as well as 
gauge invariant with respect to 
the local gauge group of QCD $\ SU3_{\ c} \ $ 

\vspace*{-0.3cm}
\begin{equation}
\label{eq:3I-22}
\begin{array}{l}
\partial^{\ \mu} \ \vartheta_{\ \mu \ \nu} \ ( \ x \ ) = \ 0
\hspace*{0.2cm} ; \hspace*{0.2cm} \partial^{\ \mu} \ = \ \eta^{\ \mu \ \beta}
\ \partial_{\ \beta} \ = \ \eta^{\ \mu \ \beta} 
\ \partial \ / \ \partial_{\ x^{\ \beta}}
\end{array}
\vspace*{-0.2cm}
\end{equation}

\noindent
The trace anomaly allows precisely the general form of the
modified Lagrangean $\ \overline{{\cal{L}}} \ $, as introduced in 
eq. \ref{eq:3A-6} 
including the a priori free constant $\ J \ $. It takes the form using
eq. \ref{eq:3I-21} , independent of $\ J \ $

\vspace*{-0.3cm}
\begin{equation}
\label{eq:3I-23}
\begin{array}{l}
\overline{{\cal{L}}} \ = \ - 
\ \left \lbrack \ \begin{array}{c}
1
\vspace*{0.2cm} \\
\hline  \vspace*{-0.3cm} \\
\overline{g}^{\ 2} \ ( \ {\cal{X}} \ )
\end{array} \hspace*{0.1cm} - \ J 
\ \right \rbrack \ {\cal{X}}
\hspace*{0.2cm} \longrightarrow
\vspace*{0.2cm} \\
\begin{array}{lll}
\vartheta^{\ \mu}_{\hspace*{0.2cm} \mu}
& = &
\frac{1}{8 \ \pi^{\ 2}} \ b_{\ 0} 
\ B \ ( \ \overline{\kappa} \ )
\ \left \lbrack \ \frac{1}{4}
\ B_{\ \mu \sigma}^{\ s} \ B^{\ \sigma \mu \ s}
\ \right \rbrack
\vspace*{0.2cm} \\
& = & - 
\ \frac{1}{8 \ \pi^{\ 2}} \ b_{\ 0} 
\ : \ B \ \left ( \ \overline{\kappa} \ ( {\cal{X}} \ ) \ \right )  
\ {\cal{X}} \ :
\end{array}
\end{array}
\end{equation}

\noindent
In the last expression in eq. \ref{eq:3I-23} the need of ( normal ) ordering
of local products of quantized fields is indicated by the : symbols , omitted
before and also hereafter whenever not explicitely necessary .

\noindent
The negative sign in the same expression is significant:
the factor \\
$\  B \ \left ( \ \overline{\kappa} \ ) \ ( \ {\cal{X}} \ \right ) \ $ is a 
positive operator at least for $\ \overline{\kappa} \ $ in the perturbative
regime, wile $\ {\cal{X}} \ $ becomes  a positive operator in the Euclidean
region

\vspace*{-0.3cm}
\begin{equation}
\label{eq:3I-24}
\begin{array}{l}
{\cal{X}} \ = \ \frac{1}{2} 
\ \left \lbrack \ \vec{B}^{\ s} \ \vec{B}^{\ s}
\ - \ \vec{E}^{\ s} \ \vec{E}^{\ s} \ \right \rbrack
\ \rightarrow  
\vspace*{0.1cm} \\
\ \rightarrow \ {\cal{X}}_{\ Euc} \ = \ \frac{1}{2} 
\ \left \lbrack \ \vec{B}^{\ s} \ \vec{B}^{\ s}
\ + \ \vec{E}^{\ s} \ \vec{E}^{\ s} \ \right \rbrack_{\ Euc}
\end{array}
\vspace*{-0.1cm}
\end{equation}

\noindent
This completes the construction  of the trace anomaly induced 
canonical variables and modified Lagrangean 
$\ \overline{{\cal{L}}} \ $.
Consequences are discussed in the next subsection.

}



{\color{blue} 

\begin{center}
\vspace*{-0.1cm}
{\bf \color{red} 3 C - Remarks and consequences arising from
derivations in the last subsection
}
 \label{'3C-I'}
 \end{center}
 \vspace*{-0.6cm}

\begin{center}
\vspace*{-0.0cm}
{\bf \color{cyan} $ \left ( \begin{tabular}{ll}
3 I - Insertion : & Energy momentum tensor density as a
\vspace*{-0.1cm} \\
& conserved generlized N\"{o}ther current
\vspace*{-0.1cm} \\
& restricted to $\ \overline{{\cal{L}}} \ $ in the absence
\vspace*{-0.1cm} \\
& of matter fields
\vspace*{-0.1cm} \\
& i.e. neglecting 
$\ {\cal{L}}_{\ \left \lbrace q \right \rbrace} \ $ 
\end{tabular}  \right ) $
}
 \end{center}

\begin{description}

\item 1) Remark relative to the symmetric energy momentum tensor 
compatible with the trace anomaly

while the energy momentum tensor as defined in eq. \ref{eq:3I-21}
is symmetric by the consistency of the embedding into gravitational
interactions leading to the variational form given in eq. \ref{eq:3I-11} ,
it is only conserved if the equations of motions or Euler-Lagrange equations
relative to a specific Lagrangean density -- $\ {\cal{L}}_{\ general} \ $ --
are satisfied, i.e. if the relations , restricted to gauge fields 
in eq. \ref{eq:3I-11} 

\vspace*{-0.3cm}
\begin{equation}
\label{eq:3I-25}
\begin{array}{l}
{\cal{E}}^{\ \sigma \ s} \ ( \ x \ ) \ =
\ \left (
\ \partial_{\ \varrho} \ \overline{{\cal{L}}}_{\ , \ \partial_{\ \varrho}
\ W_{\ \sigma}^{\ s}}
\ - 
\hspace*{0.15cm} \overline{{\cal{L}}}_{\ , \ W_{\ \sigma}^{\ s}}
\ \right ) \ ( \ x \ ) \ \rightarrow \ 0
\vspace*{0.1cm} \\
\hspace*{0.2cm} \mbox{\begin{tabular}[t]{l} 
Euler-Lagrange
\vspace*{-0.1cm} \\
equations
\end{tabular}}
\vspace*{0.1cm} \\
\overline{{\cal{L}}} \ \rightarrow \ {\cal{L}}^{\ general} 
\hspace*{0.2cm} ; \ \mbox{e.g.} \hspace*{0.2cm}
\rightarrow \ {\cal{L}}_{\ gauge} \ =
\ - 
\ \begin{array}{c}
1
\vspace*{0.2cm} \\
\hline  \vspace*{-0.3cm} \\
g^{\ 2}
\end{array} \hspace*{0.1cm} {\cal{X}}
\vspace*{0.1cm} \\
{\cal{X}} \ = \ \frac{1}{4} 
\ B_{\ \mu \nu}^{\ r} \ B^{\ \mu \nu \ r}
\end{array}
\end{equation}

are satisfied , yet for any gauge invariant Lagrangean, depending only on
gauge potentials and their first derivatives. In particular 
substituting ( back ) $\ \overline{{\cal{L}}}  \rightarrow \ {\cal{L}} \ $,
defined in eq. \ref{eq:3A-6} ,
as indicated in the last relation in eq. \ref{eq:3I-25} . 

\end{description}

\noindent
We now distinguish both the canonical energy momentum ( density ) tensors
( eq. \ref{eq:3I-10} )
and their symmetric equivalents according to the two choices for the 
gauge field Lagrangean
( density ) ( eq. \ref{eq:3I-20} ) , abbreviated to 
$\ {\cal{L}}_{\ gauge} \ = \ {\cal{L}} \ $
and $\ \overline{{\cal{L}}} \ $ respectively

\vspace*{-0.5cm}
\begin{equation}
\label{eq:3I-26}
\begin{array}{l}
T_{\ \nu}^{\hspace*{0.2cm} \varrho}
\ = \ \left ( \ \partial_{\ \nu} \ \ W_{\ \sigma}^{\ s} \ \right )
{\cal{L}}^{\ genaral}_{\ , \ \partial_{\ \varrho} \ W_{\ \sigma}^{\ s}}
\ - \ \delta_{\ \nu}^{\ \varrho} \ {\cal{L}}^{\ general}
\vspace*{0.1cm} \\
2 \ \delta \ \left ( \ \sqrt{\ g \ } \hspace*{0.15cm} {\cal{L}}^{\ general}
\ \right ) \ = \ \left . \sqrt{\ g \ } \hspace*{0.15cm} \vartheta_{\ \mu \ \nu}
\ \left ( \ \delta \ g^{\ \mu \ \nu} \ \right ) 
\ \right |_{\ g_{\ \alpha \ \beta} \ \rightarrow \ \eta_{\ \alpha \ \beta}}
\hspace*{0.2cm} \rightarrow
\vspace*{0.2cm} \\ \hline \vspace*{-0.3cm} \\
T_{\ \nu}^{\hspace*{0.2cm} \varrho} \ = 
\ T_{\ \nu}^{\hspace*{0.2cm} \varrho} \ ( \ . \ )
\hspace*{0.1cm} ; \hspace*{0.1cm}
\vartheta_{\ \mu \ \nu} \ = \ \vartheta_{\ \mu \ \nu} \ ( \ . \ )
\hspace*{0.2cm} ; \hspace*{0.2cm}
\hspace*{0.1cm} \mbox{with} \hspace*{0.25cm}
. \ = \ \left \lbrace \ {\overline{\cal{L}}} \ , {\cal{L}} \ \right \rbrace
\end{array}
\vspace*{0.1cm}
\end{equation}

\noindent
The Belinfante construction \cite{Belinf} links uniquely
$\ T_{\ \nu}^{\ \mu} \ ( \ . \ ) \ \leftrightarrow
\ \vartheta_{\ \mu \ \nu} \ ( \ . \ ) \ $, separately for both
values of $\ ( \ . \ ) \ $. The respective canonical energy momentum tensors
are neither symmetric nor gauge invariant also for both choices of 
$\ ( \ . \ ) \ $, whereas both $\ \vartheta_{\ \mu \ \nu} \ ( \ . \ ) \ $
are gauge invariant.

\noindent
Furthermore all energy momentum tensors are conserved , provided
the equations of motion are enforced

\vspace*{-0.3cm}
\begin{equation}
\label{eq:3I-27}
\begin{array}{l}
\partial^{\ \mu} \ T_{\ \nu  \mu} \ ( \ . \ ) \ = \ 0 
\hspace*{0.2cm} ; \hspace*{0.2cm} 
T_{\ \nu  \mu} \ ( \ . \ ) \ = \ \eta_{\ \mu \varrho} 
\ T_{\ \nu}^{\hspace*{0.2cm} \varrho} \ ( \ . \ )
\vspace*{0.1cm} \\
\partial^{\ \mu}  \ \vartheta_{\ \nu \mu} \ ( \ . \ ) \ = \ 0
\end{array}
\end{equation}

\noindent
Notwithstanding the two cases it is $\ \overline{{\cal{L}}} \ $ which
needs to be chosen.
With respect to the trace anomaly ( eq. \ref{eq:3I-23} ) and thereby 
with respect to dilatation
transformations, for which the symmetric tensor(s)
$\ \vartheta_{\ \nu \mu} \ ( \ . \ ) \ $ are the relevant ones, it follows

\vspace*{-0.1cm}
\begin{equation}
\label{eq:3I-28}
\begin{array}{l}
\begin{array}{lll}
\vartheta^{\ \mu}_{\hspace*{0.2cm} \mu} \ ( \ \overline{{\cal{L}}} \ )
& = &
\hspace*{0.15cm} \frac{1}{8 \ \pi^{\ 2}} \ b_{\ 0} 
\ B \ ( \ \overline{\kappa} \ )
\ \left \lbrack \ \frac{1}{4}
\ B_{\ \mu \sigma}^{\ s} \ B^{\ \sigma \mu \ s}
\ \right \rbrack
\vspace*{0.2cm} \\
& = & - 
\ \frac{1}{8 \ \pi^{\ 2}} \ b_{\ 0} 
\ : \ B \ \left ( \ \overline{\kappa} \ ( {\cal{X}} \ ) \ \right )  \ {\cal{X}}
\ : \ 
\ \neq \ 0 
\end{array}
\vspace*{0.2cm} \\
\vartheta^{\ \mu}_{\hspace*{0.2cm} \mu} \ ( \ {\cal{L}} \ ) \ = \ 0
\end{array}
\end{equation}
 
\noindent
From $\ \vartheta_{\ \nu \mu} \ ( \ . \ ) \ $ the canonical form
the local dilatation current is constructed

\vspace*{-0.3cm}
\begin{equation}
\label{eq:3I-29}
\begin{array}{l}
d_{\ \mu} \ ( \ . \ ) \ ( \ x \ ) \ = \ \left ( \ x \ - \ x_{\ (0)} 
\ \right )^{\ \nu}
\ \vartheta_{\ \nu \mu} \ ( \ . \ ) \ ( \ x \ )
\vspace*{0.1cm} \\
\ \partial^{\ \mu} \ d_{\ \mu} \ ( \ . \ ) \ ( \ x \ ) \ =
\vartheta^{\ \mu}_{\hspace*{0.2cm} \mu} \ ( \ . \ ) \ ( \ x \ )
\vspace*{0.1cm} \\
\partial^{\ \mu} \ d_{\ \mu} \ ( \ \overline{{\cal{L}}} \ ) \ \neq \ 0
\hspace*{0.2cm} ; \hspace*{0.2cm}
\partial^{\ \mu} \ d_{\ \mu} \ ( \ {\cal{L}} \ ) \ = \ 0
\end{array}
\end{equation}

\noindent
In the case of $\ {\cal{L}} \ $ the dilatation current is conserved and thus
dilatation symmetry is enforced and as a direct consequence the
entire group of conformal space time transformations becomes a symmetry group.

\noindent
It is however $\ \overline{{\cal{L}}} \ $ which has to be chosen as the only
case
compatible with the infrared regularity conditions inherent to maintaining
local gauge transformations exact through the infrared unstable
region , as discussed in ref. \cite{phaseQCD2011} .

\begin{description}

\item 2) The new content of the last subsection dates from December 2011

The material presented here mainly represents a traceback of aspects
of QCD , having been noted but not carried out in the past in any detail
and characterized ( best ) as 'work in progress' .

This topic concerns the embedding of Hamiltonian quantum mechanics
and canonically conjugate variables within local
field theory into the predominantly perturbative treatments derived
from asymptotic freedom in the ultraviolet of QCD .

Some bridges ahead of completion go back to my contributions to two events 
organized by Harald Fritzsch and the Nanyang Technological University, 
Singapore :

\end{description}

\begin{description}

\item a) Conference in Honour of Murray Gell-Mann's 80th Birthday

'Quantum Mechanics, Elementary Particles, Quantum Cosmology and Complexity', \\
24.-26. February 2010

\item b) International Conference on Flavor physics in the LHC era

8. - 12. November 2010

both held at the Nanyang Executive Centre in Singapore .

\vspace*{-0.2cm}
\end{description}\footnote{\color{red} 
\hspace*{0.1cm} \begin{tabular}{l}
My gratitude goes to Professor Kokk Koo Phua and the organizing committees
\vspace*{-0.1cm} \\
for making these events possible .
\end{tabular}}

\begin{center}
\vspace*{-0.0cm}
{\bf \color{cyan} 
3 C 1 - Equations of motion and canonically conjugate variables pertaining to
$\ \overline{{\cal{L}}} \ +
\ {\cal{L}}_{\ \left \lbrace q \right \rbrace} \ $ 
}
\label{'3C-1'}
 \end{center}

\noindent
The conflict between the canonical forms of obviously inequivalent \\ 
Lagrangean densities in the gauge field sector 

\vspace*{-0.3cm}
\begin{equation}
\label{eq:3I-30}
\begin{array}{l}
\left . \overline{{\cal{L}}}  
\hspace*{0.2cm} \longleftrightarrow \hspace*{0.2cm}
{\cal{L}}
\hspace*{0.2cm} \right |_{\ gauge fields}
\end{array}
\end{equation}

\noindent
can be traced -- in the perturbative sector -- to maintaining Poincar\'{e}
invariance and full gauge invariance , which necessitates the use of 
Fermi like gauges and thus a nontrivial coupling of ghost fields, which
do contribute to the energy momentum tensor \cite{KluZu} .

\vspace{-0.2cm}
\begin{equation}
\label{eq:C-1}
\begin{array}{c}
{\cal{L}} \ = 
\begin{array}{c}
 \left \lbrack  \begin{array}{c}
\overline{q}^{\ \dot{c}'}_{\ \dot{{\cal{A}}}' \ \dot{f}}
\ \left \lbrace \ \begin{array}{c}
\frac{i}{2} \ \stackrel{\leftharpoondown \hspace*{-0.3cm} \rightharpoonup}{\partial}_{\ \mu} 
\ \delta_{\ c' c}
\vspace*{0.1cm} \\
- \ v_{\ \mu}^{\ s} \ \left ( \ \frac{1}{2}  \lambda^{\ s} 
\ \right )_{\ c' \dot{c}}
\end{array}                     
\right \rbrace
\ \gamma^{\ \mu}_{\ \dot{{\cal{A}}}' {\cal{A}}} \ q^{\ c}_{\ {\cal{A}} \ f} 
\vspace*{0.1cm} \\
\hspace*{0.5cm}
- \ m_{\ f} \ \overline{q}^{\ \dot{c}}_{\ \dot{{\cal{A}}} \ \dot{f}} 
\  q^{\ c}_{\ {\cal{A}} \ f}
\end{array}
 \right \rbrack
\vspace*{0.1cm} \\
- \ \frac{1}  
{\mbox{\vspace*{0.05cm} \begin{tabular}{c} $4 \ \color{red} g^{\ 2} $
\end{tabular}}} 
\ \color{blue} B^{\ \mu \nu \ s} 
\ B_{\ \mu \nu}^{\ s}
\hspace*{0.1cm} 
+ \ \Delta \ {\cal{L}}
\end{array}
\hspace*{0.1cm} ; \hspace*{0.1cm} 
v_{ \mu}^{\ s} \ = \ - \ W_{ \mu}^{\ s}
\vspace*{0.1cm} \\
\color{magenta} \mbox{quarks} \ : \color{blue}  c' \ , \ c  =  1,2,3 \ \mbox{color}
\hspace*{0.2cm} , \hspace*{0.2cm}
f \ = \ 1,\cdots,6 \ \mbox{flavor}
\vspace*{0.1cm} \\
{\cal{A}}' , {\cal{ A}} \ = \ 1,\cdots,4 \ \mbox{spin} \ , \  m_{\ f} \ \mbox{mass}
\end{array}
\vspace*{-0.3cm}
\end{equation}

}



{\color{blue} 

\vspace{-0.3cm}
\begin{equation}
\label{eq:C-2}
\begin{array}{c}
\color{red} \mbox{gauge bosons} :
\vspace*{0.1cm} \\
\color{blue}
B_{\ \mu \nu}^{\ r} \ = \ \partial_{\ \mu} \ W_{\ \nu}^{\ r} \ -
\ \partial_{\ \nu} \ W_{\ \mu}^{\ r} \ + \ f_{\ r s t} \ W_{\ \mu}^{\ s} 
\ W_{\ \nu}^{\ t}
\vspace*{0.1cm} \\
r, s, t \ = \ 1,\cdots,dim \ ( \ G \ = \ SU3_{\ c} \ ) \ = \ 8
\vspace*{0.1cm} \\ 
\mbox{Lie algebra labels}  ,
 \left \lbrack  \frac{1}{2} \ \lambda^{\ r} \ , \ \frac{1}{2} \ \lambda^{\ s}
 \right \rbrack \ = \ i  f_{\ r s t} \ \frac{1}{2} \ \lambda^{\ t}
\vspace*{0.1cm} \\
\color{magenta} \mbox{perturbative rescaling :}
\vspace*{0.0cm} \\
W_{\ \mu}^{\ r} \ = \ \color{red} g \ \color{blue} W_{\ \mu \ pert}^{\ r} 
\ , \ B_{\ \mu \nu}^{\ r} \ = \ \color{red} g 
\ \color{blue} B_{\ \mu \nu \ pert}^{\ r}
\end{array}
\end{equation}

\noindent
Degrees of freedom are seen in jets , in (e.g.) the energy momentum sum rule in 
deep inelastic scattering 
but not clearly in spectroscopy.

\noindent
Completing $\Delta \ {\cal{L}}$ in Fermi gauges

\begin{equation}
\label{eq:C-3}
\begin{array}{l}
\Delta \ {\cal{L}} \ =
\ \left \lbrace \begin{array}{c}
- \ \frac{1}{2 \ \eta \ \color{red} g^{\ 2}}
\ \color{blue} \left ( \ \partial_{\ \mu} \ W^{\ \mu \ s} \ \right )^{\ 2}
\vspace*{0.1cm} \\
+ \ \partial^{\ \mu} \ \overline{c}^{\ s} \ ( \ D_{\ \mu} \ c \ )^{\ s}
\end{array}
\right \rbrace
\hspace*{0.2cm} ; \hspace*{0.2cm}
\eta \ : \ \mbox{gauge parameter}
\vspace*{0.1cm} \\
\color{magenta} \mbox{ghost fermion fields :} \ \color{blue} c \ , \ \overline{c}
\hspace*{0.2cm} ; \hspace*{0.2cm}
( \ D_{\ \mu} \ c \ )^{\ r} \ = \ \partial_{\ \mu} \ c^{\ r} \ + \ f_{\ r s t} 
\ W_{\ \mu}^{\ s}
\ c^{\ t}
\vspace*{0.1cm} \\
\color{magenta} \mbox{gauge fixing constraint :} \ \color{blue}
C^{\ r} \ = \ \partial_{\ \mu}  \ W^{\ \mu \ r} 
\end{array}
\end{equation}

}



{\color{blue} 

\noindent
After these preliminary remarks we turn to the equations of motion
for gauge fields as induced by the Lagrangean 
$\ \overline{{\cal{L}}} \ +
\ {\cal{L}}_{\ \left \lbrace q \right \rbrace} \ $ 
defined in eqs. \ref {eq:3-15} , \ref{eq:3A-6} \\
( and \ref{eq:C-1} ) .

\noindent
We merge eqs. \ref{eq:3A-6g} and \ref{eq:3I-17} 

\vspace*{-0.3cm}
\begin{equation}
\label{eq:C-4}
\begin{array}{l}
\delta \ \overline{{\cal{L}}} \ = 
\ - \ \left \lbrack
\ \left ( \ \overline{g}^{\ -2} \ - J \ \right )
\ + \ \frac{1}{32 \ \pi^{\ 2}} \ b_{\ 0} \ B \ ( \ \overline{\kappa} \ )
\ \right \rbrack
\ \delta \ {\cal{X}}
\vspace*{0.1cm} \\
\begin{array}{c}
\delta \ {\cal{X}}
\vspace*{0.2cm} \\
\hline  \vspace*{-0.3cm} \\
\delta \ \left ( \ \partial_{\ \varrho} \ W_{\ \sigma}^{\ s} \ \right )
\end{array}
\ = \ B^{\ \varrho \ \sigma \ s}
\hspace*{0.2cm} ; \hspace*{0.2cm}
\begin{array}{c}
\delta \ {\cal{X}}
\vspace*{0.2cm} \\
\hline  \vspace*{-0.3cm} \\
\delta \ W_{\ \sigma}^{\ s}
\end{array} \ =
\ - \ f_{\ s t r} \ W_{\ \varrho}^{\ t} \ B^{\ \varrho \ \sigma \ r}
\end{array}
\end{equation}

\noindent
It follows for the partial derivatives of $\ \overline{{\cal{L}}} \ $

\vspace*{-0.3cm}
\begin{equation}
\label{eq:C-5}
\begin{array}{l@{\hspace*{0.0cm}}l@{\hspace*{0.0cm}}l}
\overline{{\cal{L}}}_{\ , \ \partial_{\ \varrho}
\ W_{\ \sigma}^{\ s}} & = 
& + \ \left \lbrack
 \left ( \ \overline{g}^{\ -2} \ - J \ \right )
\ + \ \frac{1}{32 \ \pi^{\ 2}} \ b_{\ 0} \ B \ ( \ \overline{\kappa} \ )
 \right \rbrack
 B^{\ \sigma \ \varrho \ s}
\vspace*{0.1cm} \\
\overline{{\cal{L}}}_{\ , \ W_{\ \sigma}^{\ s}} & =
& - \ \left \lbrack
\ \left ( \ \overline{g}^{\ -2} \ - J \ \right )
\ + \ \frac{1}{32 \ \pi^{\ 2}} \ b_{\ 0} \ B \ ( \ \overline{\kappa} \ )
 \right \rbrack
\begin{array}[t]{l}
f_{\ s t r} \ \times
\vspace*{0.1cm} \\
\ \times \ W_{\ \varrho}^{\ t} \ B^{\ \sigma \varrho \ \ r}
\end{array}
\end{array}
\end{equation}

\noindent
Using eq. \ref{eq:3A-9}

\vspace*{-0.3cm}
\begin{equation}
\label{eq:C-6}
\begin{array}{l}
\begin{array}[t]{l}
\ \left ( \ D_{\ \varrho} \ ( \ ad \ ) \ \right )_{\ s r} 
\ \left \lbrace 
\ B_{\ \sigma \tau}^{\ r} \right \rbrace \ =
\ \partial_{\ \varrho} 
\ \left \lbrace
\hspace*{0.1cm} B_{\ \sigma \tau}^{\ s} \ \right \rbrace
\ +
\ f_{\ s t r } \ \ W_{\ \varrho}^{\ t}
\ \left \lbrace
\ B_{\ \sigma \tau}^{\ r} \ \right \rbrace 
\end{array}
\end{array}
\end{equation}

\noindent
the Euler-Lagrange derivative of $\ \overline{{\cal{L}}} \ $ on the
left hand side of eq. \ref{eq:3I-7} becomes

\vspace*{-0.3cm}
\begin{equation}
\label{eq:C-7}
\begin{array}{l}
\left (
\ \partial_{\ \varrho} \ \overline{{\cal{L}}}_{\ , \ \partial_{\ \varrho}
\ W_{\ \sigma}^{\ s}}
\ - 
\hspace*{0.15cm} \overline{{\cal{L}}}_{\ , \ W_{\ \sigma}^{\ s}}
\ \right ) \ ( \ x \ )
\ = 
\vspace*{0.1cm} \\
\hspace*{0.5cm} =
D_{\ \varrho} \ ( \ ad \ ) 
\ \left \lbrace \ \left \lbrack
\ \left ( \ \overline{g}^{\ -2} \ - J \ \right )
\ + \ \frac{1}{32 \ \pi^{\ 2}} \ b_{\ 0} \ B \ ( \ \overline{\kappa} \ )
\ \right \rbrack
\ B^{\ \sigma \ \varrho} \ \right \rbrace^{\ s} \ ( \ x \ )
\end{array}
\end{equation}

\noindent
The Euler-Lagrange equations including quark flavors according to 
$\ {\cal{L}}_{\ \left \lbrace q \right \rbrace} \ $ ( eqs. \ref{eq:3-13} ,
\ref{eq:3-18} , \ref{eq:C-1} ) take the form

\vspace*{-0.5cm}
\begin{equation}
\label{eq:C-8}
\begin{array}{l}
D_{\ \varrho} \ ( \ ad \ ) 
\ \left \lbrace \ \left \lbrack
\ \left ( \ \overline{g}^{\ -2} \ - J \ \right )
\ + \ \frac{1}{32 \ \pi^{\ 2}} \ b_{\ 0} \ B \ ( \ \overline{\kappa} \ )
\ \right \rbrack
\ B^{\ \sigma \ \varrho} \ \right \rbrace^{\ s} \ ( \ x \ )
\ = 
\vspace*{0.2cm} \\
\hspace*{1.0cm} = \ \left ( \hspace*{0.1cm} j^{\ \sigma \ s} 
\ \right )_{\ \left \lbrace q \right \rbrace} \ ( \ x \ )
\vspace*{0.2cm} \\ \hline \vspace*{-0.3cm} \\
\left ( \hspace*{0.1cm} j^{\ \sigma \ s} 
\ \right )_{\ \left \lbrace q \right \rbrace} \ =
\ \sum_{\ q-fl} \ \overline{q}^{\ \dot{c}'}
\ \left \lbrace \ \gamma^{\ \sigma}
\ \left ( \ \frac{1}{2} \ \lambda^{\ s} \ \right )_{\ c' \dot{c}}
\ \right \rbrace \ q^{\ c}
\ = \ \left ( \ {\cal{L}}_{\ \left \lbrace q \right \rbrace} 
\ \right )_{\ , \ W_{\ \sigma}^{\ s}}
\vspace*{-1.5cm} \\ 
\end{array}
\vspace*{0.3cm}
\end{equation}
\vspace*{0.5cm}

\noindent
The quantity on the gauge field side of the 'divergence' on the
right hand side of eq. \ref{eq:C-7} and on the
left hand side
of the first relation in eq. \ref{eq:C-8} contrasts in an essential way with
the bare Lagrangean expressions corresponding to $\ {\cal{L}}_{\ gauge} \ $
( eqs. \ref{eq:3-18} , \ref{eq:3-19} )
This is the main and new result of these last subsections . 

}



{\color{blue} 

\noindent
To make this more transparent lets introduce the notation

\vspace*{-0.3cm}
\begin{equation}
\label{eq:C-9}
\begin{array}{l}
\begin{array}{clc}
\overline{{\cal{L}}} 
\hspace*{0.1cm} \leftrightarrow \hspace*{0.1cm} 
\overline{G} 
& \leftrightarrow &
{\cal{L}} 
\hspace*{0.1cm} \leftrightarrow \hspace*{0.1cm} 
G
\vspace*{0.1cm} \\
\overline{G} \ =
\begin{array}[t]{c}
\mbox{local, scalar, color neutral field}
\end{array}
& \leftrightarrow &
G \ = \begin{array}[t]{c}
g^{\ -2} 
\vspace*{0.0cm} \\
\mbox{const. c-number}
\end{array}
\end{array}
\vspace*{0.1cm} \\
\overline{G} \ =
: \left \lbrack \ \left ( \ \overline{g}^{\ -2} \ - J \ \right )
\ + \ \frac{1}{32 \ \pi^{\ 2}} \ b_{\ 0} \ B \ ( \ \overline{\kappa} \ )
\ \right \rbrack : \ ( \ x \ )
\end{array}
\end{equation}

\noindent
The precise way , $\ \overline{G} \ $ and associated quantities,
in particular the energy momentum (density) tensor pertaining to gauge fields, 
defined in eq. \ref{eq:3I-21}, are determined as local field operatoras ,
shall be (re-) specified below

\vspace*{-0.3cm}
\begin{equation}
\label{eq:C-10}
\begin{array}{l}
(a) \ : \ \overline{G} \ = \ \overline{G} \ ( \ l \ ) \ ;
\ l \ = \ \log \ \left ( \ \overline{\mu} \ / \ \mu \ \right )
\hspace*{0.2cm} \rightarrow
\vspace*{0.1cm} \\
(b) \ : \ l \ \rightarrow \ \frac{1}{8} \ \log 
\ \left ( \left ( \ {\cal{X}} \ ( \ x \ ) \ / \ \mu^{\ 4} 
\ \right )^{\ 2} \ \right )
\hspace*{0.1cm} \leftrightarrow \hspace*{0.1cm} 
e^{\ 8 \hspace*{0.05cm} l} \ = \ \left ( 
\ {\cal{X}} \ ( \ x \ ) \ / \ \mu^{\ 4} \ \right )^{\ 2}
\vspace*{0.2cm} \\ \hline \vspace*{-0.3cm} \\
{\cal{X}} \ ( \ x \ ) \ = \ : \ \frac{1}{4} 
\ B_{\ \mu \nu}^{\ r} \ B^{\ \mu \nu \ r} \ : \ ( \ x \ )
\end{array}
\end{equation}

\noindent
The functional dependence $\ \overline{G} \ = \ \overline{G} \ ( \ l \ ) \ $
in step (a) in eq. \ref{eq:C-10} can be determined in the perturbative regime,
through the renormalization group equation(s) , e.g. in the 
$\ \overline{MS} \ $ renormalization scheme .

}



{\color{blue} 

\noindent
Through 4 loop order $\ \overline{MS} \ $ is renormalization group invariant,
through the substiturions $\ \mu \ = \ \Lambda_{QCD} \ $ and  
e.g. through the moments of deep inelastic scattering amplitudes ,
$\ \overline{\mu}^{\ 2} \ = \ Q^{\ 2} \ $, where
$\ Q^{2} \ $ is the (positive) momentum transfer square in the deep inelastic 
reaction studied \cite{Gorishnii,RitVermLar} . 

\noindent
In any renormalization scheme, where at least in the perturbative regime 
full renormalization group invariance is verified , it follows
that the substitution in the second step (b) in eq. \ref{eq:C-10}

\vspace*{-0.3cm}
\begin{equation}
\label{eq:C-11}
\begin{array}{l}
l \ \rightarrow \ \frac{1}{8} \ \log 
\ \left ( \ \left ( \ {\cal{X}} \ ( \ x \ ) \ / \ \mu^{\ 4} 
\ \right )^{\ 2} \ \right )
\hspace*{0.1cm} \leftrightarrow \hspace*{0.1cm} 
e^{\ 8 \hspace*{0.05cm} l} \ = \ \left ( 
\ {\cal{X}} \ ( \ x \ ) \ / \ \mu^{\ 4} \ \right )^{\ 2}
\end{array}
\end{equation}

\noindent
can equally be performed in a renormalization group invariant manner .
This is tantamount to resolve all ambiguities in the definition 
of the composite local field $\ {\cal{X}} \ ( \ x \ ) \ $ by a normalization in
terms of renormalization group invariant , i.e. measurable quantities .

\noindent
Within QCD sum rules introduced by Shifman, Vainshtain and Zakharov 
\cite{ShiVaiZakh} , the vacuum expected value
of the multiplicatively related field , denoted $\ \alpha_{\ s} \ G^{\ 2} \ $  
has been intensively studied .

\noindent
I cite here a recent paper and result(s) by
Stephan Narison \cite{SNari} in partcular with respect to the
renormalization group invariant setting -- in principle -- of composite local
field normalization 

\vspace*{-0.3cm}
\begin{equation}
\label{eq:C-12}
\begin{array}{l}
\alpha_{\ s} \ G^{\ 2} \ = \ \pi^{\ -1} \ {\cal{X}}
\vspace*{0.1cm} \\
\begin{array}{lll}
\left \langle \ \Omega \ \right | \alpha_{\ s} \ G^{\ 2} \ \left |
\ \Omega \ \right \rangle  & = & ( 7.0 \ \pm \ 1.3 \ ) \ 10^{\ -2} 
\ \mbox{GeV}^{\ 4} 
\vspace*{0.1cm} \\
& = &
\pi^{\ -1} \ \left ( \ 0.22 \ \pm \ 0.04 \ \right )
\ \mbox{GeV}^{\ 4}
\end{array}
\end{array}
\end{equation}

}



{\color{blue} 

\noindent
Having defined the local field structure of two inequivalent such fields
$ {\cal{X}} \ ( \ x \ ) $ and {\it separately} 
$\ \overline{G} \ ( \ x \ ) \ $ in eqs. \ref{eq:C-9} and \ref{eq:C-10},
the symmetric, gauge invariant energy momentum
tensor incompletely defined in eqs. \ref{eq:3I-20} and \ref{eq:3I-21}
is represented as follows

\vspace*{-0.3cm}
\begin{equation}
\label{eq:C-13}
\begin{array}{l}
\begin{array}{lll}
\vartheta_{\ \mu \ \nu} \ ( \ x ) 
& = &
\left \lbrack \begin{array}{c}
\ : \ \overline{G}
\ \left \lbrack 
\ B_{\ \mu \sigma}^{\ s} \ B_{\hspace*{0.2cm} \nu}^{\ \sigma \ s}
\ + \ g_{\ \mu \nu} \ {\cal{X}} 
\ \right \rbrack \ :
\vspace*{0.1cm} \\
-
\hspace*{0.15cm} g_{\ \mu \nu}
\ : \ \frac{1}{32 \ \pi^{\ 2}} \ b_{\ 0} 
\ B \ \left ( \ \overline{\kappa} \ ( \ {\cal{X}} \ \right )
\ {\cal{X}} \ :
\end{array} \ \right \rbrack
\ ( \ x \ )
\vspace*{0.2cm} \\
\vartheta^{\ \mu}_{\hspace*{0.2cm} \mu} \ ( \ x \ )
& = & 
- 
\ \frac{1}{8 \ \pi^{\ 2}} \ b_{\ 0} 
\ : \ B \ \left ( \ \overline{\kappa} \ ( {\cal{X}} \ ) \ \right )  
\ {\cal{X}} \ : \ ( \ x \ )
\vspace*{0.2cm} \\ \hline \vspace*{-0.3cm} \\ 
\overline{G} \ ( \ x \ ) 
& = &
: \ \left \lbrack \ \left ( \ \overline{g}^{\ -2} \ - J \ \right )
\ + \ \frac{1}{32 \ \pi^{\ 2}} \ b_{\ 0} 
\ B \ \left ( \ \overline{\kappa} \ ( \ {\cal{X}} \ ) \ \right )
\ \right \rbrack : \ ( \ x \ )
\end{array}
\vspace*{0.2cm} \\
B \ \left ( \ \overline{\kappa} \ ( \ {\cal{X}} \ ) \ \right ) \ ( \ x \ )
\ \rightarrow \
\equiv
\ \overline{B} \ \left ( \ \cal{X} \ \right ) \ ( \ x \ ) 
\end{array}
\end{equation}

\noindent
In eq. \ref{eq:C-13} 
$\ B \ \left ( \ \overline{\kappa} \ ( \ {\cal{X}} \ ) \ \right ) \ $
indicates that through the substitution in step (b) of $\ \overline{G} \ $
in eq. \ref{eq:C-10} , each of the additive terms of $\ \overline{G} \ $
and thus also $\ B \ ( \ \overline{\kappa} \ ) \ \rightarrow
\ \overline{B} \ \left ( \ \cal{X} \ \right ) \ $ becomes
implicitly dependent on $\ {\cal{X}} \ $ .

}



{\color{blue} 

\noindent
Next we complete the Euler-Lagrange equations of motion defined in
eqs. \ref{eq:C-7} and \ref{eq:C-8} , which introduce 
a local, color neutral, hermitian scalar field , denoted 
$\ \varphi \ ( \ x \ ) \ $ below , depending implicitely on the gauge field
strengths bilinear $\ {\cal{X}} \ ( \ x \ ) \ $ as specified
in eqs. \ref{eq:C-9} - \ref{eq:C-11}

\vspace*{-0.5cm}
\begin{equation}
\label{eq:C-14}
\begin{array}{l}
\left (
\ \partial_{\ \varrho} \ \overline{{\cal{L}}}_{\ , \ \partial_{\ \varrho}
\ W_{\ \sigma}^{\ s}}
\ - 
\hspace*{0.15cm} \overline{{\cal{L}}}_{\ , \ W_{\ \sigma}^{\ s}}
\ \right ) \ ( \ x \ )
\ = 
\vspace*{0.1cm} \\
\hspace*{0.1cm} =
\ D_{\ \varrho} \ ( \ ad \ ) 
\ : \ \left \lbrace \left \lbrack
\ \left ( \begin{array}{l} \overline{g}^{\ -2} \ - J 
\ + 
\vspace*{0.1cm} \\
+ \ \frac{1}{32 \ \pi^{\ 2}} \ b_{\ 0} \ \overline{B}
\end{array}
 \right ) \ ( \ {\cal{X}} \ )  \right \rbrack
\ B^{\ \sigma \ \varrho} \ \right \rbrace^{\ s} \ : \ ( \ x \ )
\hspace*{0.15cm} \rightarrow
\vspace*{0.2cm} \\
\varphi \ ( \ x \ ) \ = \ 
\ : \ \left ( \ \overline{g}^{\ -2} \ - J 
\ + \ \frac{1}{32 \ \pi^{\ 2}} \ b_{\ 0} \ \overline{B}
\ \right ) \ ( \ {\cal{X}} \ ) \ : \ ( \ x \ )
\ \equiv
\ \overline{G} \ ( \ x \ ) 
\end{array}
\end{equation}

\noindent
The field $\ \varphi \ $ introduced in eq. \ref{eq:C-14} is 
{\color{cyan} dimensionless} ,
which in itself is a consequence of the violation of dilatation invariance, or 
the trace anomaly.

\noindent
The equations of motion for the gauge fields ( eq. \ref{eq:C-8} ) become ,
suppressing the ordering signs $ : \ : \ $

\vspace*{-0.5cm}
\begin{equation}
\label{eq:C-15}
\begin{array}{l}
D_{\ \varrho} \ ( \ ad \ ) 
\ \left \lbrace 
\ {\color{red} \varphi} \ \ B^{\ \sigma \ \varrho} 
\ \right \rbrace^{\ s} \ ( \ x \ )
\ = \ \left ( \hspace*{0.1cm} j^{\ \sigma \ s} 
\ \right )_{\ \left \lbrace q \right \rbrace} \ ( \ x \ )
\vspace*{0.2cm} \\ \hline \vspace*{-0.3cm} \\
\left ( \hspace*{0.1cm} j^{\ \sigma \ s} 
\ \right )_{\ \left \lbrace q \right \rbrace} \ =
\ \sum_{\ q-fl} \ \overline{q}^{\ \dot{c}'}
\ \left \lbrace \ \gamma^{\ \sigma}
\ \left ( \ \frac{1}{2} \ \lambda^{\ s} \ \right )_{\ c' \dot{c}}
\ \right \rbrace \ q^{\ c}
\ = \ \left ( \ {\cal{L}}_{\ \left \lbrace q \right \rbrace} 
\ \right )_{\ , \ W_{\ \sigma}^{\ s}}
\vspace*{0.2cm} \\
\begin{array}{lll}
\begin{array}{l}
D_{\ \varrho} \ ( \ ad \ ) 
\vspace*{0.1cm} \\
\ \left \lbrace 
\ {\color{red} \varphi} \ \ B^{\ \sigma \ \varrho} 
\ \right \rbrace^{\ s} 
\vspace*{0.1cm} \\
\hspace*{0.7cm} ( \ x \ )
\end{array}
& = & 
\left \lbrack \begin{array}{l}
\hspace*{0.3cm}
\partial_{\ x \ \varrho}
\ \left \lbrace 
\ {\color{red} \varphi \ ( \ x \ )} \ B^{\ \sigma \ \varrho \ s} \ ( \ x \ ) 
\ \right \rbrace \ +
\vspace*{0.1cm} \\
+ \ f_{\ s t r} 
\ \left \lbrace \ W_{\ \varrho}^{\ t} \ ( \ x \ ) 
\ {\color{red} \varphi \ ( \ x \ )} 
\ B^{\ \sigma \varrho \ \ r} \ ( \ x \ )
\ \right \rbrace
\end{array} 
\right \rbrack
\vspace*{0.2cm} \\
& = &
\left \lbrack \begin{array}{l}
\hspace*{0.3cm}
\left \lbrace
\ \partial_{\ x \ \varrho} \ {\color{red} \varphi \ ( \ x \ )} 
\ \right \rbrace
\ \left \lbrace
B^{\ \sigma \ \varrho \ s} \ ( \ x \ )
\ \right \rbrace
\ +
\vspace*{0.1cm} \\
+ \ {\color{red} \varphi \ ( \ x \ )}
\ \left \lbrace
\ D_{\ \varrho} \ ( \ ad \ )  \  B^{\ \sigma \ \varrho} 
\ \right \rbrace^{\ s} \ ( \ x \ )
\end{array} 
\right \rbrack
\end{array}
\vspace*{0.2cm} \\ \hline \vspace*{-0.3cm} \\
{\color{red} \varphi \ ( \ x \ )} \ = \ 
\ : \ \left ( \ \overline{g}^{\ -2} \ - J 
\ + \ \frac{1}{32 \ \pi^{\ 2}} \ b_{\ 0} \ \overline{B}
\ \right ) \ ( \ {\cal{X}} \ ) \ : \ ( \ x \ )
\vspace*{0.2cm} \\ \hline
\end{array}
\end{equation}

\noindent
In deriving eq. \ref{eq:C-15} we assumed that chain rules for normal partial
derivatives and covariant derivatives are maintained through the 
ordering processes , suppressed for simplicity of notation .

}



{\color{blue} 

\begin{center}
\vspace*{-0.0cm}
{\bf \color{cyan} 
3 CD - Equations of motion modify the canonically conjugate variables 
pertaining to
$\ \overline{{\cal{L}}} \ $, beyond the reduction to consider 
exclusively the composite local field $\ {\cal{X}} \ ( \ x \ ) \ $
}
\label{'3C-D'}
 \end{center}

\noindent
Finally we turn to the consequences elaborated in the previous subsection ,
as they arise for the structure of canonically conjugate variables ,
in an axial gauge $\ W_{\ 0}^{\ s} \ ( \ x \ ) \ = \ 0 \ $, in 
{\it essential} contrast
to the structure pertaining to the bare Lagrangen $\ {\cal{L}} \ $,
as discussed in the section
{\bf \color{cyan} 3 a - Bare Lagrangean density
and equations of motion in unconstrained gauges} .

\noindent
Thus we consider the canonically conjugate variables pertaining to 
$\ \overline{{\cal{L}}} \ $

\vspace*{-0.3cm}
\begin{equation}
\label{eq:CD-1}
\begin{array}{l}
W_{\ \sigma}^{\ s} \ \leftrightarrow 
\ \overline{{\cal{L}}}_{\ , \ \partial_{\ 0}
\ W_{\ \sigma}^{\ s}}
\hspace*{0.2cm} \rightarrow \hspace*{0.2cm}
\sigma \ = \ m \ = \ 1,2,3
\hspace*{0.2cm} \mbox{for} \hspace*{0.2cm}
W_{\ 0}^{\ s} \ = \ 0
\end{array}
\end{equation}

\noindent
The derivatives of $\ \overline{{\cal{L}}} \ $ are given in eq. \ref{eq:C-5} .
Using the scalar field $\ \varphi \ $,
defined in eq. \ref{eq:C-14} , eq. \ref{eq:C-5} takes the form 

\vspace*{-0.3cm}
\begin{equation}
\label{eq:CD-2}
\begin{array}{l}
\begin{array}{lll}
\overline{{\cal{L}}}_{\ , \ \partial_{\ \varrho}
\ W_{\ \sigma}^{\ s}} & = 
& {\color{red} \varphi} \ B^{\ \sigma \ \varrho \ s} 
\vspace*{0.1cm} \\
\overline{{\cal{L}}}_{\ , \ W_{\ \sigma}^{\ s}} & =
& - \ f_{\ s t r} \ W_{\ \varrho}^{\ t} \ {\color{red} \varphi}
\ B^{\ \sigma \varrho \ \ r} 
\end{array}
\vspace*{0.2cm} \\ \hline \vspace*{-0.3cm} \\
{\color{red} \varphi \ ( \ x \ )} \ = \ 
\ : \ \left ( \ \overline{g}^{\ -2} \ - J 
\ + \ \frac{1}{32 \ \pi^{\ 2}} \ b_{\ 0} \ \overline{B}
\ \right ) \ ( \ {\cal{X}} \ ) \ : \ ( \ x \ )
\end{array}
\end{equation}

}



{\color{blue} 

\noindent
From eq. \ref{eq:CD-2} we obain the canonically conjugate momentum fields
relative to $\ W_{\ m = \sigma}^{\ s} \ $

\vspace*{-0.3cm}
\begin{equation}
\label{eq:CD-3}
\begin{array}{ll}
{\cal{A}} \ : &
\overline{\Pi}^{\ m \ s} \ = \ \varphi \ B^{\ m \ 0 \ s} \ = \ - 
\ \varphi \ \left ( \ \vec{E}^{\ s} \ \right )^{\ m}
\hspace*{0.2cm} ; \hspace*{0.2cm} m \ = \ 1,2,3
\end{array}
\end{equation}

\noindent
to be compared with eq. \ref{eq:3A-4} relative to $\ {\cal{L}} \ $.

\noindent
Using the substitution $\ \overline{\Pi} \ \leftrightarrow \
\overline{{\cal{L}}} \ $, the Ansatz for equal time commutation relations 
in eq. \ref{eq:3A-5} remains the same 

\vspace*{-0.5cm}
\begin{equation}
\label{eq:CD-4}
\begin{array}{ll}
{\cal{A}} \ : &
\left \lbrack 
\ W^{\ r}_{\ m} \ ( \ t \ , \ \vec{x} \ ) \ ,
\ \overline{\Pi}^{\ s \ n} \ ( \ t \ , \ \vec{y} \ ) \ \right \rbrack
\ = \ i \ \delta^{\ r \ s} \ \delta_{\ m}^{\ n} 
\ \delta^{\ (3)} \ ( \ \vec{x} \ - \ \vec{y} \ ) \ \P  
\vspace*{0.1cm} \\
& \left \lbrack
\ W^{\ r}_{\ m} \ ( \ t \ , \ \vec{x} \ ) \ ,
\ W^{\ s}_{\ n} \ ( \ t \ , \ \vec{y} \ ) \ \right \rbrack \ = \ 0
\vspace*{0.1cm} \\
& \left \lbrack
\ \overline{\Pi}^{\ r \ m} \ ( \ t \ , \ \vec{x} \ ) \ ,
\ \overline{\Pi}^{\ s \ n} \ ( \ t \ , \ \vec{y} \ ) \ \right \rbrack \ = \ 0
\end{array}
\end{equation}

\noindent
yet the intervention of the nontrivial scalar field $\ \varphi \ $
-- very inequivalent from the bare Lagrangean counterpart
$\ g^{\ -2} \ $ -- shows that consistency of the Euler-Lagrange equations
takes its toll, in an interesting way. This result is the key advance 
I can report here and now .

\noindent
In this and the next subsection it remains to draw some conclusions and sketch eventual future advances .

}



{\color{blue} 

\begin{center}
\vspace*{-0.0cm}
{\bf \color{red} 3 D - Remarks and consequences arising from
derivations in the last subsections : {\color{cyan} 3I} , 3C , 
{\color{cyan} 3C1 , 3CD}
}
 \label{'3D'}
 \end{center}
 \vspace*{-0.1cm}

\begin{description}

\item 1) The orignal derivation of the trace anomaly was done in the
perturbative region in the ultraviolet for QCD and in the infrared for QED .
This was sufficient for QCD , to establish through dominance of the leading
contribution $\ 2 \ \beta \ ( \ g \ ) \ / \ g \ 
\ \left ( \ \frac{1}{4} \ B_{\ \mu \nu}^{\ s \ pert.} \ B^{\ \mu \nu \ s \
pert.} \ \right ) \ $ over quark mass terms , 
the breaking of dilatation invariance for vanishing quark masses.

Nevertheless -- as far as the overall dominance is concerned --
for {\it all} scales , 
extending to the
nonperturbative and unstable
regions , I have not checked before , whether the unique renormalization and
normalization of the field strength bilinear $\ {\cal{X}} \ = 
\ \frac{1}{4} \ B_{\ \mu \nu}^{\ s} \ B^{\ \mu \nu \ s} \ $ was sufficient to
yield the ultraviolet limiting form
of the trace of the symmetric energy momentum tensor 
$\ \vartheta^{\ \mu}_{\hspace*{0.2cm} \mu} \ $ modulo a numerical constant
related to the rescaling function $\ \beta \ ( \ g \ ) \ $ and vanishing only
for identically vanishing $\ \beta \ $.

However there persisted an inconsistency between singling out the composite
local field $\ {\cal{X}} \ $ as the only contribution to the trance anomaly,
since the so restricted Euler-Lagrange equations of motion did not satisfy
the anomalous trace Ward identities, rather yield a traceless energy momentum
tensor .

This leads as shown in the subsections {\color{cyan} 3I} , 3C ,
{\color{cyan} 3C1 , 3CD} to a very different structure of the full trace
anomaly, whereby the above inconsistency is resolved.

In the language of e.g. the QCD sum rules, in which the perturbative ordering 
of composite local operators according to their mass  ($ \ M^{\ n} \ $) 
dimension -- n -- is essential in the perturbative regime, this ordering is
upset in the full trace anomaly , whence extended to all physical scales,
and mixing of essentially all n dimensional operators, remaining local does
occur. The high n fields do not enter with arbitrarily large numbers of
derivatives, endangering locality, but in a canonical Hamiltonian sense through
high powers of only zero'th and first ( covariant ) derivatives 
of the base fields of QCD .

This is, a new result, borne out in the appearance in the equations of
motion of the compsite scalar and gauge invariant local field 
$\ \varphi \ ( \ x \ ) \ $ as defined in eqs. \ref{eq:C-5} -  \ref{eq:CD-2} ,
in the subsections {\color{cyan} 3C1 , 3CD} .
$\ \varphi \ $ is dependent on th basic fields of QCD and carries mass
dimension 0 .

\item 2) Generating consistent second order ( Hamiltonian ) equations through
a variational principle 

In a way the use of generalized N\"{o}ther currents \cite{EmmaN}
modulo Euler-Lagrange equations is here reversed . The relations derived
are consistency equations between the trace of the energy momentum density
tensor and the local field operators equal to it. They derive from 
a Hamiltonian system and
its Euler-Lagrange equations only , in contrast to quark current algebra
relations for additional flavor symmetries .

\end{description}

}



{\color{blue} 

\begin{description}

\item 3) In the construction of the full symmetric and gauge invariant 
energy momentum density tensor components , also gravitational interactions 
enter, albeit only in the uncurved space limit 
$\ g_{\ \mu \nu} \ \rightarrow \ \eta_{\ \mu \nu} \ $ as shown in 
eq. \ref{eq:3I-11} .
The complications in enforcing strict local gauge invarince 
-- with respect to $\ SU3_{\ c} \ $ here -- have a clear basis in
the Becchi, Rouet, Stora and Tyutin construction of BRST transfromation
rules for ghost fields \cite{BRS,Tyutin} , necessarily present in 
Lorentz covariant gauges .

The nonperturbative completion of the full field theories in question
is a seperate far goal, beyond the scope of the present discussion .

\end{description}

\begin{center}
\vspace*{-0.0cm}
{\bf \color{magenta} --- \hspace*{0.3cm} Thank you \hspace*{0.3cm} ---
}

\end{center}
\vspace*{0.2cm}

{\color{cyan} Acknowledgement : I would like to thank Raymond Stora ,
Rod Crewther and Andrei Kataev for discussions , even if in the late stages
of this work they were only 'preliminary' as far as the topics presented here
are concerned .}

}



{\color{blue} 

\begin{center}
\vspace*{-0.0cm}
{\bf \color{cyan} 3 b - Dimension 
$\ \left \lbrack \  M^{\ 4} \ \right \rbrack \ $ equations for field
strengths in unconstrained gauges
}
 \label{'3b'}
 \end{center}

\noindent
{\color{cyan} The following maerial was derived before the main discussion of
the insertions : sections 3I to} {\color{red} 3D .}

\noindent
The equations of motion ( eq. \ref{eq:3-18} ) as well as the Bianchi identities
to which we turn below are of engineering dimension 
$\ \left \lbrack \  M^{\ 3} \ \right \rbrack \ $.
In order to derive the dimension 
$\ \left \lbrack \  M^{\ 4} \ \right \rbrack \ $ equations we introduce
the matrix notations as adapted to the adjoint representation and convert
Lorentz indices freely between covariant and contravariant ones

\vspace*{-0.3cm}
\begin{equation}
\label{eq:3-19}
\begin{array}{l}
D_{\ \varrho} \ = \ \partial_{\ \varrho} \ + \ {\cal{W}}_{\ \varrho}
\hspace*{0.2cm} ; \hspace*{0.2cm}
{\cal{W}}_{\ \varrho} \ = \ W^{\ r}_{\ \varrho} \ ad_{\ r}
\vspace*{0.1cm} \\
D^{\ \varrho} \ \left \lbrace \ \begin{array}{c}
1
\vspace*{0.2cm} \\
\hline  \vspace*{-0.3cm} \\
g^{\ 2}
\end{array} \hspace*{0.1cm}
\ B_{\ \sigma \varrho} \ \right \rbrace^{\ s} 
\ = \ j_{\ \sigma}^{\ s}
\end{array}
\end{equation}

}



{\color{blue} 

\noindent
Then we form the antisymmetric covariant ( electric- and magnetic- )
dipole current densities and suppress the
adjoint component $\ ^{s} \ $ and abbreviate the inverse square bare
coupling constant by $\ G \ = \ g^{\ -2} \ $ 

\vspace*{-0.3cm}
\begin{equation}
\label{eq:3-20}
\begin{array}{l}
G \ = \ g^{\ -2} 
\vspace*{0.1cm} \\
D_{\ \tau} \ D_{\ \varrho}
\ \left \lbrace
\ G \ B_{\ \sigma}^{\hspace*{0.2cm} \varrho} \ \right \rbrace
\ - \ D_{\ \sigma} \ D_{\ \varrho}
\ \left \lbrace
\ G \ B_{\ \tau}^{\hspace*{0.2cm} \varrho} \ \right \rbrace
\ = \ D_{\ \tau} \ j_{\ \sigma} \ - \ D_{\ \sigma} \ j_{\ \tau}
\end{array}
\end{equation}

\noindent
Next we move the covariant derivatives $\ D_{\ \tau} \ , \ D_{\ \sigma} \ $
to the right using the identity

\vspace*{-0.3cm}
\begin{equation}
\label{eq:3-21}
\begin{array}{l}
D_{\ \tau} \ D_{\ \varrho} \ = 
\ \left \lbrack \ D_{\ \tau} \ , \ D_{\ \varrho} \ \right \rbrack
\ + \ D_{\ \varrho} \ D_{\ \tau}
\end{array}
\end{equation}

\noindent
Eq. \ref{eq:3-20} thus takes the form

\vspace*{-0.3cm}
\begin{equation}
\label{eq:3-22}
\begin{array}{l}
\left \lbrace \ \left \lbrack \ D_{\ \tau} \ , \ D_{\ \varrho} \ \right \rbrack
\ + \ D_{\ \varrho} \ D_{\ \tau} \ \right \rbrace
\ \left \lbrace
\ G \ B_{\ \sigma}^{\hspace*{0.2cm} \varrho} \ \right \rbrace 
\ - \ \left ( \ \tau \ \leftrightarrow \ \sigma \ \right )
\ = 
\vspace*{0.1cm} \\
\ = \ D_{\ \tau} \ j_{\ \sigma} \ - \ D_{\ \sigma} \ j_{\ \tau}
\end{array}
\end{equation}

\noindent
The commutators of covariant derivatives reduce to the adjoint field strength
matrix valued form

\vspace*{-0.3cm}
\begin{equation}
\label{eq:3-23}
\begin{array}{l}
\left \lbrack \ D_{\ \tau} \ , \ D_{\ \varrho} \ \right \rbrack
\ = \ {\cal{B}}_{\ \tau \varrho} \ = \ B_{\ \tau \varrho}^{\ t} \ ad_{\ t} 
\hspace*{0.2cm} ; \hspace*{0.2cm}
\tau \ \varrho \ \rightarrow \ \sigma \ \varrho
\end{array}
\end{equation}

\noindent
which yields , substituded into eq. \ref{eq:3-22} 

}



{\color{blue} 

\vspace*{-0.3cm}
\begin{equation}
\label{eq:3-24}
\begin{array}{l}
\left \lbrack
\begin{array}{c}
D^{\ \varrho} 
\ \left ( \ D_{\ \tau} \ B_{\ \sigma \varrho} 
\ -
\ D_{\ \sigma} \ B_{\ \tau \varrho} 
\ \right )
\ +
\vspace*{0.1cm} \\
+ \ {\cal{B}}_{\ \tau \varrho} \ B_{\ \sigma}^{\hspace*{0.2cm} \varrho}
\ - \ {\cal{B}}_{\ \sigma \varrho} \ B_{\ \tau}^{\hspace*{0.2cm} \varrho}
\end{array} \right \rbrack
\ =
\ g^{\ 2} \ \left ( \ 
\ D_{\ \tau} \ j_{\ \sigma} \ - \ D_{\ \sigma} \ j_{\ \tau} \ \right )
\end{array}
\end{equation}

\noindent
At this stage whence factoring out G beyond partial derivatives in 
eq. \ref{eq:3-24} we assume that G 
is not space-time dependent, and come back at a later stage to  consider
an arbitrary space-time dependent extension

\vspace*{-0.3cm}
\begin{equation}
\label{eq:3-25}
\begin{array}{l}
\partial_{\ \mu} G \ = 0 
\hspace*{0.2cm} \mbox{with the extension}
\ G \ \rightarrow \ \widetilde{G} \ ( \ x \ ) 
\hspace*{0.1cm} \mbox{with} \hspace*{0.1cm}
\lim_{\ x \rightarrow \infty} \widetilde{G} \ ( \ x \ ) \ = \ G
\end{array}
\vspace*{-0.1cm}
\end{equation}

\noindent
as an external source with appropriate bondary conditions
for $\ x \ \rightarrow \ \infty \ $.

\noindent
The qantity in brackets in the first line of the left hand side of 
eq. \ref{eq:3-24} can be transformed , using the Bianchi identity
\cite{phaseQCD2011}
for the covariant derivatives of field strengths

\vspace*{-0.3cm}
\begin{equation}
\label{eq:3-26}
\begin{array}{l}
D_{\ \tau} \ B_{\ \sigma \varrho} \ -
\ D_{\ \sigma} \ B_{\ \tau \varrho} \ =
\ D_{\ \tau} \ B_{\ \sigma \varrho} \ +
\ D_{\ \sigma} \ B_{\ \varrho \tau} \ = 
\ - \ D_{\ \varrho} \ B_{\ \tau \sigma}
\hspace*{0.2cm} \longrightarrow
\vspace*{0.1cm} \\
\mbox{Bianchi identity :} \hspace*{0.25cm}
D_{\ \tau} \ B_{\ \sigma \varrho} \ +
\ D_{\ \sigma} \ B_{\ \varrho \tau} \ +
\ D_{\ \varrho} \ B_{\ \tau \sigma} \ = \ 0
\end{array}
\end{equation}

\noindent
The dipole density second order differential equation thus can be brought 
to the form

\vspace*{-0.3cm}
\begin{equation}
\label{eq:3-27}
\begin{array}{l}
\left \lbrack
\begin{array}{c}
D^{\ \varrho} \ D_{\ \varrho} \ B_{\ \sigma \tau}
\ +
\vspace*{0.1cm} \\
+ \ {\cal{B}}_{\ \tau \varrho} \ B_{\ \sigma}^{\hspace*{0.2cm} \varrho}
\ - \ {\cal{B}}_{\ \sigma \varrho} \ B_{\ \tau}^{\hspace*{0.2cm} \varrho}
\end{array} \right \rbrack
\ =
\ g^{\ 2} \ \left ( \ D_{\ \tau} \ j_{\ \sigma} 
\ - \ D_{\ \sigma} \ j_{\ \tau} \ \right )
\end{array}
\end{equation}

}



{\color{blue} 

\noindent
We summarize the equations of motion -- eqs. \ref{eq:3-17} , \ref{eq:3-19} --
and the chromo-dipole-density 
$\ dim \ \left \lbrack \ M^{\ 4} \ \right \rbrack \ $ derived equations
-- eq. \ref{eq:3-27} -- below

\vspace*{-0.3cm}
\begin{equation}
\label{eq:3-28}
\begin{array}{|c|}
\hline \\
\partial_{\ \varrho}
\ \left \lbrace 
\begin{array}{c}
\delta \ {\cal{L}}
\vspace*{0.2cm} \\
\hline  \vspace*{-0.3cm} \\
\delta \ B_{\ \mu \ \nu}^{\ r} 
\end{array} 
\hspace*{0.1cm} \begin{array}{c}
\delta \ B_{\ \mu \nu}^{\ r}
\vspace*{0.2cm} \\
\hline  \vspace*{-0.3cm} \\
\delta \ \left ( \ \partial_{\ \varrho} \ W_{\ \sigma}^{\ s} \ \right )
\end{array} \right \rbrace
\ - \ \begin{array}{c}
\delta \ {\cal{L}}
\vspace*{0.2cm} \\
\hline  \vspace*{-0.3cm} \\
\delta \ B_{\ \mu \ \nu}^{\ r}
\end{array}
\hspace*{0.1cm} \begin{array}{c}
\delta \ B_{\ \mu \nu}^{\ r}
\vspace*{0.2cm} \\
\hline  \vspace*{-0.3cm} \\
\delta \ W_{\ \sigma}^{\ s}
\end{array}
\ = \ \begin{array}{c}
\delta \ {\cal{L}}_{\ \left \lbrace q \right \rbrace}
\vspace*{0.2cm} \\
\hline  \vspace*{-0.3cm} \\
\delta \ W_{\ \sigma}^{\ s}
\end{array}
\vspace*{0.0cm} \\ \vspace*{-0.2cm} \\
\begin{array}{c}
\delta \ {\cal{L}}_{\ \left \lbrace q \right \rbrace}
\vspace*{0.2cm} \\
\hline  \vspace*{-0.3cm} \\
\delta \ W_{\ \sigma}^{\ s}
\end{array} \hspace*{0.1cm} = 
\ \sum_{\ q-fl} \ \overline{q}^{\ \dot{c}'}
\ \left \lbrace \ \gamma^{\ \sigma}
\ \left ( \ \frac{1}{2} \ \lambda^{\ s} \ \right )_{\ c' \dot{c}}
\ \right \rbrace \ q^{\ c}
\ = \ \left ( \hspace*{0.1cm} j^{\ \sigma \ s} 
\ \right )_{\ \left \lbrace q \right \rbrace}
\vspace*{-0.2cm} \\ \\
\hline \vspace*{-0.35cm} \\
D_{\ \varrho} \ = \ \partial_{\ \varrho} \ + \ {\cal{W}}_{\ \varrho}
\hspace*{0.2cm} ; \hspace*{0.2cm}
{\cal{W}}_{\ \varrho} \ = \ W^{\ r}_{\ \varrho} \ ad_{\ r}
\vspace*{-0.1cm} \\ \vspace*{-0.3cm} \\
D^{\ \varrho} \ \left \lbrace 
\ B_{\ \sigma \varrho} \ \right \rbrace^{\ s} 
\ = \ g^{\ 2} \ j_{\ \sigma}^{\ s}
\vspace*{-0.3cm} \\ \\
\hline \vspace*{-0.25cm} \\
\left \lbrack
\begin{array}{c}
D^{\ \varrho} \ D_{\ \varrho} \ B_{\ \sigma \tau}
\ +
\vspace*{0.1cm} \\
+ \ {\cal{B}}_{\ \tau \varrho} \ B_{\ \sigma}^{\hspace*{0.2cm} \varrho}
\ - \ {\cal{B}}_{\ \sigma \varrho} \ B_{\ \tau}^{\hspace*{0.2cm} \varrho}
\end{array} \right \rbrack
\ =
\ g^{\ 2} \ \left ( \ D_{\ \tau} \ j_{\ \sigma} 
\ - \ D_{\ \sigma} \ j_{\ \tau} \ \right )
\vspace*{-0.3cm} \\ \\ \hline
\end{array}
\vspace*{0.5cm}
\end{equation}

\begin{center}
\vspace*{-0.0cm}
{\bf \color{red} 3 c - QED : bare Lagrangean density
and equations of motion in unconstrained -- abelian -- gauges
}
 \label{'3c'}
 \end{center}

\noindent
I consider it worth the effort to align all conventions to the (an) extension
of QCD to the only other unbroken charge like gauge field theory : QED .

\noindent
To this end we extend the equations of motion accordingly starting with
connections in eq. \ref{eq:2-1} in section 2 and eq. \ref{eq:3-10} in 
subsection 3a , as follows

\vspace*{-0.3cm}
\begin{equation}
\label{eq:3-29}
\begin{array}{l}
{\cal{G}} \ \rightarrow \ SU3_{\ c} \ \times \ U1_{\ em}
\hspace*{0.2cm} ; \hspace*{0.2cm}
q \ \rightarrow \ f \ = \ \left ( \begin{array}{l} \updownarrow \ q_{\ c} 
\vspace*{0.1cm} \\
\updownarrow \ \ell 
\end{array}
\right ) 
\vspace*{0.1cm} \\
{\cal{W}}_{\ \mu} \ \rightarrow \ W_{\ \mu}^{\ r} \ d_{\ r} \ ( \ {\cal{D}} \ )
\ + \ W_{\ \mu}^{\ em} \ d_{\ em}
\vspace*{0.1cm} \\
{\cal{D}} \ : 
\ \left \lbrack \mbox{\begin{tabular}{c} irreducible
\vspace*{-0.1cm} \\
representation
\vspace*{-0.1cm} \\
of
\ $\ SU3_{\ c} \ $
\vspace*{-0.1cm} \\
acting on q flavors only
\end{tabular}
}
\right \rbrack
\hspace*{0.2cm} ; \hspace*{0.2cm}
d_{\ em} \ : \ \left \lbrack \ \mbox{diagonal matrix} \ \right \rbrack
\end{array}
\end{equation}

\noindent
The $\ \updownarrow \ $ arrows in the column vector $\ f_{\ \bullet} \ $
in eq. \ref{eq:3-29} shall indicate that quark- and charged lepton flavor- 
entries -- tricolored for quarks and colorless for charged leptons --
are arranged sequentially . Taking e.g. up and down qark flavors and 
$\ e^{\ -} \ , \ \mu^{\ -} \ $ charged lepton flavors the transposed 
row vector $\ f_{\ \bullet}^{\ T} \ $

\vspace*{-0.3cm}
\begin{equation}
\label{eq:3-30}
\begin{array}{l}
f^{\ T} \ = \ \left \lbrace
\ u_{\ r} \ , \ u_{\ g} \ , \ u_{\ b} \ ; \ d_{\ r} \ , \ d_{\ g} \ , \ d_{\ b}
\ ; \ e^{\ -} \ , \ \mu^{\ -}
\ \right \rbrace
\end{array}
\vspace*{-0.3cm}
\end{equation}

}



{\color{blue} 

\noindent
Generalizations beyond the six known color triplet quark flavors \\
$\ u \ , \ d \ ; \ c \ , \ s \ ; \ t \ , \ b \ $ and the three 
equally charged leptons $\ e^{\ -} \ , \ \mu^{\ -} \ , \ \tau^{\ -} \ $
to colored flavors in other irreducible representations of $\ SU3_{\ c} \ $,
as well as to charged leptons with charges different from 
$\ e^{\ -} \ , \ \mu^{\ -} \ , \ \tau^{\ -} \ $ are then straightforward ,
whereby the spins shall be restricted to $\ \frac{1}{2} \ $ ( fermions )
and the union of color representations is limited such as not to upset 
asymptotic freedom in the ultraviolet for $\ SU3_{\ c} \ $.\footnote{
{\color{blue} \begin{tabular}[t]{l} 
The ultraviolet stability of QCD , 
whence isolated from QED
is obviously not \vspace*{-0.1cm} \\
sufficient to establish
its asymptotic ultraviolet behavior 
in the context of \vspace*{-0.1cm} \\
QCD-QED,
because of the 
opposite
stability - instability
behaviour of QCD 
\vspace*{-0.1cm} \\
relative to QED, considered in
isolation
one from the other. In the present sub-
\vspace*{-0.1cm} \\
section
we are not concerned with the full QCD-QED complex .
\end{tabular}}}

\noindent
This said, the matrices $\ d_{\ r} \ , \ d_{\ em} \ $ in eq. \ref{eq:3-29}
over the union of irreducible color representations $\ \bigcup \ {\cal{D}} \ $
is suitably extended , are block diagonal for 
$\ d_{\ r} \ \left ( \ \begin{array}{l} \bigcup \ {\cal{D}} \ \end{array} 
\right ) \ $ and $\ d_{\ em} 
\ \left ( \ \begin{array}{l} \bigcup \ \ell \ \end{array} \right ) \ $,
with

\vspace*{-0.3cm}
\begin{equation}
\label{eq:3-31}
\begin{array}{l}
\left \lbrack d_{\ r} , d_{\ em} \right \rbrack \ = \ 0
\hspace*{0.1cm} ; \hspace*{0.1cm} 
r \ = \ 1, \cdots , 8
\end{array}
\end{equation}

}



{\color{blue} 

\noindent
For the flavor set in eq. \ref{eq:3-30} the 9 matrices 
( $\ 8 \ \times \ 8 \ $) 
$\ d_{\ r} , \ d_{\ em} \ $ are 

\vspace*{-0.3cm}
\begin{equation}
\label{eq:3-32}
\hspace*{0.0cm} \begin{array}{l}
d_{\ r} \ =
\ \frac{1}{i} \hspace*{0.15cm} \begin{array}{|c|c|c|}
\hline  \vspace*{-0.47cm} \\ & & \vspace*{-0.33cm} \\ 
\frac{1}{2} \ \lambda_{\ r} & 0 & 0
\vspace*{-0.3cm} \\ & & \vspace*{-0.2cm} \\
0 & \frac{1}{2} \ \lambda_{\ r} & 0
\vspace*{-0.3cm} \\ \vspace*{-0.2cm} \\
0 & 0 & 0
\\ \hline
\end{array}
\hspace*{0.1cm} ; \hspace*{0.1cm}
d_{\ em} =
 \frac{1}{i} \hspace*{0.15cm} \begin{array}{|c|c|c|}
\hline  \vspace*{-0.47cm} \\ & & \vspace*{-0.33cm} \\
\frac{2}{3} \ \P_{\ 3 \times \ 3} & 0_{\ 3 \times 3} & 0_{\ 3
\times 2}
\vspace*{-0.3cm} \\ & & \vspace*{-0.2cm} \\
0_{\ 3 \times 3} & - \ \frac{1}{3} \ \P_{\ 3 \times 3} & 0_{\ 3
\times 2}
\vspace*{-0.3cm} \\ \vspace*{-0.2cm} \\
0_{\ 2 \times 3} & 0_{\ 2 \times 3} & - \ \P_{\ 2 \times 2}
\\ \hline
\end{array}
\end{array} 
\end{equation}

\noindent
The sub-block sizes are indicated in the entries for $\ d_{\ em} \ $ in
eq. \ref{eq:3-32} . We retain the form of the electromagnetic potentials
( eq. \ref{eq:3-29} )

\vspace*{-0.3cm}
\begin{equation}
\label{eq:3-33}
\begin{array}{l}
{\cal{W}}^{\ em}_{\ \mu} \ = \ W^{\ em}_{\ \mu} \ d_{\ em}
\hspace*{0.2cm} ; \hspace*{0.2cm}
d_{\ em} \ = \ \frac{1}{i} \ Q_{\ em}
\vspace*{0.1cm} \\
Q_{\ em} \ = \ diag \ ( \ Q_{\ f_{\ 1}} \ , \ \cdots \ Q_{\ f_{\ N}} \ )
\hspace*{0.2cm} ; \hspace*{0.2cm}
Q_{\ f_{\ \bullet}} \ = \ e \ ( \ f_{\ \bullet} \ ) \ / \ e
\end{array}
\end{equation}

\noindent
The diagonal elements of the ( hermitian ) matrix $\ Q_{\ em} \ $ as defined in
eq. \ref{eq:3-33} are the relative electric charges of the individal fermions
to the elementary charge , of the proton say , identical for all members
transforming under an irreducible representation of $\ SU3_{\ c} \ $.

}



{\color{blue} 

\noindent
In a first step we extend the Lagrangean density 
$\ {\cal{L}}_{\ \left \lbrace q \right \rbrace} \ $ in eq. \ref{eq:3-14}
for color triplet quark flavors, to which we restrict the $\ f_{\ \bullet} \ $
components considered here, to QCD-QED

\vspace*{-0.3cm}
\begin{equation}
\label{eq:3-34}
\begin{array}{l}
{\cal{L}}_{\ \left \lbrace q \right \rbrace}
 =  \sum_{ q-fl}  \overline{q}^{\ \dot{c}'}
\ \left \lbrace \frac{i}{2}
\ \gamma^{\ \mu} \ \left \lbrack
\begin{array}{c}
 \left (  
\begin{array}{c}
\vspace*{-0.5cm} \\
\rightarrow
\vspace*{-0.25cm} \\
D
\end{array}_{ \mu} \ \left ( \ 3 \ \right ) \ \right )_{\ c' \dot{c}}
\ - 
\vspace*{0.1cm} \\
- \ \left ( \
\begin{array}{c}
\vspace*{-0.5cm} \\
\leftarrow
\vspace*{-0.25cm} \\
D
\end{array}_{ \mu} \ \left ( \overline{3} \right ) 
\ \right )_{\ \dot{c} c'}
\end{array} \right \rbrack
\ - \ m_{\ q} \ \delta_{ c' \dot{c}} \right \rbrace
 q^{\ c}
\vspace*{0.1cm} \\
\begin{array}{lll}
\left ( \
\begin{array}{c}
\vspace*{-0.5cm} \\
\rightarrow
\vspace*{-0.25cm} \\
D
\end{array}_{\ \mu} \ \left ( \ 3 \ \right ) \ \right )_{\ c' \dot{c}}
& = & \begin{array}[t]{l} \begin{array}{c}
\vspace*{-0.5cm} \\
\rightharpoonup
\vspace*{-0.25cm} \\
\partial_{\ \mu}
\end{array} \ \delta_{\ c' \dot{c}} \ + \ W_{\ \mu}^{\ r} \ \frac{1}{i} 
\ \left ( \ \frac{1}{2} \ \lambda^{\ r} \ \right )_{\ c' \dot{c}} \ +
\vspace*{0.1cm} \\
\hspace*{1.5cm} + \hspace*{0.15cm} 
W_{\ \mu}^{\ em} \ \frac{1}{i} \ Q_{\ q} \  \delta_{\ c' \dot{c}}
\end{array}
\vspace*{0.1cm} \\
\left ( \
\begin{array}{c}
\vspace*{-0.5cm} \\
\leftarrow
\vspace*{-0.25cm} \\
D
\end{array}_{\ \mu} \ \left ( \ \overline{3} \ \right )
\ \right )_{\ \dot{c} c'} & =
& \begin{array}[t]{l} \begin{array}{c}
\vspace*{-0.5cm} \\
\leftharpoonup
\vspace*{-0.25cm} \\
\partial_{\ \mu}
\end{array} \ \delta_{\ \dot{c} c'} \ - \ W_{\ \mu}^{\ r} \ \frac{1}{i}
\ \left ( \ \frac{1}{2} \ \overline{\lambda}^{\ r} \ \right )_{\ \dot{c} c'}
\ -
\vspace*{0.1cm} \\
\hspace*{1.5cm} - \hspace*{0.15cm} W_{\ \mu}^{\ em} \ \frac{1}{i} 
\ Q_{\ q} \ \delta_{\ c' \dot{c}} 
\end{array}
\vspace*{0.1cm} \\
& = &
\begin{array}[t]{l} \begin{array}{c}
\vspace*{-0.5cm} \\
\leftharpoonup
\vspace*{-0.25cm} \\
\partial_{\ \mu}
\end{array} \ \delta_{\ c' \dot{c}} \ - \ W_{\ \mu}^{\ r} \ \frac{1}{i}
\ \left ( \ \frac{1}{2} \ \lambda^{\ r} \ \right )_{\ c' \dot{c}} \ -
\vspace*{0.1cm} \\
\hspace*{1.5cm} - \hspace*{0.15cm} W_{\ \mu}^{\ em} \ \frac{1}{i}
\ Q_{\ q} \ \delta_{\ c' \dot{c}}
\end{array}
\end{array}
\end{array}
\end{equation}

}



{\color{blue} 

\noindent
Eq. \ref{eq:3-15} becomes extending to all fermion flavors 
$\ f \ = \ \left ( \ q \ , \ \ell^{\ \alpha} \ \right ) \ $

\vspace*{-0.3cm}
\begin{equation}
\label{eq:3-35}
\begin{array}{l}
{\cal{L}}_{\ \left \lbrace q \right \rbrace}
\ = \ \sum_{\ q-fl} \ \overline{q}^{\ \dot{c}'}
 \left \lbrace
\ \gamma^{\ \mu} 
 \left \lbrack 
\begin{array}{l} \frac{i}{2} \ \begin{array}{c}
\vspace*{-0.5cm} \\
\rightleftharpoons
\vspace*{-0.20cm} \\
\partial \vspace*{-0.08cm}
\end{array}_{\mu} \hspace*{0.1cm}  \delta_{\ c' \dot{c}} 
\vspace*{0.1cm} \\
+ \hspace*{0.1cm} W_{\ \mu}^{\ r}
\ \left ( \ \frac{1}{2} \ \lambda^{\ r} \ \right )_{\ c' \dot{c}}
\vspace*{0.1cm} \\
+ \hspace*{0.1cm} \ W_{\ \mu}^{\ em} \ Q_{\ q} \ \delta_{\ c' \dot{c}} 
\end{array} \right \rbrack
\ - \ m_{\ q} \ \ \delta_{\ c' \dot{c}}
 \right \rbrace \ q^{\ c}
\vspace*{0.2cm} \\
{\cal{L}}_{\ \left \lbrace \ell \right \rbrace}
\ = \ \sum_{ \ell-fl} \ \overline{\ell}^{\ \dot{\alpha}'}
 \left \lbrace
\ \gamma^{ \mu} 
 \left \lbrack 
\begin{array}{l} \frac{i}{2} \ \begin{array}{c}
\vspace*{-0.5cm} \\
\rightleftharpoons
\vspace*{-0.20cm} \\
\partial \vspace*{-0.08cm}
\end{array}_{\mu} \hspace*{0.1cm}  \delta_{\ \alpha' \dot{\alpha}}
\vspace*{0.1cm} \\
+ \hspace*{0.1cm} W_{\ \mu}^{\ em} \ Q_{\ \ell_{\ \alpha}} 
\ \delta_{\ \alpha' \dot{\alpha}} 
\end{array} \right \rbrack
\ - \ m_{\ \ell_{ \alpha}} \ \ \delta_{\ \alpha' \dot{\alpha}}
 \right \rbrace \ \ell^{\ \alpha}
\vspace*{0.2cm} \\
{\cal{L}}_{\ \left \lbrace f \right \rbrace} \ =
\ {\cal{L}}_{\ \left \lbrace q \right \rbrace} \ +
\ {\cal{L}}_{\ \left \lbrace \ell \right \rbrace}
\end{array}
\end{equation}

\noindent
For the photon potentials (connection) and fielstrengths we perform
the analogous steps as for the nonabelian counterparts, repeating
for the latter eq. \ref{eq:3-3}

\vspace*{-0.3cm}
\begin{equation}
\label{eq:3-36}
\begin{array}{l}
{\cal{W}}^{\ (2)} \ ( \ {\cal{D}} \ ) \ \rightarrow \ {\cal{W}}^{\ (2)}
\ = \ \partial \ {\cal{W}}^{\ (1)} \ + \ \left ( \ {\cal{W}}^{\ (1)} 
\ \right )^{\ 2}
\hspace*{0.2cm} ; \hspace*{0.2cm}
\partial \ \equiv \ d \ x^{\ \mu} \ \partial_{\ x \ \mu}
\vspace*{0.1cm} \\
\left ( \ {\cal{W}}^{\ (2)} \ \right )_{\ \alpha \beta}
\ = \ \frac{1}{2} \ W^{\ r}_{\ \mu \nu} \ \left ( \ d_{\ r} 
\ \right )_{\ \alpha \beta} \ d \ x^{\ \mu} \ \wedge \ d \ x^{\ \nu} 
\vspace*{0.1cm} \\
d_{\ r} \ \in \ Lie \ ( \ {\cal{D}} \ )
\hspace*{0.2cm} \rightarrow
\vspace*{0.1cm} \\
{\cal{W}}^{\ (2)}_{\ \mu \nu} \ = \ \partial_{\ \mu} 
\ {\cal{W}}^{\ (1)}_{\ \nu}
\ - \ \partial_{\ \nu} \ {\cal{W}}^{\ (1)}_{\ \mu}
\ + \ \left \lbrack \ {\cal{W}}^{\ (1)}_{\ \mu} \ , \ {\cal{W}}^{\ (1)}_{\ \nu} 
\ \right \rbrack
\vspace*{0.1cm} \\
W^{\ r}_{\ \mu \nu} \ = \ - \ W^{\ r}_{\ \nu \mu} \ =
\ \partial_{\ \mu} \ W^{\ r}_{\ \nu} \ - \ \partial_{\ \nu} \ W^{\ r}_{\ \mu}
\ + \ f_{r p q} \ W^{\ p}_{\ \mu} \ W^{\ q}_{\ \nu}
\vspace*{0.2cm} \\ \hline \vspace*{-0.1cm}
{\cal{W}}^{\ (2)} \ ( \ {\cal{D}} \ ) \ \equiv \ {\cal{B}}^{\ (2)} 
\ ( \ {\cal{D}} \ )
\hspace*{0.2cm} ; \hspace*{0.2cm}
W^{\ r}_{\ \mu \nu} \ \equiv \ B^{\ r}_{\ \mu \nu}
\hspace*{0.1cm} \mbox{\begin{tabular}{c} 
components of field
\vspace*{-0.1cm} \\
strengths independent 
\vspace*{-0.1cm} \\
of $\ {\cal{D}} \ $
\end{tabular}
} 
\end{array}
\end{equation}

\noindent
The electromagnetic counterparts are ( eqs. \ref{eq:3-34} , \ref{eq:3-35} )

\vspace*{-0.3cm}
\begin{equation}
\label{eq:3-37}
\begin{array}{l}
{\cal{W}}^{\ em}_{\ \mu} \ = \ W^{\ em}_{\ \mu} \ d_{\ em} \ ( \ f \ )
\vspace*{0.1cm} \\
\left ( \ d_{\ em} \ \right )_{\ \alpha' \alpha}\ = \ \frac{1}{i} 
\ Q_{\ f_{\ \alpha}} \ \delta_{\ \alpha' \alpha}
\hspace*{0.2cm} ; \hspace*{0.2cm}
\left \lbrack \ d_{\ em} \ , \ d_{\ r} \ \left ( \ \bigcup 
\ {\cal{D}} \ ( \ f \ ) \ \right ) 
\ \right \rbrack \ = \ 0
\vspace*{0.2cm} \\
W_{\ \mu \nu}^{\ em} \ \equiv \ B^{\ em}_{\ \mu \nu}
\ = \ \partial_{\ \mu} \ W_{\ \nu}^{\ em}
\ - \ \partial_{\ \nu} \ W_{\ \mu}^{\ em} 
\vspace*{0.2cm} \\ \hline \vspace*{-0.3cm} \\
L^{\ em} \ = \ - \ \frac{1}{4 \ e^{\ 2}} \ B_{\ \mu \nu}^{\ em} 
\ B^{\ \mu \nu \ em}
\ + \ {\cal{L}}_{\ \left \lbrace f \right \rbrace}
\end{array}
\end{equation}

}



{\color{blue} 

\noindent
The electromagnetic Euler-Lagrange equations extending eqs. 
\ref{eq:3-11}, \ref{eq:3-12},
\ref{eq:3-17}, \ref{eq:3-19} and \ref{eq:3-28} become

\vspace*{-0.2cm}
\begin{equation}
\label{eq:3-38}
\begin{array}{c}
\partial_{\ \varrho}
\ \left \lbrace 
\begin{array}{c}
\delta \ {\cal{L}}
\vspace*{0.2cm} \\
\hline  \vspace*{-0.3cm} \\
\delta \ B_{\ \mu \ \nu}^{\ em} 
\end{array} 
\hspace*{0.1cm} \begin{array}{c}
\delta \ B_{\ \mu \nu}^{\ em}
\vspace*{0.2cm} \\
\hline  \vspace*{-0.3cm} \\
\delta \ \left ( \ \partial_{\ \varrho} \ W_{\ \sigma}^{\ em} \ \right )
\end{array} \right \rbrace
\ = \ \begin{array}{c}
\delta \ {\cal{L}}_{\ \left \lbrace f \right \rbrace}
\vspace*{0.2cm} \\
\hline  \vspace*{-0.3cm} \\
\delta \ W_{\ \sigma}^{\ em}
\end{array}
\ = \ j^{\ \sigma \ em}
\vspace*{0.3cm} \\
\begin{array}{c}
\delta \ {\cal{L}}
\vspace*{0.2cm} \\
\hline  \vspace*{-0.3cm} \\
\delta \ B_{\ \mu \ \nu}^{\ em}
\end{array}
\ = \ \begin{array}{c}
1
\vspace*{0.2cm} \\
\hline  \vspace*{-0.3cm} \\
2 \ e^{\ 2}
\end{array} \hspace*{0.1cm} B^{\ \nu \mu \ em}
\hspace*{0.2cm} ; \hspace*{0.2cm}
\begin{array}{c}
\delta \ B_{\ \mu \nu}^{\ em}
\vspace*{0.2cm} \\
\hline  \vspace*{-0.3cm} \\
\delta \ \left ( \ \partial_{\ \varrho} \ W_{\ \sigma}^{\ em} \ \right )
\end{array} \ =
\ \delta_{\ \mu}^{\ \varrho} 
\ \delta_{\ \nu}^{\ \sigma} \ - \ \delta_{\ \nu}^{\ \varrho}
\ \delta_{\ \mu}^{\ \sigma} 
\end{array}
\end{equation}

\noindent
The elecromagnetic Euler-Lagrange equations, i.e. the inhomogeneous \\
Maxwell equations, thus take the form

\vspace*{-0.3cm}
\begin{equation}
\label{eq:3-39}
\begin{array}{l}
\partial_{\ \varrho} \ E \ B^{\ \sigma \varrho \ em}
\ = \ j^{\ \sigma \ em} \ = \ \sum_{\ \alpha} 
\ \overline{f}_{\ \alpha} \ \gamma^{\ \sigma} \ Q_{\ f_{\ \alpha}} 
\ f_{\ \alpha}
\vspace*{0.1cm} \\
E \ = \ 1 \ / \ e^{\ 2}
\end{array}
\end{equation}

\noindent
We are now in the position to make contact with the 
$\ dim \ \left \lbrack \ M^{\ 4} \ \right \rbrack \ $ equations for QCD
-- eq. \ref{eq:3-20} repeated below

\vspace*{-0.3cm}
\begin{equation}
\label{eq:3-40}
\begin{array}{l}
G \ = \ g^{\ -2} 
\vspace*{0.1cm} \\
D_{\ \tau} \ D_{\ \varrho}
\ \left \lbrace
\ G \ B_{\ \sigma}^{\hspace*{0.2cm} \varrho} \ \right \rbrace
\ - \ D_{\ \sigma} \ D_{\ \varrho}
\ \left \lbrace
\ G \ B_{\ \tau}^{\hspace*{0.2cm} \varrho} \ \right \rbrace
\ = \ D_{\ \tau} \ j_{\ \sigma} \ - \ D_{\ \sigma} \ j_{\ \tau}
\end{array}
\end{equation}

\noindent
The QED analogous equations are derived from the equations of motion \\
( eq. \ref{eq:3-39} )

\vspace*{-0.3cm}
\begin{equation}
\label{eq:3-41}
\begin{array}{l}
\partial^{\ \varrho} \ \left ( \ \partial_{\ \tau} 
\ B_{\ \sigma \varrho}^{\ em} \ - \ \partial_{\ \sigma} 
\ B_{\ \tau \varrho}^{\ em}
\ \right ) \ = e^{\ 2} \ \left ( \ \partial_{\ \tau} \ j_{\ \sigma}^{\ em}
\ - \ \partial_{\ \sigma} \ j_{\ \tau}^{\ em} \ \right )
\end{array}
\end{equation}

\noindent
The essential difference between the nonabelian chromo-dipole equations
( eq. \ref{eq:3-40} ) and their $\ U1^{\ em} \ $ counterparts 
( eq. \ref{eq:3-41} ) is that the former are $\ SU3_{\ c} \ $ covariant,
both sides of eq. \ref{eq:3-40} transforming according to the 
adjoint- ( i.e. octet- ) representation of the local $\ SU3_{\ c} \ $ 
gauge group, whereas 
the electromagnetic counterpart as displayed in eq. \ref{eq:3-41} is 
gauge invariant under the local $\ U1^{\ em} \ $ gauge group, and thus
independent of the electromagnetic potentials 
$\ W_{\ \mu}^{\ em} \ \equiv \ - \ w_{\ \mu}^{\ em} \ $.

\noindent
Nevertheless in both cases the homogeneous Bianchi identities for 
( covariant ) derivatives of the field strength are used as analog
identities to the case of QCD in eq. \ref{eq:3-26}, 
repeated below

\vspace*{-0.3cm}
\begin{equation}
\label{eq:3-42}
\begin{array}{l}
D_{\ \tau} \ B_{\ \sigma \varrho} \ -
\ D_{\ \sigma} \ B_{\ \tau \varrho} \ =
\ D_{\ \tau} \ B_{\ \sigma \varrho} \ +
\ D_{\ \sigma} \ B_{\ \varrho \tau} \ = 
\ - \ D_{\ \varrho} \ B_{\ \tau \sigma}
\hspace*{0.2cm} \longrightarrow
\vspace*{0.1cm} \\
\mbox{Bianchi identity :} \hspace*{0.25cm}
D_{\ \tau} \ B_{\ \sigma \varrho} \ +
\ D_{\ \sigma} \ B_{\ \varrho \tau} \ +
\ D_{\ \varrho} \ B_{\ \tau \sigma} \ = \ 0
\end{array}
\end{equation}

\noindent
and the QED ( abelian ) analog Bianchi identity , i.e. the homogeneous Maxwell
equations

\vspace*{-0.3cm}
\begin{equation}
\label{eq:3-43}
\begin{array}{l}
\partial_{\ \tau} \ B_{\ \sigma \varrho}^{\ em} \ -
\ \partial_{\ \sigma} \ B_{\ \tau \varrho}^{\ em} \ =
\ \partial_{\ \tau} \ B_{\ \sigma \varrho}^{\ em} \ +
\ \partial_{\ \sigma} \ B_{\ \varrho \tau}^{\ em} \ = 
\ - \ \partial_{\ \varrho} \ B_{\ \tau \sigma}^{\ em}
\hspace*{0.2cm} \longrightarrow
\vspace*{0.1cm} \\
\mbox{Bianchi identity :} \hspace*{0.25cm}
\partial_{\ \tau} \ B_{\ \sigma \varrho}^{\ em} \ +
\ \partial_{\ \sigma} \ B_{\ \varrho \tau}^{\ em} \ +
\ \partial_{\ \varrho} \ B_{\ \tau \sigma}^{\ em} \ = \ 0
\end{array}
\end{equation}

\noindent
As in the pair of eqations \ref{eq:3-40} ( $\ SU3_{\ c} \ $ ) and
\ref{eq:3-41} ( $\ U1^{\ em} \ $ ) , eq. \ref{eq:3-42} is gauge covariant
with respect to tho local $\ SU3_{\ c} \ $ gauge group, whereas
eq. \ref{eq:3-43} is gauge invariant with respect to the local
$\ U1^{\ em} \ $ gauge group .

\noindent
Substituting eq. \ref{eq:3-43} in eq. \ref{eq:3-41} we obtain 
the electromagnetic dipole equations for the electrodynamic field strengths

\vspace*{-0.3cm}
\begin{equation}
\label{eq:3-44}
\begin{array}{l}
\Box \ B_{\ \sigma \ \tau}^{\ em} 
\ = e^{\ 2} \ \left ( \ \partial_{\ \tau} \ j_{\ \sigma}^{\ em}
\ - \ \partial_{\ \sigma} \ j_{\ \tau}^{\ em} \ \right )
\vspace*{0.1cm} \\
\Box \ = \ \partial^{\ \varrho} \ \partial_{\ \varrho}
\end{array}
\end{equation}

\noindent
Eq. \ref{eq:3-44} is readily compared with its nonabelian analog ,
eq. \ref{eq:3-27} , repeated below

\vspace*{-0.5cm}
\begin{equation}
\label{eq:3-45}
\begin{array}{l}
\left \lbrack 
 \begin{array}{c}
D^{\ \varrho} \ D_{\ \varrho} \ B_{\ \sigma \tau} 
\ +
\vspace*{0.1cm} \\
+ \ {\cal{B}}_{\ \tau \varrho} \ B_{\ \sigma}^{\hspace*{0.2cm} 
\varrho}
\ - \ {\cal{B}}_{\ \sigma \varrho} \ B_{\ \tau}^{\hspace*{0.2cm} \varrho}
\end{array} \right \rbrack^{\ s}
\ =
\ g^{\ 2} \ \left ( \ D_{\ \tau} \ j_{\ \sigma} 
\ - \ D_{\ \sigma} \ j_{\ \tau} \ \right )^{\ s}
\hspace*{0.1cm} ; \hspace*{0.1cm}
\vspace*{0.2cm} \\ \hline \vspace*{-0.3cm} \\
\begin{array}[t]{c} ^{s} \ = \ 1,\cdots,8
\mbox{ : color octet label}
\end{array}
\end{array}
\end{equation}

\noindent
We conclude this section repeating the definitions in eq. \ref{eq:3-13}
relative to the adjoint covariant derivative and the field strength components
pertaining to the adjoint representation , appearing in shorthand notation in
eqs. \ref{eq:3-27} and \ref{eq:3-45}

\vspace*{-0.3cm}
\begin{equation}
\label{eq:3-46}
\begin{array}{l}
\left ( \ D_{\ \varrho} \ ( \ ad \ ) \ \right )_{\ s r} \ =
\ \partial_{\ \varrho} \ \delta_{\ s r} \ +
\ W_{\ \varrho}^{\ t} \ \left ( \ ad_{\ t} \ \right )_{\ s r}
\vspace*{0.1cm} \\
B_{\ \sigma \tau} \ = \ \left ( B_{\ \sigma \tau} \ \right )^{\ t} \ = 
\ B_{\ \sigma \tau}^{\ t}
\vspace*{0.1cm} \\
\left ( \ {\cal{B}}_{\ \sigma \tau} \ \right )_{\ s r} \ = 
\ B_{\ \sigma \tau}^{\ t} \ \left ( \ ad_{\ t} \ \right )_{\ sr}
\hspace*{0.2cm} ; \hspace*{0.2cm} \left ( \ ad_{\ t} \ \right )_{\ sr}
\ = \ f_{\ s t r}
\end{array}
\end{equation}

\noindent
The chromo-dipole equation ( eqs. \ref{eq:3-45} - \ref{eq:3-46} ) was derived
-- but considering the Lagrangean 
$\ {\cal{L}} \ \sim \ \frac{1}{4} \ B_{\ \mu \nu}^{\ s} \ B^{\ \mu \nu \ s} \ $
{\it not} \ $\ \overline{{\cal{L}}} \ $ as given in eq. \ref{eq:3I-21} --
only recently  by the author of these notes. 
To my knowledge it represents a new element.

}



{\color{blue}

}

\end{document}